\documentclass[twocolumn,traditabstract,longauth]{aa}
\usepackage{graphicx}
\usepackage{amsmath,amsfonts,amssymb}
\usepackage{txfonts}
\usepackage[breaklinks,colorlinks,citecolor=blue]{hyperref}
\usepackage{color}
\usepackage{fixltx2e}
\usepackage{natbib}
\usepackage{caption,subcaption}
\bibpunct{(}{)}{;}{a}{}{,}

\def\setsymbol#1#2{\expandafter\def\csname #1\endcsname{#2}}
\def\getsymbol#1{\csname #1\endcsname}

\def\Planck{\textit{Planck}}

\newbox\tablebox    \newdimen\tablewidth
\def\leaderfil{\leaders\hbox to 5pt{\hss.\hss}\hfil}
\def\endPlancktablewide{\tablewidth=\textwidth 
    $$\hss\copy\tablebox\hss$$
    \vskip-\lastskip\vskip -2pt}
\def\tablenote#1 #2\par{\begingroup \parindent=0.8em
    \abovedisplayshortskip=0pt\belowdisplayshortskip=0pt
    \noindent
    $$\hss\vbox{\hsize\tablewidth \hangindent=\parindent \hangafter=1 \noindent
    \hbox to \parindent{$^#1$\hss}\strut#2\strut\par}\hss$$
    \endgroup}
\def\doubleline{\vskip 3pt\hrule \vskip 1.5pt \hrule \vskip 5pt}

\def\L2{\ifmmode L_2\else $L_2$\fi}

\def\DeltaT{\ifmmode \Delta T\else $\Delta T$\fi}
\def\deltat{\ifmmode \Delta t\else $\Delta t$\fi}
\def\fknee{\ifmmode f_{\rm knee}\else $f_{\rm knee}$\fi}
\def\Fmax{\ifmmode F_{\rm max}\else $F_{\rm max}$\fi}
\def\solar{\ifmmode{\rm M}_{\mathord\odot}\else${\rm M}_{\mathord\odot}$\fi}
\def\Msolar{\ifmmode{\rm M}_{\mathord\odot}\else${\rm M}_{\mathord\odot}$\fi}
\def\Lsolar{\ifmmode{\rm L}_{\mathord\odot}\else${\rm L}_{\mathord\odot}$\fi}

\def\inv{\ifmmode^{-1}\else$^{-1}$\fi}
\def\mo{\ifmmode^{-1}\else$^{-1}$\fi}
\def\sup#1{\ifmmode ^{\rm #1}\else $^{\rm #1}$\fi}
\def\expo#1{\ifmmode \times 10^{#1}\else $\times 10^{#1}$\fi}
\def\,{\thinspace}
\def\lsim{\mathrel{\raise .4ex\hbox{\rlap{$<$}\lower 1.2ex\hbox{$\sim$}}}}
\def\gsim{\mathrel{\raise .4ex\hbox{\rlap{$>$}\lower 1.2ex\hbox{$\sim$}}}}

\def\simprop{\mathrel{\raise .4ex\hbox{\rlap{$\propto$}\lower 1.2ex\hbox{$\sim$}}}}
\def\deg{\ifmmode^\circ\else$^\circ$\fi}
\def\pdeg{\ifmmode $\setbox0=\hbox{$^{\circ}$}\rlap{\hskip.11\wd0 .}$^{\circ}
          \else \setbox0=\hbox{$^{\circ}$}\rlap{\hskip.11\wd0 .}$^{\circ}$\fi}
\def\arcs{\ifmmode {^{\scriptstyle\prime\prime}}
          \else $^{\scriptstyle\prime\prime}$\fi}
\def\arcm{\ifmmode {^{\scriptstyle\prime}}
          \else $^{\scriptstyle\prime}$\fi}
\newdimen\sa  \newdimen\sb
\def\parcs{\sa=.07em \sb=.03em
     \ifmmode \hbox{\rlap{.}}^{\scriptstyle\prime\kern -\sb\prime}\hbox{\kern -\sa}
     \else \rlap{.}$^{\scriptstyle\prime\kern -\sb\prime}$\kern -\sa\fi}
\def\parcm{\sa=.08em \sb=.03em
     \ifmmode \hbox{\rlap{.}\kern\sa}^{\scriptstyle\prime}\hbox{\kern-\sb}
     \else \rlap{.}\kern\sa$^{\scriptstyle\prime}$\kern-\sb\fi}
\def\ra[#1 #2 #3.#4]{#1\sup{h}#2\sup{m}#3\sup{s}\llap.#4}
\def\dec[#1 #2 #3.#4]{#1\deg#2\arcm#3\arcs\llap.#4}
\def\deco[#1 #2 #3]{#1\deg#2\arcm#3\arcs}
\def\rra[#1 #2]{#1\sup{h}#2\sup{m}}

\def\dots{\relax\ifmmode \ldots\else $\ldots$\fi}
\def\WHzsr{\ifmmode $W\,Hz\mo\,sr\mo$\else W\,Hz\mo\,sr\mo\fi}
\def\mHz{\ifmmode $\,mHz$\else \,mHz\fi}
\def\GHz{\ifmmode $\,GHz$\else \,GHz\fi}
\def\mKs{\ifmmode $\,mK\,s$^{1/2}\else \,mK\,s$^{1/2}$\fi}
\def\muKs{\ifmmode \,\mu$K\,s$^{1/2}\else \,$\mu$K\,s$^{1/2}$\fi}
\def\muKRJs{\ifmmode \,\mu$K$_{\rm RJ}$\,s$^{1/2}\else \,$\mu$K$_{\rm RJ}$\,s$^{1/2}$\fi}
\def\muKHz{\ifmmode \,\mu$K\,Hz$^{-1/2}\else \,$\mu$K\,Hz$^{-1/2}$\fi}
\def\MJysr{\ifmmode \,$MJy\,sr\mo$\else \,MJy\,sr\mo\fi}
\def\MJysrmK{\ifmmode \,$MJy\,sr\mo$\,mK$_{\rm CMB}\mo\else \,MJy\,sr\mo\,mK$_{\rm CMB}\mo$\fi}
\def\microns{\ifmmode \,\mu$m$\else \,$\mu$m\fi}
\def\micron{\microns}
\def\muK{\ifmmode \,\mu$K$\else \,$\mu$\hbox{K}\fi}
\def\microK{\ifmmode \,\mu$K$\else \,$\mu$\hbox{K}\fi}
\def\muW{\ifmmode \,\mu$W$\else \,$\mu$\hbox{W}\fi}
\def\kms{\ifmmode $\,km\,s$^{-1}\else \,km\,s$^{-1}$\fi}
\def\kmsMpc{\ifmmode $\,\kms\,Mpc\mo$\else \,\kms\,Mpc\mo\fi}
\setsymbol{LFI:center:frequency:70GHz:units}{70.3\,GHz}
\setsymbol{LFI:center:frequency:44GHz:units}{44.1\,GHz}
\setsymbol{LFI:center:frequency:30GHz:units}{28.5\,GHz}

\setsymbol{LFI:center:frequency:70GHz}{70.3}
\setsymbol{LFI:center:frequency:44GHz}{44.1}
\setsymbol{LFI:center:frequency:30GHz}{28.5}

\setsymbol{LFI:center:frequency:LFI18:Rad:M:units}{71.7\GHz}
\setsymbol{LFI:center:frequency:LFI19:Rad:M:units}{67.5\GHz}
\setsymbol{LFI:center:frequency:LFI20:Rad:M:units}{69.2\GHz}
\setsymbol{LFI:center:frequency:LFI21:Rad:M:units}{70.4\GHz}
\setsymbol{LFI:center:frequency:LFI22:Rad:M:units}{71.5\GHz}
\setsymbol{LFI:center:frequency:LFI23:Rad:M:units}{70.8\GHz}
\setsymbol{LFI:center:frequency:LFI24:Rad:M:units}{44.4\GHz}
\setsymbol{LFI:center:frequency:LFI25:Rad:M:units}{44.0\GHz}
\setsymbol{LFI:center:frequency:LFI26:Rad:M:units}{43.9\GHz}
\setsymbol{LFI:center:frequency:LFI27:Rad:M:units}{28.3\GHz}
\setsymbol{LFI:center:frequency:LFI28:Rad:M:units}{28.8\GHz}
\setsymbol{LFI:center:frequency:LFI18:Rad:S:units}{70.1\GHz}
\setsymbol{LFI:center:frequency:LFI19:Rad:S:units}{69.6\GHz}
\setsymbol{LFI:center:frequency:LFI20:Rad:S:units}{69.5\GHz}
\setsymbol{LFI:center:frequency:LFI21:Rad:S:units}{69.5\GHz}
\setsymbol{LFI:center:frequency:LFI22:Rad:S:units}{72.8\GHz}
\setsymbol{LFI:center:frequency:LFI23:Rad:S:units}{71.3\GHz}
\setsymbol{LFI:center:frequency:LFI24:Rad:S:units}{44.1\GHz}
\setsymbol{LFI:center:frequency:LFI25:Rad:S:units}{44.1\GHz}
\setsymbol{LFI:center:frequency:LFI26:Rad:S:units}{44.1\GHz}
\setsymbol{LFI:center:frequency:LFI27:Rad:S:units}{28.5\GHz}
\setsymbol{LFI:center:frequency:LFI28:Rad:S:units}{28.2\GHz}

\setsymbol{LFI:center:frequency:LFI18:Rad:M}{71.7}
\setsymbol{LFI:center:frequency:LFI19:Rad:M}{67.5}
\setsymbol{LFI:center:frequency:LFI20:Rad:M}{69.2}
\setsymbol{LFI:center:frequency:LFI21:Rad:M}{70.4}
\setsymbol{LFI:center:frequency:LFI22:Rad:M}{71.5}
\setsymbol{LFI:center:frequency:LFI23:Rad:M}{70.8}
\setsymbol{LFI:center:frequency:LFI24:Rad:M}{44.4}
\setsymbol{LFI:center:frequency:LFI25:Rad:M}{44.0}
\setsymbol{LFI:center:frequency:LFI26:Rad:M}{43.9}
\setsymbol{LFI:center:frequency:LFI27:Rad:M}{28.3}
\setsymbol{LFI:center:frequency:LFI28:Rad:M}{28.8}
\setsymbol{LFI:center:frequency:LFI18:Rad:S}{70.1}
\setsymbol{LFI:center:frequency:LFI19:Rad:S}{69.6}
\setsymbol{LFI:center:frequency:LFI20:Rad:S}{69.5}
\setsymbol{LFI:center:frequency:LFI21:Rad:S}{69.5}
\setsymbol{LFI:center:frequency:LFI22:Rad:S}{72.8}
\setsymbol{LFI:center:frequency:LFI23:Rad:S}{71.3}
\setsymbol{LFI:center:frequency:LFI24:Rad:S}{44.1}
\setsymbol{LFI:center:frequency:LFI25:Rad:S}{44.1}
\setsymbol{LFI:center:frequency:LFI26:Rad:S}{44.1}
\setsymbol{LFI:center:frequency:LFI27:Rad:S}{28.5}
\setsymbol{LFI:center:frequency:LFI28:Rad:S}{28.2}

\setsymbol{LFI:white:noise:sensitivity:70GHz:units}{134.7\muKs}
\setsymbol{LFI:white:noise:sensitivity:44GHz:units}{164.7\muKs}
\setsymbol{LFI:white:noise:sensitivity:30GHz:units}{143.4\muKs}

\setsymbol{LFI:white:noise:sensitivity:70GHz}{134.7}
\setsymbol{LFI:white:noise:sensitivity:44GHz}{164.7}
\setsymbol{LFI:white:noise:sensitivity:30GHz}{143.4}

\setsymbol{LFI:white:noise:sensitivity:LFI18:Rad:M:units}{512.0\muKs}
\setsymbol{LFI:white:noise:sensitivity:LFI19:Rad:M:units}{581.4\muKs}
\setsymbol{LFI:white:noise:sensitivity:LFI20:Rad:M:units}{590.8\muKs}
\setsymbol{LFI:white:noise:sensitivity:LFI21:Rad:M:units}{455.2\muKs}
\setsymbol{LFI:white:noise:sensitivity:LFI22:Rad:M:units}{492.0\muKs}
\setsymbol{LFI:white:noise:sensitivity:LFI23:Rad:M:units}{507.7\muKs}
\setsymbol{LFI:white:noise:sensitivity:LFI24:Rad:M:units}{462.2\muKs}
\setsymbol{LFI:white:noise:sensitivity:LFI25:Rad:M:units}{413.6\muKs}
\setsymbol{LFI:white:noise:sensitivity:LFI26:Rad:M:units}{478.6\muKs}
\setsymbol{LFI:white:noise:sensitivity:LFI27:Rad:M:units}{277.7\muKs}
\setsymbol{LFI:white:noise:sensitivity:LFI28:Rad:M:units}{312.3\muKs}
\setsymbol{LFI:white:noise:sensitivity:LFI18:Rad:S:units}{465.7\muKs}
\setsymbol{LFI:white:noise:sensitivity:LFI19:Rad:S:units}{555.6\muKs}
\setsymbol{LFI:white:noise:sensitivity:LFI20:Rad:S:units}{623.2\muKs}
\setsymbol{LFI:white:noise:sensitivity:LFI21:Rad:S:units}{564.1\muKs}
\setsymbol{LFI:white:noise:sensitivity:LFI22:Rad:S:units}{534.4\muKs}
\setsymbol{LFI:white:noise:sensitivity:LFI23:Rad:S:units}{542.4\muKs}
\setsymbol{LFI:white:noise:sensitivity:LFI24:Rad:S:units}{399.2\muKs}
\setsymbol{LFI:white:noise:sensitivity:LFI25:Rad:S:units}{392.6\muKs}
\setsymbol{LFI:white:noise:sensitivity:LFI26:Rad:S:units}{418.6\muKs}
\setsymbol{LFI:white:noise:sensitivity:LFI27:Rad:S:units}{302.9\muKs}
\setsymbol{LFI:white:noise:sensitivity:LFI28:Rad:S:units}{285.3\muKs}

\setsymbol{LFI:white:noise:sensitivity:LFI18:Rad:M}{512.0}
\setsymbol{LFI:white:noise:sensitivity:LFI19:Rad:M}{581.4}
\setsymbol{LFI:white:noise:sensitivity:LFI20:Rad:M}{590.8}
\setsymbol{LFI:white:noise:sensitivity:LFI21:Rad:M}{455.2}
\setsymbol{LFI:white:noise:sensitivity:LFI22:Rad:M}{492.0}
\setsymbol{LFI:white:noise:sensitivity:LFI23:Rad:M}{507.7}
\setsymbol{LFI:white:noise:sensitivity:LFI24:Rad:M}{462.2}
\setsymbol{LFI:white:noise:sensitivity:LFI25:Rad:M}{413.6}
\setsymbol{LFI:white:noise:sensitivity:LFI26:Rad:M}{478.6}
\setsymbol{LFI:white:noise:sensitivity:LFI27:Rad:M}{277.7}
\setsymbol{LFI:white:noise:sensitivity:LFI28:Rad:M}{312.3}
\setsymbol{LFI:white:noise:sensitivity:LFI18:Rad:S}{465.7}
\setsymbol{LFI:white:noise:sensitivity:LFI19:Rad:S}{555.6}
\setsymbol{LFI:white:noise:sensitivity:LFI20:Rad:S}{623.2}
\setsymbol{LFI:white:noise:sensitivity:LFI21:Rad:S}{564.1}
\setsymbol{LFI:white:noise:sensitivity:LFI22:Rad:S}{534.4}
\setsymbol{LFI:white:noise:sensitivity:LFI23:Rad:S}{542.4}
\setsymbol{LFI:white:noise:sensitivity:LFI24:Rad:S}{399.2}
\setsymbol{LFI:white:noise:sensitivity:LFI25:Rad:S}{392.6}
\setsymbol{LFI:white:noise:sensitivity:LFI26:Rad:S}{418.6}
\setsymbol{LFI:white:noise:sensitivity:LFI27:Rad:S}{302.9}
\setsymbol{LFI:white:noise:sensitivity:LFI28:Rad:S}{285.3}

\setsymbol{LFI:knee:frequency:70GHz:units}{29.5\mHz}
\setsymbol{LFI:knee:frequency:44GHz:units}{56.2\mHz}
\setsymbol{LFI:knee:frequency:30GHz:units}{113.7\mHz}

\setsymbol{LFI:knee:frequency:70GHz}{29.5}
\setsymbol{LFI:knee:frequency:44GHz}{56.2}
\setsymbol{LFI:knee:frequency:30GHz}{113.7}

\setsymbol{LFI:knee:frequency:LFI18:Rad:M:units}{16.3\mHz}
\setsymbol{LFI:knee:frequency:LFI19:Rad:M:units}{15.1\mHz}
\setsymbol{LFI:knee:frequency:LFI20:Rad:M:units}{18.7\mHz}
\setsymbol{LFI:knee:frequency:LFI21:Rad:M:units}{37.2\mHz}
\setsymbol{LFI:knee:frequency:LFI22:Rad:M:units}{12.7\mHz}
\setsymbol{LFI:knee:frequency:LFI23:Rad:M:units}{34.6\mHz}
\setsymbol{LFI:knee:frequency:LFI24:Rad:M:units}{46.2\mHz}
\setsymbol{LFI:knee:frequency:LFI25:Rad:M:units}{24.9\mHz}
\setsymbol{LFI:knee:frequency:LFI26:Rad:M:units}{67.6\mHz}
\setsymbol{LFI:knee:frequency:LFI27:Rad:M:units}{187.4\mHz}
\setsymbol{LFI:knee:frequency:LFI28:Rad:M:units}{122.2\mHz}
\setsymbol{LFI:knee:frequency:LFI18:Rad:S:units}{17.7\mHz}
\setsymbol{LFI:knee:frequency:LFI19:Rad:S:units}{22.0\mHz}
\setsymbol{LFI:knee:frequency:LFI20:Rad:S:units}{8.7\mHz}
\setsymbol{LFI:knee:frequency:LFI21:Rad:S:units}{25.9\mHz}
\setsymbol{LFI:knee:frequency:LFI22:Rad:S:units}{15.8\mHz}
\setsymbol{LFI:knee:frequency:LFI23:Rad:S:units}{129.8\mHz}
\setsymbol{LFI:knee:frequency:LFI24:Rad:S:units}{100.9\mHz}
\setsymbol{LFI:knee:frequency:LFI25:Rad:S:units}{38.9\mHz}
\setsymbol{LFI:knee:frequency:LFI26:Rad:S:units}{58.9\mHz}
\setsymbol{LFI:knee:frequency:LFI27:Rad:S:units}{104.4\mHz}
\setsymbol{LFI:knee:frequency:LFI28:Rad:S:units}{40.7\mHz}

\setsymbol{LFI:knee:frequency:LFI18:Rad:M}{16.3}
\setsymbol{LFI:knee:frequency:LFI19:Rad:M}{15.1}
\setsymbol{LFI:knee:frequency:LFI20:Rad:M}{18.7}
\setsymbol{LFI:knee:frequency:LFI21:Rad:M}{37.2}
\setsymbol{LFI:knee:frequency:LFI22:Rad:M}{12.7}
\setsymbol{LFI:knee:frequency:LFI23:Rad:M}{34.6}
\setsymbol{LFI:knee:frequency:LFI24:Rad:M}{46.2}
\setsymbol{LFI:knee:frequency:LFI25:Rad:M}{24.9}
\setsymbol{LFI:knee:frequency:LFI26:Rad:M}{67.6}
\setsymbol{LFI:knee:frequency:LFI27:Rad:M}{187.4}
\setsymbol{LFI:knee:frequency:LFI28:Rad:M}{122.2}
\setsymbol{LFI:knee:frequency:LFI18:Rad:S}{17.7}
\setsymbol{LFI:knee:frequency:LFI19:Rad:S}{22.0}
\setsymbol{LFI:knee:frequency:LFI20:Rad:S}{8.7}
\setsymbol{LFI:knee:frequency:LFI21:Rad:S}{25.9}
\setsymbol{LFI:knee:frequency:LFI22:Rad:S}{15.8}
\setsymbol{LFI:knee:frequency:LFI23:Rad:S}{129.8}
\setsymbol{LFI:knee:frequency:LFI24:Rad:S}{100.9}
\setsymbol{LFI:knee:frequency:LFI25:Rad:S}{38.9}
\setsymbol{LFI:knee:frequency:LFI26:Rad:S}{58.9}
\setsymbol{LFI:knee:frequency:LFI27:Rad:S}{104.4}
\setsymbol{LFI:knee:frequency:LFI28:Rad:S}{40.7}

\setsymbol{LFI:slope:70GHz:units}{$-1.03$\mHz}
\setsymbol{LFI:slope:44GHz:units}{$-0.89$\mHz}
\setsymbol{LFI:slope:30GHz:units}{$-0.87$\mHz}

\setsymbol{LFI:slope:70GHz}{$-1.03$}
\setsymbol{LFI:slope:44GHz}{$-0.89$}
\setsymbol{LFI:slope:30GHz}{$-0.87$}

\setsymbol{LFI:slope:LFI18:Rad:M:units}{$-1.04$\mHz}
\setsymbol{LFI:slope:LFI19:Rad:M:units}{$-1.09$\mHz}
\setsymbol{LFI:slope:LFI20:Rad:M:units}{$-0.69$\mHz}
\setsymbol{LFI:slope:LFI21:Rad:M:units}{$-1.56$\mHz}
\setsymbol{LFI:slope:LFI22:Rad:M:units}{$-1.01$\mHz}
\setsymbol{LFI:slope:LFI23:Rad:M:units}{$-0.96$\mHz}
\setsymbol{LFI:slope:LFI24:Rad:M:units}{$-0.83$\mHz}
\setsymbol{LFI:slope:LFI25:Rad:M:units}{$-0.91$\mHz}
\setsymbol{LFI:slope:LFI26:Rad:M:units}{$-0.95$\mHz}
\setsymbol{LFI:slope:LFI27:Rad:M:units}{$-0.87$\mHz}
\setsymbol{LFI:slope:LFI28:Rad:M:units}{$-0.88$\mHz}
\setsymbol{LFI:slope:LFI18:Rad:S:units}{$-1.15$\mHz}
\setsymbol{LFI:slope:LFI19:Rad:S:units}{$-1.00$\mHz}
\setsymbol{LFI:slope:LFI20:Rad:S:units}{$-0.95$\mHz}
\setsymbol{LFI:slope:LFI21:Rad:S:units}{$-0.92$\mHz}
\setsymbol{LFI:slope:LFI22:Rad:S:units}{$-1.01$\mHz}
\setsymbol{LFI:slope:LFI23:Rad:S:units}{$-0.95$\mHz}
\setsymbol{LFI:slope:LFI24:Rad:S:units}{$-0.73$\mHz}
\setsymbol{LFI:slope:LFI25:Rad:S:units}{$-1.16$\mHz}
\setsymbol{LFI:slope:LFI26:Rad:S:units}{$-0.79$\mHz}
\setsymbol{LFI:slope:LFI27:Rad:S:units}{$-0.82$\mHz}
\setsymbol{LFI:slope:LFI28:Rad:S:units}{$-0.91$\mHz}

\setsymbol{LFI:slope:LFI18:Rad:M}{$-1.04$}
\setsymbol{LFI:slope:LFI19:Rad:M}{$-1.09$}
\setsymbol{LFI:slope:LFI20:Rad:M}{$-0.69$}
\setsymbol{LFI:slope:LFI21:Rad:M}{$-1.56$}
\setsymbol{LFI:slope:LFI22:Rad:M}{$-1.01$}
\setsymbol{LFI:slope:LFI23:Rad:M}{$-0.96$}
\setsymbol{LFI:slope:LFI24:Rad:M}{$-0.83$}
\setsymbol{LFI:slope:LFI25:Rad:M}{$-0.91$}
\setsymbol{LFI:slope:LFI26:Rad:M}{$-0.95$}
\setsymbol{LFI:slope:LFI27:Rad:M}{$-0.87$}
\setsymbol{LFI:slope:LFI28:Rad:M}{$-0.88$}
\setsymbol{LFI:slope:LFI18:Rad:S}{$-1.15$}
\setsymbol{LFI:slope:LFI19:Rad:S}{$-1.00$}
\setsymbol{LFI:slope:LFI20:Rad:S}{$-0.95$}
\setsymbol{LFI:slope:LFI21:Rad:S}{$-0.92$}
\setsymbol{LFI:slope:LFI22:Rad:S}{$-1.01$}
\setsymbol{LFI:slope:LFI23:Rad:S}{$-0.95$}
\setsymbol{LFI:slope:LFI24:Rad:S}{$-0.73$}
\setsymbol{LFI:slope:LFI25:Rad:S}{$-1.16$}
\setsymbol{LFI:slope:LFI26:Rad:S}{$-0.79$}
\setsymbol{LFI:slope:LFI27:Rad:S}{$-0.82$}
\setsymbol{LFI:slope:LFI28:Rad:S}{$-0.91$}

\setsymbol{LFI:FWHM:70GHz:units}{13\parcm01}
\setsymbol{LFI:FWHM:44GHz:units}{27\parcm92}
\setsymbol{LFI:FWHM:30GHz:units}{32\parcm65}

\setsymbol{LFI:FWHM:70GHz}{13.01}
\setsymbol{LFI:FWHM:44GHz}{27.92}
\setsymbol{LFI:FWHM:30GHz}{32.65}

\setsymbol{LFI:FWHM:LFI18:units}{13\parcm39}
\setsymbol{LFI:FWHM:LFI19:units}{13\parcm01}
\setsymbol{LFI:FWHM:LFI20:units}{12\parcm75}
\setsymbol{LFI:FWHM:LFI21:units}{12\parcm74}
\setsymbol{LFI:FWHM:LFI22:units}{12\parcm87}
\setsymbol{LFI:FWHM:LFI23:units}{13\parcm27}
\setsymbol{LFI:FWHM:LFI24:units}{22\parcm98}
\setsymbol{LFI:FWHM:LFI25:units}{30\parcm46}
\setsymbol{LFI:FWHM:LFI26:units}{30\parcm31}
\setsymbol{LFI:FWHM:LFI27:units}{32\parcm65}
\setsymbol{LFI:FWHM:LFI28:units}{32\parcm66}

\setsymbol{LFI:FWHM:LFI18}{13.39}
\setsymbol{LFI:FWHM:LFI19}{13.01}
\setsymbol{LFI:FWHM:LFI20}{12.75}
\setsymbol{LFI:FWHM:LFI21}{12.74}
\setsymbol{LFI:FWHM:LFI22}{12.87}
\setsymbol{LFI:FWHM:LFI23}{13.27}
\setsymbol{LFI:FWHM:LFI24}{22.98}
\setsymbol{LFI:FWHM:LFI25}{30.46}
\setsymbol{LFI:FWHM:LFI26}{30.31}
\setsymbol{LFI:FWHM:LFI27}{32.65}
\setsymbol{LFI:FWHM:LFI28}{32.66}

\setsymbol{LFI:FWHM:uncertainty:LFI18:units}{0.170\arcm}
\setsymbol{LFI:FWHM:uncertainty:LFI19:units}{0.174\arcm}
\setsymbol{LFI:FWHM:uncertainty:LFI20:units}{0.170\arcm}
\setsymbol{LFI:FWHM:uncertainty:LFI21:units}{0.156\arcm}
\setsymbol{LFI:FWHM:uncertainty:LFI22:units}{0.164\arcm}
\setsymbol{LFI:FWHM:uncertainty:LFI23:units}{0.171\arcm}
\setsymbol{LFI:FWHM:uncertainty:LFI24:units}{0.652\arcm}
\setsymbol{LFI:FWHM:uncertainty:LFI25:units}{1.075\arcm}
\setsymbol{LFI:FWHM:uncertainty:LFI26:units}{1.131\arcm}
\setsymbol{LFI:FWHM:uncertainty:LFI27:units}{1.266\arcm}
\setsymbol{LFI:FWHM:uncertainty:LFI28:units}{1.287\arcm}

\setsymbol{LFI:FWHM:uncertainty:LFI18}{0.170}
\setsymbol{LFI:FWHM:uncertainty:LFI19}{0.174}
\setsymbol{LFI:FWHM:uncertainty:LFI20}{0.170}
\setsymbol{LFI:FWHM:uncertainty:LFI21}{0.156}
\setsymbol{LFI:FWHM:uncertainty:LFI22}{0.164}
\setsymbol{LFI:FWHM:uncertainty:LFI23}{0.171}
\setsymbol{LFI:FWHM:uncertainty:LFI24}{0.652}
\setsymbol{LFI:FWHM:uncertainty:LFI25}{1.075}
\setsymbol{LFI:FWHM:uncertainty:LFI26}{1.131}
\setsymbol{LFI:FWHM:uncertainty:LFI27}{1.266}
\setsymbol{LFI:FWHM:uncertainty:LFI28}{1.287}

\setsymbol{HFI:center:frequency:100GHz:units}{100\,GHz}
\setsymbol{HFI:center:frequency:143GHz:units}{143\,GHz}
\setsymbol{HFI:center:frequency:217GHz:units}{217\,GHz}
\setsymbol{HFI:center:frequency:353GHz:units}{353\,GHz}
\setsymbol{HFI:center:frequency:545GHz:units}{545\,GHz}
\setsymbol{HFI:center:frequency:857GHz:units}{857\,GHz}

\setsymbol{HFI:center:frequency:100GHz}{100}
\setsymbol{HFI:center:frequency:143GHz}{143}
\setsymbol{HFI:center:frequency:217GHz}{217}
\setsymbol{HFI:center:frequency:353GHz}{353}
\setsymbol{HFI:center:frequency:545GHz}{545}
\setsymbol{HFI:center:frequency:857GHz}{857}

\setsymbol{HFI:Ndetectors:100GHz}{8}
\setsymbol{HFI:Ndetectors:143GHz}{11}
\setsymbol{HFI:Ndetectors:217GHz}{12}
\setsymbol{HFI:Ndetectors:353GHz}{12}
\setsymbol{HFI:Ndetectors:545GHz}{3}
\setsymbol{HFI:Ndetectors:857GHz}{4}

\setsymbol{HFI:FWHM:Maps:100GHz:units}{9\parcm88}
\setsymbol{HFI:FWHM:Maps:143GHz:units}{7\parcm18}
\setsymbol{HFI:FWHM:Maps:217GHz:units}{4\parcm87}
\setsymbol{HFI:FWHM:Maps:353GHz:units}{4\parcm65}
\setsymbol{HFI:FWHM:Maps:545GHz:units}{4\parcm72}
\setsymbol{HFI:FWHM:Maps:857GHz:units}{4\parcm39}
\setsymbol{HFI:FWHM:Maps:100GHz}{9.88}
\setsymbol{HFI:FWHM:Maps:143GHz}{7.18}
\setsymbol{HFI:FWHM:Maps:217GHz}{4.87}
\setsymbol{HFI:FWHM:Maps:353GHz}{4.65}
\setsymbol{HFI:FWHM:Maps:545GHz}{4.72}
\setsymbol{HFI:FWHM:Maps:857GHz}{4.39}

\setsymbol{HFI:beam:ellipticity:Maps:100GHz}{1.15}
\setsymbol{HFI:beam:ellipticity:Maps:143GHz}{1.01}
\setsymbol{HFI:beam:ellipticity:Maps:217GHz}{1.06}
\setsymbol{HFI:beam:ellipticity:Maps:353GHz}{1.05}
\setsymbol{HFI:beam:ellipticity:Maps:545GHz}{1.14}
\setsymbol{HFI:beam:ellipticity:Maps:857GHz}{1.19}

\setsymbol{HFI:FWHM:Mars:100GHz:units}{9\parcm37}
\setsymbol{HFI:FWHM:Mars:143GHz:units}{7\parcm04}
\setsymbol{HFI:FWHM:Mars:217GHz:units}{4\parcm68}
\setsymbol{HFI:FWHM:Mars:353GHz:units}{4\parcm43}
\setsymbol{HFI:FWHM:Mars:545GHz:units}{3\parcm80}
\setsymbol{HFI:FWHM:Mars:857GHz:units}{3\parcm67}

\setsymbol{HFI:FWHM:Mars:100GHz}{9.37}
\setsymbol{HFI:FWHM:Mars:143GHz}{7.04}
\setsymbol{HFI:FWHM:Mars:217GHz}{4.68}
\setsymbol{HFI:FWHM:Mars:353GHz}{4.43}
\setsymbol{HFI:FWHM:Mars:545GHz}{3.80}
\setsymbol{HFI:FWHM:Mars:857GHz}{3.67}

\setsymbol{HFI:beam:ellipticity:Mars:100GHz}{1.18}
\setsymbol{HFI:beam:ellipticity:Mars:143GHz}{1.03}
\setsymbol{HFI:beam:ellipticity:Mars:217GHz}{1.14}
\setsymbol{HFI:beam:ellipticity:Mars:353GHz}{1.09}
\setsymbol{HFI:beam:ellipticity:Mars:545GHz}{1.25}
\setsymbol{HFI:beam:ellipticity:Mars:857GHz}{1.03}

\setsymbol{HFI:CMB:relative:calibration:100GHz}{$\lsim 1\%$}
\setsymbol{HFI:CMB:relative:calibration:143GHz}{$\lsim 1\%$}
\setsymbol{HFI:CMB:relative:calibration:217GHz}{$\lsim 1\%$}
\setsymbol{HFI:CMB:relative:calibration:353GHz}{$\lsim 1\%$}
\setsymbol{HFI:CMB:relative:calibration:545GHz}{}
\setsymbol{HFI:CMB:relative:calibration:857GHz}{}

\setsymbol{HFI:CMB:absolute:calibration:100GHz}{$\lsim 2\%$}
\setsymbol{HFI:CMB:absolute:calibration:143GHz}{$\lsim 2\%$}
\setsymbol{HFI:CMB:absolute:calibration:217GHz}{$\lsim 2\%$}
\setsymbol{HFI:CMB:absolute:calibration:353GHz}{$\lsim 2\%$}
\setsymbol{HFI:CMB:absolute:calibration:545GHz}{}
\setsymbol{HFI:CMB:absolute:calibration:857GHz}{}

\setsymbol{HFI:FIRAS:gain:calibration:accuracy:statistical:100GHz}{}
\setsymbol{HFI:FIRAS:gain:calibration:accuracy:statistical:143GHz}{}
\setsymbol{HFI:FIRAS:gain:calibration:accuracy:statistical:217GHz}{}
\setsymbol{HFI:FIRAS:gain:calibration:accuracy:statistical:353GHz}{2.5\%}
\setsymbol{HFI:FIRAS:gain:calibration:accuracy:statistical:545GHz}{1\%}
\setsymbol{HFI:FIRAS:gain:calibration:accuracy:statistical:857GHz}{0.5\%}

\setsymbol{HFI:FIRAS:gain:calibration:accuracy:systematic:100GHz}{}
\setsymbol{HFI:FIRAS:gain:calibration:accuracy:systematic:143GHz}{}
\setsymbol{HFI:FIRAS:gain:calibration:accuracy:systematic:217GHz}{}
\setsymbol{HFI:FIRAS:gain:calibration:accuracy:systematic:353GHz}{}
\setsymbol{HFI:FIRAS:gain:calibration:accuracy:systematic:545GHz}{7\%}
\setsymbol{HFI:FIRAS:gain:calibration:accuracy:systematic:857GHz}{7\%}

\setsymbol{HFI:FIRAS:zero:point:accuracy:100GHz:units}{0.8\MJysr}
\setsymbol{HFI:FIRAS:zero:point:accuracy:143GHz:units}{}
\setsymbol{HFI:FIRAS:zero:point:accuracy:217GHz:units}{}
\setsymbol{HFI:FIRAS:zero:point:accuracy:353GHz:units}{1.4\MJysr}
\setsymbol{HFI:FIRAS:zero:point:accuracy:545GHz:units}{2.2\MJysr}
\setsymbol{HFI:FIRAS:zero:point:accuracy:857GHz:units}{1.7\MJysr}

\setsymbol{HFI:FIRAS:zero:point:accuracy:100GHz}{0.8}
\setsymbol{HFI:FIRAS:zero:point:accuracy:143GHz}{}
\setsymbol{HFI:FIRAS:zero:point:accuracy:217GHz}{}
\setsymbol{HFI:FIRAS:zero:point:accuracy:353GHz}{1.4}
\setsymbol{HFI:FIRAS:zero:point:accuracy:545GHz}{2.2}
\setsymbol{HFI:FIRAS:zero:point:accuracy:857GHz}{1.7}

\setsymbol{HFI:unit:conversion:100GHz:units}{0.2415\MJysrmK}
\setsymbol{HFI:unit:conversion:143GHz:units}{0.3694\MJysrmK}
\setsymbol{HFI:unit:conversion:217GHz:units}{0.4811\MJysrmK}
\setsymbol{HFI:unit:conversion:353GHz:units}{0.2883\MJysrmK}
\setsymbol{HFI:unit:conversion:545GHz:units}{0.05826\MJysrmK}
\setsymbol{HFI:unit:conversion:857GHz:units}{0.002238\MJysrmK}

\setsymbol{HFI:unit:conversion:100GHz}{0.2415}
\setsymbol{HFI:unit:conversion:143GHz}{0.3694}
\setsymbol{HFI:unit:conversion:217GHz}{0.4811}
\setsymbol{HFI:unit:conversion:353GHz}{0.2883}
\setsymbol{HFI:unit:conversion:545GHz}{0.05826}
\setsymbol{HFI:unit:conversion:857GHz}{0.002238}

\setsymbol{HFI:colour:correction:alpha=-2:V1.01:100GHz}{0.9893}
\setsymbol{HFI:colour:correction:alpha=-2:V1.01:143GHz}{0.9759}
\setsymbol{HFI:colour:correction:alpha=-2:V1.01:217GHz}{1.0007}
\setsymbol{HFI:colour:correction:alpha=-2:V1.01:353GHz}{1.0028}
\setsymbol{HFI:colour:correction:alpha=-2:V1.01:545GHz}{1.0019}
\setsymbol{HFI:colour:correction:alpha=-2:V1.01:857GHz}{0.9889}

\setsymbol{HFI:colour:correction:alpha=0:V1.01:100GHz}{1.0008}
\setsymbol{HFI:colour:correction:alpha=0:V1.01:143GHz}{1.0148}
\setsymbol{HFI:colour:correction:alpha=0:V1.01:217GHz}{0.9909}
\setsymbol{HFI:colour:correction:alpha=0:V1.01:353GHz}{0.9888}
\setsymbol{HFI:colour:correction:alpha=0:V1.01:545GHz}{0.9878}
\setsymbol{HFI:colour:correction:alpha=0:V1.01:857GHz}{1.0014}

\def\LCDM{$\Lambda$CDM}
\def\ffp{FFP6}
\def\unionmask{U73}
\def\nside{N_{\mathrm{side}}}

\def\healpix{\texttt{HEALPix}}
\def\commander{\texttt{Commander}}
\def\ruler{\texttt{Ruler}}
\def\comrul{\texttt{Commander-Ruler}}
\def\CR{\texttt{C-R}}
\def\nilc{\texttt{NILC}}
\def\sevem{\texttt{SEVEM}}
\def\smica{\texttt{SMICA}}
\def\CamSpec{\texttt{CamSpec}}
\def\Plik{\texttt{Plik}}
\def\XFaster{\texttt{XFaster}}

\def\adj{^{\dagger}}
\def\inv{^{-1}}
\def\lm{{\ell m}}

\title{\Planck\ 2013 results. XII. Component separation}
\author{\small
Planck Collaboration:
P.~A.~R.~Ade\inst{90}
\and
N.~Aghanim\inst{63}
\and
C.~Armitage-Caplan\inst{95}
\and
M.~Arnaud\inst{77}
\and
M.~Ashdown\inst{74, 6}\thanks{Corresponding author: M.~Ashdown, \url{maja1@mrao.cam.ac.uk}.}
\and
F.~Atrio-Barandela\inst{20}
\and
J.~Aumont\inst{63}
\and
C.~Baccigalupi\inst{89}
\and
A.~J.~Banday\inst{98, 11}
% ...
}
\institute{\small
APC, AstroParticule et Cosmologie, Universit\'{e} Paris Diderot, CNRS/IN2P3, CEA/lrfu, Observatoire de Paris, Sorbonne Paris Cit\'{e}, 10, rue Alice Domon et L\'{e}onie Duquet, 75205 Paris Cedex 13, France\\
\and
Aalto University Mets\"{a}hovi Radio Observatory, Mets\"{a}hovintie 114, FIN-02540 Kylm\"{a}l\"{a}, Finland\\
\and
African Institute for Mathematical Sciences, 6-8 Melrose Road, Muizenberg, Cape Town, South Africa\\
\and
Agenzia Spaziale Italiana Science Data Center, c/o ESRIN, via Galileo Galilei, Frascati, Italy\\
\and
Agenzia Spaziale Italiana, Viale Liegi 26, Roma, Italy\\
\and
Astrophysics Group, Cavendish Laboratory, University of Cambridge, J J Thomson Avenue, Cambridge CB3 0HE, U.K.\\
\and
Astrophysics \& Cosmology Research Unit, School of Mathematics, Statistics \& Computer Science, University of KwaZulu-Natal, Westville Campus, Private Bag X54001, Durban 4000, South Africa\\
\and
Atacama Large Millimeter/submillimeter Array, ALMA Santiago Central Offices, Alonso de Cordova 3107, Vitacura, Casilla 763 0355, Santiago, Chile\\
\and
CITA, University of Toronto, 60 St. George St., Toronto, ON M5S 3H8, Canada\\
\and
CNR - ISTI, Area della Ricerca, via G. Moruzzi 1, Pisa, Italy\\
\and
CNRS, IRAP, 9 Av. colonel Roche, BP 44346, F-31028 Toulouse cedex 4, France\\
\and
California Institute of Technology, Pasadena, California, U.S.A.\\
\and
Centre for Theoretical Cosmology, DAMTP, University of Cambridge, Wilberforce Road, Cambridge CB3 0WA U.K.\\
\and
Centro de Estudios de F\'{i}sica del Cosmos de Arag\'{o}n (CEFCA), Plaza San Juan, 1, planta 2, E-44001, Teruel, Spain\\
\and
Computational Cosmology Center, Lawrence Berkeley National Laboratory, Berkeley, California, U.S.A.\\
\and
Consejo Superior de Investigaciones Cient\'{\i}ficas (CSIC), Madrid, Spain\\
\and
DSM/Irfu/SPP, CEA-Saclay, F-91191 Gif-sur-Yvette Cedex, France\\
\and
DTU Space, National Space Institute, Technical University of Denmark, Elektrovej 327, DK-2800 Kgs. Lyngby, Denmark\\
\and
D\'{e}partement de Physique Th\'{e}orique, Universit\'{e} de Gen\`{e}ve, 24, Quai E. Ansermet,1211 Gen\`{e}ve 4, Switzerland\\
\and
Departamento de F\'{\i}sica Fundamental, Facultad de Ciencias, Universidad de Salamanca, 37008 Salamanca, Spain\\
\and
Departamento de F\'{\i}sica, Universidad de Oviedo, Avda. Calvo Sotelo s/n, Oviedo, Spain\\
\and
Departamento de Matem\'{a}ticas, Estad\'{\i}stica y Computaci\'{o}n, Universidad de Cantabria, Avda. de los Castros s/n, Santander, Spain\\
\and
Department of Astronomy and Astrophysics, University of Toronto, 50 Saint George Street, Toronto, Ontario, Canada\\
\and
Department of Astrophysics/IMAPP, Radboud University Nijmegen, P.O. Box 9010, 6500 GL Nijmegen, The Netherlands\\
\and
Department of Electrical Engineering and Computer Sciences, University of California, Berkeley, California, U.S.A.\\
\and
Department of Physics \& Astronomy, University of British Columbia, 6224 Agricultural Road, Vancouver, British Columbia, Canada\\
\and
Department of Physics and Astronomy, Dana and David Dornsife College of Letter, Arts and Sciences, University of Southern California, Los Angeles, CA 90089, U.S.A.\\
\and
Department of Physics and Astronomy, University College London, London WC1E 6BT, U.K.\\
\and
Department of Physics, Gustaf H\"{a}llstr\"{o}min katu 2a, University of Helsinki, Helsinki, Finland\\
\and
Department of Physics, Princeton University, Princeton, New Jersey, U.S.A.\\
\and
Department of Physics, University of California, One Shields Avenue, Davis, California, U.S.A.\\
\and
Department of Physics, University of California, Santa Barbara, California, U.S.A.\\
\and
Department of Physics, University of Illinois at Urbana-Champaign, 1110 West Green Street, Urbana, Illinois, U.S.A.\\
\and
Dipartimento di Fisica e Astronomia G. Galilei, Universit\`{a} degli Studi di Padova, via Marzolo 8, 35131 Padova, Italy\\
\and
Dipartimento di Fisica e Scienze della Terra, Universit\`{a} di Ferrara, Via Saragat 1, 44122 Ferrara, Italy\\
\and
Dipartimento di Fisica, Universit\`{a} La Sapienza, P. le A. Moro 2, Roma, Italy\\
\and
Dipartimento di Fisica, Universit\`{a} degli Studi di Milano, Via Celoria, 16, Milano, Italy\\
\and
Dipartimento di Fisica, Universit\`{a} degli Studi di Trieste, via A. Valerio 2, Trieste, Italy\\
\and
Dipartimento di Fisica, Universit\`{a} di Roma Tor Vergata, Via della Ricerca Scientifica, 1, Roma, Italy\\
\and
Discovery Center, Niels Bohr Institute, Blegdamsvej 17, Copenhagen, Denmark\\
\and
Dpto. Astrof\'{i}sica, Universidad de La Laguna (ULL), E-38206 La Laguna, Tenerife, Spain\\
\and
European Southern Observatory, ESO Vitacura, Alonso de Cordova 3107, Vitacura, Casilla 19001, Santiago, Chile\\
\and
European Space Agency, ESAC, Planck Science Office, Camino bajo del Castillo, s/n, Urbanizaci\'{o}n Villafranca del Castillo, Villanueva de la Ca\~{n}ada, Madrid, Spain\\
\and
European Space Agency, ESTEC, Keplerlaan 1, 2201 AZ Noordwijk, The Netherlands\\
\and
Haverford College Astronomy Department, 370 Lancaster Avenue, Haverford, Pennsylvania, U.S.A.\\
\and
Helsinki Institute of Physics, Gustaf H\"{a}llstr\"{o}min katu 2, University of Helsinki, Helsinki, Finland\\
\and
INAF - Osservatorio Astrofisico di Catania, Via S. Sofia 78, Catania, Italy\\
\and
INAF - Osservatorio Astronomico di Padova, Vicolo dell'Osservatorio 5, Padova, Italy\\
\and
INAF - Osservatorio Astronomico di Roma, via di Frascati 33, Monte Porzio Catone, Italy\\
\and
INAF - Osservatorio Astronomico di Trieste, Via G.B. Tiepolo 11, Trieste, Italy\\
\and
INAF/IASF Bologna, Via Gobetti 101, Bologna, Italy\\
\and
INAF/IASF Milano, Via E. Bassini 15, Milano, Italy\\
\and
INFN, Sezione di Bologna, Via Irnerio 46, I-40126, Bologna, Italy\\
\and
INFN, Sezione di Roma 1, Universit\`{a} di Roma Sapienza, Piazzale Aldo Moro 2, 00185, Roma, Italy\\
\and
INFN/National Institute for Nuclear Physics, Via Valerio 2, I-34127 Trieste, Italy\\
\and
IPAG: Institut de Plan\'{e}tologie et d'Astrophysique de Grenoble, Universit\'{e} Joseph Fourier, Grenoble 1 / CNRS-INSU, UMR 5274, Grenoble, F-38041, France\\
\and
ISDC Data Centre for Astrophysics, University of Geneva, ch. d'Ecogia 16, Versoix, Switzerland\\
\and
IUCAA, Post Bag 4, Ganeshkhind, Pune University Campus, Pune 411 007, India\\
\and
Imperial College London, Astrophysics group, Blackett Laboratory, Prince Consort Road, London, SW7 2AZ, U.K.\\
\and
Infrared Processing and Analysis Center, California Institute of Technology, Pasadena, CA 91125, U.S.A.\\
\and
Institut N\'{e}el, CNRS, Universit\'{e} Joseph Fourier Grenoble I, 25 rue des Martyrs, Grenoble, France\\
\and
Institut Universitaire de France, 103, bd Saint-Michel, 75005, Paris, France\\
\and
Institut d'Astrophysique Spatiale, CNRS (UMR8617) Universit\'{e} Paris-Sud 11, B\^{a}timent 121, Orsay, France\\
\and
Institut d'Astrophysique de Paris, CNRS (UMR7095), 98 bis Boulevard Arago, F-75014, Paris, France\\
\and
Institute for Space Sciences, Bucharest-Magurale, Romania\\
\and
Institute of Astronomy and Astrophysics, Academia Sinica, Taipei, Taiwan\\
\and
Institute of Astronomy, University of Cambridge, Madingley Road, Cambridge CB3 0HA, U.K.\\
\and
Institute of Theoretical Astrophysics, University of Oslo, Blindern, Oslo, Norway\\
\and
Instituto de Astrof\'{\i}sica de Canarias, C/V\'{\i}a L\'{a}ctea s/n, La Laguna, Tenerife, Spain\\
\and
Instituto de F\'{\i}sica de Cantabria (CSIC-Universidad de Cantabria), Avda. de los Castros s/n, Santander, Spain\\
\and
Istituto di Fisica del Plasma, CNR-ENEA-EURATOM Association, Via R. Cozzi 53, Milano, Italy\\
\and
Jet Propulsion Laboratory, California Institute of Technology, 4800 Oak Grove Drive, Pasadena, California, U.S.A.\\
\and
Jodrell Bank Centre for Astrophysics, Alan Turing Building, School of Physics and Astronomy, The University of Manchester, Oxford Road, Manchester, M13 9PL, U.K.\\
\and
Kavli Institute for Cosmology Cambridge, Madingley Road, Cambridge, CB3 0HA, U.K.\\
\and
LAL, Universit\'{e} Paris-Sud, CNRS/IN2P3, Orsay, France\\
\and
LERMA, CNRS, Observatoire de Paris, 61 Avenue de l'Observatoire, Paris, France\\
\and
Laboratoire AIM, IRFU/Service d'Astrophysique - CEA/DSM - CNRS - Universit\'{e} Paris Diderot, B\^{a}t. 709, CEA-Saclay, F-91191 Gif-sur-Yvette Cedex, France\\
\and
Laboratoire Traitement et Communication de l'Information, CNRS (UMR 5141) and T\'{e}l\'{e}com ParisTech, 46 rue Barrault F-75634 Paris Cedex 13, France\\
\and
Laboratoire de Physique Subatomique et de Cosmologie, Universit\'{e} Joseph Fourier Grenoble I, CNRS/IN2P3, Institut National Polytechnique de Grenoble, 53 rue des Martyrs, 38026 Grenoble cedex, France\\
\and
Laboratoire de Physique Th\'{e}orique, Universit\'{e} Paris-Sud 11 \& CNRS, B\^{a}timent 210, 91405 Orsay, France\\
\and
Lawrence Berkeley National Laboratory, Berkeley, California, U.S.A.\\
\and
Max-Planck-Institut f\"{u}r Astrophysik, Karl-Schwarzschild-Str. 1, 85741 Garching, Germany\\
\and
McGill Physics, Ernest Rutherford Physics Building, McGill University, 3600 rue University, Montr\'{e}al, QC, H3A 2T8, Canada\\
\and
MilliLab, VTT Technical Research Centre of Finland, Tietotie 3, Espoo, Finland\\
\and
Niels Bohr Institute, Blegdamsvej 17, Copenhagen, Denmark\\
\and
Observational Cosmology, Mail Stop 367-17, California Institute of Technology, Pasadena, CA, 91125, U.S.A.\\
\and
Optical Science Laboratory, University College London, Gower Street, London, U.K.\\
\and
SB-ITP-LPPC, EPFL, CH-1015, Lausanne, Switzerland\\
\and
SISSA, Astrophysics Sector, via Bonomea 265, 34136, Trieste, Italy\\
\and
School of Physics and Astronomy, Cardiff University, Queens Buildings, The Parade, Cardiff, CF24 3AA, U.K.\\
\and
School of Physics and Astronomy, University of Nottingham, Nottingham NG7 2RD, U.K.\\
\and
Space Research Institute (IKI), Russian Academy of Sciences, Profsoyuznaya Str, 84/32, Moscow, 117997, Russia\\
\and
Space Sciences Laboratory, University of California, Berkeley, California, U.S.A.\\
\and
Stanford University, Dept of Physics, Varian Physics Bldg, 382 Via Pueblo Mall, Stanford, California, U.S.A.\\
\and
Sub-Department of Astrophysics, University of Oxford, Keble Road, Oxford OX1 3RH, U.K.\\
\and
Theory Division, PH-TH, CERN, CH-1211, Geneva 23, Switzerland\\
\and
UPMC Univ Paris 06, UMR7095, 98 bis Boulevard Arago, F-75014, Paris, France\\
\and
Universit\'{e} de Toulouse, UPS-OMP, IRAP, F-31028 Toulouse cedex 4, France\\
% ...
}

\authorrunning{\Planck\ Collaboration}  

\begin{document}

\abstract{\Planck\ has produced detailed all-sky observations over
  nine frequency bands between 30 and 857\,GHz. These observations
  allow robust reconstruction of the primordial cosmic microwave
  background (CMB) temperature fluctuations over nearly the full sky,
  as well as new constraints on Galactic foregrounds, including
  thermal dust and line emission from molecular carbon monoxide
  (CO). This paper describes the component separation framework
  adopted by \Planck\ for many cosmological analyses, including CMB
  power spectrum determination and likelihood construction on large
  angular scales, studies of primordial non-Gaussianity and
  statistical isotropy, the integrated Sachs-Wolfe effect (ISW), and
  gravitational lensing, and searches for topological defects.  We
  test four foreground-cleaned CMB maps derived using qualitatively
  different component separation algorithms.  The quality of our
  reconstructions is evaluated through detailed simulations and
  internal comparisons, and shown through various tests to be
  internally consistent and robust for CMB power spectrum and
  cosmological parameter estimation up to $\ell=2000$.  The parameter
  constraints on $\Lambda$CDM cosmologies derived from these maps are
  consistent with those presented in the cross-spectrum based
  \Planck\ likelihood analysis.  We choose two of the CMB maps for
  specific scientific goals.  We also present maps and frequency
  spectra of the Galactic low-frequency, CO, and thermal dust
  emission.  The component maps are found to provide a faithful
  representation of the sky, as evaluated by simulations, with the
  largest bias seen in the CO component at 3\,\%. For the
  low-frequency component, the spectral index varies widely over the
  sky, ranging from about $\beta = -4$ to $-2$. Considering both
  morphology and prior knowledge of the low frequency components, the
  index map allows us to associate a steep spectral index ($\beta <
  -3.2$) with strong anomalous microwave emission, corresponding to a
  spinning dust spectrum peaking below 20\,GHz, a flat index of $\beta
  > -2.3$ with strong free-free emission, and intermediate values with
  synchrotron emission.}


\maketitle

\clearpage

\section{Introduction}
\label{sec:introduction}

\begin{figure*}
\begin{center}
\includegraphics[width=\textwidth]{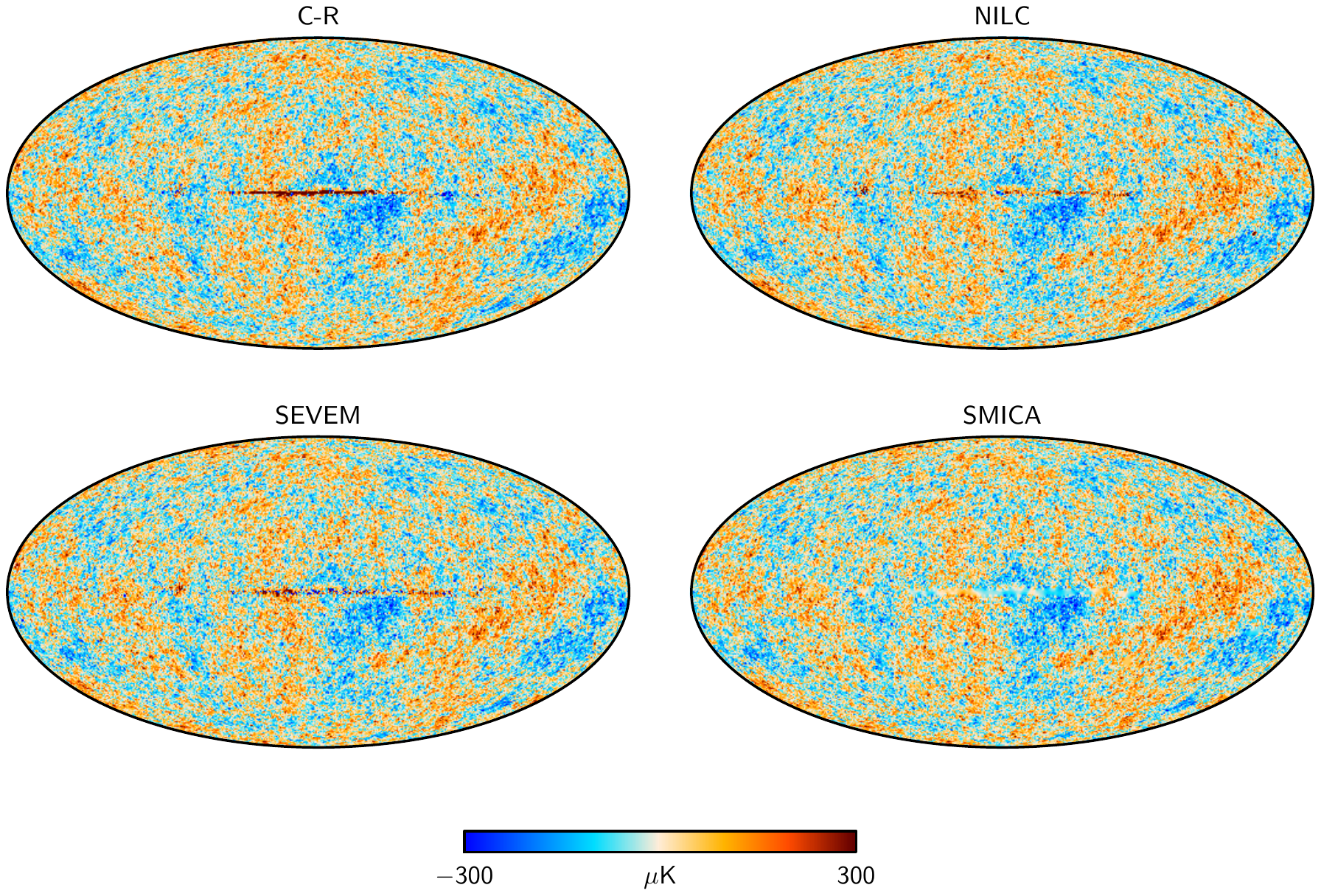}
\end{center}
\caption{Foreground-cleaned CMB maps derived by \comrul, \nilc,
  \sevem\ and \smica.  Note that the \smica\ map has been filled in
  smoothly inside a 3\,\% Galactic mask.}
\label{fig:cmb_map}
\end{figure*}

This paper, one of a set associated with the 2013 release of data from
the \Planck\footnote{\Planck\ (\url{http://www.esa.int/Planck}) is a
  project of the European Space Agency (ESA) with instruments provided
  by two scientific consortia funded by ESA member states (in
  particular the lead countries France and Italy), with contributions
  from NASA (USA) and telescope reflectors provided by a collaboration
  between ESA and a scientific consortium led and funded by Denmark.}
mission \citep{planck2013-p01}, describes the component separation
techniques applied to the \Planck\ data to produce maps of the cosmic
microwave background (CMB) temperature anisotropies (see
Fig.~\ref{fig:cmb_map}) and of diffuse foregrounds.

The sky at millimetre and sub-millimetre wavelengths contains a wealth
of cosmological and astrophysical information.  Accessing it is an
inversion process, known as component separation, to extract the
sources of emission contributing to a set of maps observed at
different frequencies.  \Planck\ gives us a powerful data set to
unlock new information in this manner by observing the entire sky from
30 to 857\GHz\ in nine frequency bands at higher angular resolution
and sensitivity than its predecessors.  Accurate and detailed
component separation is a central objective of the mission.

We divide the foregrounds into two distinct categories: diffuse
emission from the Galaxy and compact sources.  The Galactic
foregrounds are the principal source of contamination of the CMB on
large angular scales, with fluctuation power decreasing roughly as a
power law towards higher multipoles \citep{bennett2003b}.  They are
dominated by synchrotron, free-free and Anomalous Microwave Emission
(AME, ascribed to spinning dust grains, at frequencies below 70\GHz)
and by rotational line emission from carbon monoxide (CO) molecules
and thermal dust emission at frequencies above 100\GHz.  Extragalactic
foregrounds, on the other hand, dominate the small-scale contamination
of the CMB.  They arise from discrete, individually detectable compact
sources and the collective emission from unresolved radio and infrared
(IR) sources, and also from the Sunyaev-Zeldovich (SZ) effect in
galaxy clusters
\citep{planck2011-6.6,planck2013-p05,planck2013-p05a}.

In the \Planck\ analyses, these foregrounds are dealt with in a
variety of ways.  At the power spectrum and likelihood level, the
extragalactic foregrounds are modelled with parameterized power
spectra, appropriate to their statistical isotropy, over regions
restricted to low Galactic emission \citep{planck2013-p08}.  Component
separation as described in the present paper aims at removing Galactic
emission to produce CMB maps covering the largest possible sky area
for studies of the large-scale properties and higher-order statistics
of the CMB.  In addition, this component separation provides a
reconstruction of the diffuse emission from our Galaxy.  Detailed
studies of specific extragalactic foregrounds, such as the cosmic
infrared background (CIB) \citep{planck2013-p13} and the diffuse
Sunyaev-Zeldovich (SZ) signal \citep{planck2013-p05b}, employ methods
tailored to their particular needs.

Building on previous work \citep{leach2008}, we approach CMB
extraction with a philosophy designed to ensure robustness by applying
four distinct algorithms based on two different methodologies.  The
first avoids any assumptions concerning the foregrounds and relies
solely on a minimum variance criterion for the data component
possessing a blackbody spectrum (i.e., the CMB), while the second
methodology relies on parametric modelling of the foregrounds in
either real or harmonic space.  We evaluate the performance of these
component separation algorithms through detailed simulations, and we
examine the robustness of the recovered CMB maps by comparing them,
their power spectra, and their resulting cosmological constraints.  As
a diagnostic, we also briefly examine their higher-order statistics.

The CMB results presented in this work serve a number applications.
We use the real-space modelling to produce a clean CMB map and power
spectra on large angular scales, where diffuse Galactic emission is
the main contaminant, to construct the likelihood function at low
multipoles; this is then combined with the high multipole likelihood
function that models extragalactic foregrounds with power spectra
\citep{planck2013-p08}.  The high resolution CMB
maps are used as a check on primary cosmological constraints (see
below), for lensing studies \citep{planck2013-p12}, studies of the
integrated Sachs-Wolfe effect \citep{planck2013-p14}, of the isotropy
of the CMB \citep{planck2013-p09}, of non-Gaussian statistics
\citep{planck2013-p09a}, in searches for topological defects
\citep{planck2013-p20}, and for examination of the geometry and
topology of the Universe \citep{planck2013-p19}.

In addition, we present maps of diffuse Galactic emission divided into
low- and high-frequency components, as well as a molecular CO
component.  We judge the adequacy of this reconstruction through
simulations and by comparison with known properties of the diffuse
Galactic foregrounds.

The paper is organized as follows.  In Sect.~\ref{sec:sky_emissions}
we discuss the expected sources of sky emission over the
\Planck\ frequency interval and how they are modelled. Then in
Sect.~\ref{sec:comp_sep_approach} we detail the overall approach and
introduce the four component separation methods.  In
Sect.~\ref{sec:data} we present the \Planck\ data set and
pre-processing procedure, and we describe our simulations.  This is
followed by a presentation of the derived CMB maps and their
characterization in Sect.~\ref{sec:cmb_maps}.
Section~\ref{sec:powspec_params} is dedicated to power spectra and
cosmological parameter constraints obtained from these maps, and
Sect.~\ref{sec:ng} to studies of higher-order statistics.
Section~\ref{sec:foreground_components} presents a reconstruction of
the diffuse Galactic foregrounds, and Sect.~\ref{sec:conclusions}
concludes.  We relegate details of the algorithms to appendices.

\section{The sky at \Planck\ frequencies}
\label{sec:sky_emissions}

The properties of Galactic emission vary significantly across the
\Planck\ frequency range from 30 to 857\GHz. At frequencies below
70\GHz, the dominant radiation processes are: synchrotron emission
from cosmic ray electrons interacting with the Galactic magnetic field
\citep[e.g.,][]{haslam1982,reich&reich1988,broad1989,davies1996,platania2003,bennett2003b,gold2010};
thermal Bremsstrahlung (or free-free emission) from electron-electron
and electron-ion scattering
\citep[e.g.,][]{banday2003,dickinson2003,davies2006,ghosh2011,planck2013-XII,planck2011-7.2};
and AME from dust grains
\citep{kogut1996,leitch1997,banday2003,lagache2003,oliveira2004,fink2004,davies2006,bonaldi2007,dobler2008,mamd,ysard2010,gold2010,planck2011-7.2},
possibly due to their rotational line emission
\citep{DL1998,spdust1,ysard-ves2010,hoang2012}.  Over the frequency
range covered by \Planck, both synchrotron and free-free spectra are
well approximated by power laws in brightness temperature,
$T_{\mathrm{B}}\propto \nu^{\,\beta}$, with the synchrotron index,
$\beta_{\textrm{synch}}$, ranging from $-3.2$ to $-2.8$
\citep{davies1996} and the free-free index, $\beta_{\textrm{ff}}$,
lying between $-2.2$ and $-2.1$.  Less is known about the AME
spectrum, but spinning dust models with a spectrum peaking at
frequencies below 20\GHz\ (in brightness temperature units) adequately
describe current observations\footnote{Note that we adopt brightness
temperature for AME in this paper, while many other publications adopt
flux density. When comparing peak frequencies, it is useful to note
that that a spectrum that has a maximum at 30\GHz\ in flux density
peaks at 17\GHz\ in brightness temperature.}. Above the peak, the
spectrum appears consistent with a power-law
\citep[e.g.,][]{banday2003,davies2006,dobler2008,ghosh2011}.  In
addition to these three, the existence of a fourth low-frequency
foreground component, known as the ``Galactic haze'', has been
claimed, possibly due to a hard-spectrum synchrotron population near
the Galactic centre
\citep[e.g.,][]{finkbeiner2004,dobler2008,pietrobon_et_al_2011,planck2012-IX}.

At frequencies higher than 100\GHz, thermal dust emission dominates
over most of the sky and is commonly described by a modified blackbody spectrum
with power-law emissivity, $\epsilon_\nu\propto
\nu^{\,\beta_{\mathrm{d}}}$, and temperature, $T_{\mathrm{d}}$. Both
the temperature and spectral index, $\beta_{\mathrm{d}}$, vary
spatially. Prior to \Planck, the best-fitting single component dust
model had a temperature $T_{\mathrm{d}}\approx 18$\,K and spectral
index $\beta_{\mathrm{d}}\approx 1.7$
\citep{finkbeiner1999,bennett2003b,gold2010}, although there is
evidence of flattening of the spectral index from around 1.8 in the
far-infrared to 1.55 in the microwave region \citep{marta}, the
interpretation of which is still under study.

In addition to these diffuse Galactic components, extragalactic
emission contributes at \Planck\ frequencies.  In particular, a large
number of radio and far-infrared (FIR; \citealp{planck2011-6.1})
galaxies, clusters of galaxies and the Cosmic Infrared Background
(CIB; \citealp{planck2011-6.6}) produce a statistically isotropic
foreground, with frequency spectra well approximated by models similar
to those applicable to the Galactic foregrounds (modified blackbody
spectra, power laws, etc.).  Except for a frequency-dependent absolute
offset, which may be removed as part of the overall offset removal
procedure, these extragalactic components are therefore typically
absorbed by either the low-frequency or thermal dust components during
component separation. No special treatment is given here to
extragalactic foregrounds, beyond the masking of bright objects.
Dedicated scientific analyses of these sources are described in detail
in \citet{planck2011-6.6}, \citet{planck2013-p05},
and \citet{planck2013-p05a}. In the \Planck\ likelihood, extragalactic
sources are modelled in terms of power spectrum templates at high
$\ell$ \citep{planck2013-p08}.

\begin{table*}[tmb] 
  \begingroup
  \newdimen\tblskip \tblskip=5pt
  \caption{Overview and comparison of component separation algorithms.}
  \label{tab:algorithms}
  \vskip -3mm
  \footnotesize
  \setbox\tablebox=\vbox{
  \newdimen\digitwidth
  \setbox0=\hbox{\rm 0}
  \digitwidth\wd0
  \catcode`*=\active
  \def*{\kern\digitwidth}
  \newdimen\signwidth
  \setbox0=\hbox{+}
  \signwidth=\wd0
  \catcode`!=\active
  \def!{\kern\signwidth}
  \halign{#\hfil\tabskip=2em& #\hfil& #\hfil& #\hfil& #\hfil\tabskip=0pt\cr
  \noalign{\doubleline}
  \omit\hfil Characteristic\hfil&\omit\hfil\comrul\hfil&\omit\hfil\nilc\hfil&\omit\hfil\sevem\hfil&\omit\hfil\smica\hfil\cr
      \noalign{\vskip 3pt\hrule\vskip 5pt}
      Method\dotfill& Bayesian parameter& Internal linear& Internal template&  Spectral parameter\cr
      & estimation& combination& fitting& estimation\cr
      Domain\dotfill& Pixel&  Needlet& Pixel& Spherical harmonic\cr
      Channels [GHz]\dotfill& 30--353& 44--857& 30--857& 30--857\cr
      Effective beam FWHM [arcmin]\dotfill& $\sim$7.4& 5.0& 5.0& 5.0\cr
      $\ell_\mathrm{max}$\dotfill& none& 3200& 3100& 4000\cr
  \noalign{\vskip 5pt\hrule\vskip 3pt}}}
  \endPlancktablewide 
  \endgroup
\end{table*}

\begin{figure}
\begin{center}
\includegraphics[width=\columnwidth]{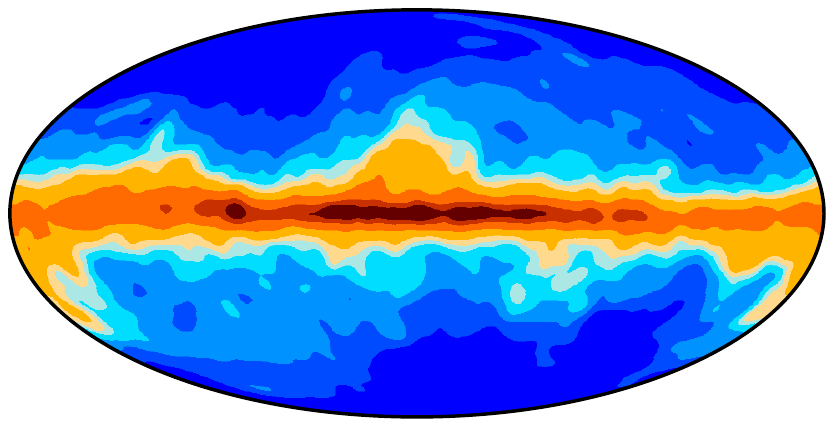}
\end{center}
\caption{Combined Galactic (CG) emission masks for the \Planck\ data,
  corresponding to sky fractions of 20, 40, 60, 70, 75, 80, 90, 97,
  and 99\,\%. The masks are named CG20, etc.}
\label{fig:masks}
\end{figure}

Other relevant sources include emission from molecular clouds,
supernova remnants, and compact \ion{H}{ii} regions inside our own
Galaxy, as well as the thermal and kinetic SZ effects, due to inverse
Compton scattering of CMB photons off free electrons in ionized media.
\Planck\ provides new and important information on all these
processes, as described both in the following and in the companion
papers \citet{planck2013-p03a}, \citet{planck2013-p05b},
and \citet{planck2013-p06b}. In particular, \Planck's frequency range,
angular resolution and sensitivity make it a powerful probe of thermal
dust, resulting in new and tight constraints on dust temperature and
emissivity. The same frequencies also allow extraction of the first
ever full-sky maps of the emission resulting from the CO
$J$=1$\rightarrow$0, $J$=2$\rightarrow$1 and $J$=3$\rightarrow$2
rotational transitions at 115, 230 and 345\GHz,
respectively \citep{planck2013-p03a}.

The focus of this paper is to reconstruct the CMB anisotropies over a
large sky fraction, exploiting only the \Planck\ frequency bands. We
also present a detailed reconstruction of the thermal dust emission at
high frequencies, as well as CO emission lines.  At low frequencies
and over the region used for CMB analysis, the total foreground
contribution is well approximated by a single power law (see
Sect.~\ref{sec:foreground_components}).  We therefore model the sum of
all low-frequency foregrounds by a power law with spatially varying
spectral index whose numerical value in any pixel results from the
influence of the dominant foreground component at that location.  The
full analysis of diffuse foregrounds, using ancillary data to resolve
the individual components at low frequencies, will be presented in a
forthcoming publication.

\section{Approach to component separation}
\label{sec:comp_sep_approach}

The rich content of the \Planck\ data encourages application of
several component separation techniques.  We consider four, as
summarized in Table \ref{tab:algorithms}, which we classify according
to one of two different general methodologies.  The first makes
minimal assumptions concerning the foregrounds and seeks only to
minimize the variance of the CMB, i.e., the sky component possessing a
blackbody spectrum.  We implement this approach with a needlet
(wavelet on the sphere) version of the internal linear combination
(ILC) algorithm (\nilc; \citealp{nilc2009}), and also with a
template-based method to remove foreground contamination from the
CMB-dominant bands.  These foreground templates are constructed from
the lowest and highest frequency channels (Spectral Estimation Via
Expectation Maximization, \sevem; \citealp{Fernandezcobos2012}).

The second methodology uses parametric modelling of the foregrounds.
In our real space implementation, we explore model parameters through
Bayesian parameter estimation techniques, fitting a parametric signal
model per pixel (\commander;
\citealp{Eriksen2006ApJ641,Eriksen2008ApJ676}); a similar
implementation is presented by \citet{Stompor2009MNRAS392}.  To
estimate spectral indices robustly in pixel space, this procedure
requires identical angular resolution across all frequencies included
in the analysis, and is therefore limited in resolution by the 30\GHz\
LFI channel. However, this is sufficient to generate the
low-resolution CMB map and power spectrum samples required for the low
multipole part of the \Planck\ likelihood function for cosmological
parameters \citep{planck2013-p08}.  To produce full resolution maps,
we use the resulting low-resolution spectral parameter samples to
solve for the component amplitudes, in an extension to the method
known as \ruler\ (we refer to the combined method as \comrul, or \CR).
In our fourth technique, we implement a CMB-oriented parametric
approach that fits the amplitude and spectral parameters of CMB and
foregrounds in the harmonic domain (Spectral Matching Independent
Component Analysis, \smica; \citealp{smicaIEEE}).

Details of each algorithm are given in the appendices.  We now turn to
their application to the data and evaluate their performance using
simulations.

\section{Data, simulations and masks}
\label{sec:data}

We use the data set from the first 15.5 months of
\Planck\ observations, corresponding to 2.6 sky surveys, from both the
Low Frequency Instrument (LFI) and High Frequency Instrument (HFI).
The primary inputs for component separation are the frequency channel
maps, including half-ring maps, bandpasses, and beam characteristics;
a full description of these products is given in
\citet{planck2013-p02} and \citet{planck2013-p03}.  No special
corrections are made for zodiacal light emission (ZLE;
\citealp{planck2013-p03}) in the analyses presented here. The ZLE is
not stationary on the sky, since it depends on \Planck's position and
scanning strategy.  Therefore the frequency maps contain a projected
version of the emission averaged over the nominal mission.  Despite
this, a series of exploratory analyses showed that our algorithms
naturally correct for this component within their existing model
space.  It was also found that larger CMB residuals were induced when
applying a correction based on a ZLE model than when applying no
correction, most likely due to uncertainties in the model itself.

To evaluate and validate our algorithms, we analyse a large suite of
realistic simulations, the so-called Full Focal Plane (FFP)
simulations, based on detailed models of the instrument and sky. The
version used for this data release is denoted \ffp, and is described
in \citet{planck2013-p28}.  The simulation procedure generates time
streams for each detector, incorporating the satellite pointing, the
individual detector beams, bandpasses, noise properties, and data
flags, and then produces simulated frequency channel maps through the
mapmaking process.  For the input sky, we use the \Planck\ Sky Model
(PSM), which includes the CMB, diffuse Galactic emission (synchrotron,
free-free, thermal dust, AME, and molecular CO lines), and compact
sources (thermal and kinetic SZ effects, radio sources, infrared
sources, the CIB, and ultra-compact \ion{H}{ii} regions).  The
pre-launch version of the PSM is described by
\citet{delabrouille2012}, and has been modified for the present work
as described in \citet{planck2013-p28}.  Each FFP data set consists of
three parts: the simulated observations, Monte Carlo realizations of
the CMB, and Monte Carlo realizations of the instrumental noise.

For both the data and the simulations, we reconstruct the CMB and
foregrounds from the full frequency channel maps and the corresponding
half-ring maps, which are made from the data in the first half or
second half of each stable pointing period.  The half-ring maps can be
used to obtain an estimate of the noise in each channel by taking half
of the difference between the two maps, thereby normalizing the noise
level to that of the full map.  This is referred to as the half-ring
half-difference (HRHD) map.  The signals fixed to the sky will be
cancelled leaving only the noise contribution.  The HRHD map can be
treated as a realization of the same underlying noise processes and it
can be used to estimate the power spectrum, and other properties, of
the noise.  If there are noise correlations between the half-ring
maps, then the estimates of the noise properties thus obtained can be
biased.  This is the case for HFI channels; the cosmic ray glitch
removal \citep{planck2013-p03,planck2013-p03e} induces correlations
that lead to the noise power spectrum being underestimated by a few
percent at high $\ell$ when using the HRHD maps.

\begin{figure}
\begin{center}
\includegraphics[width=\columnwidth]{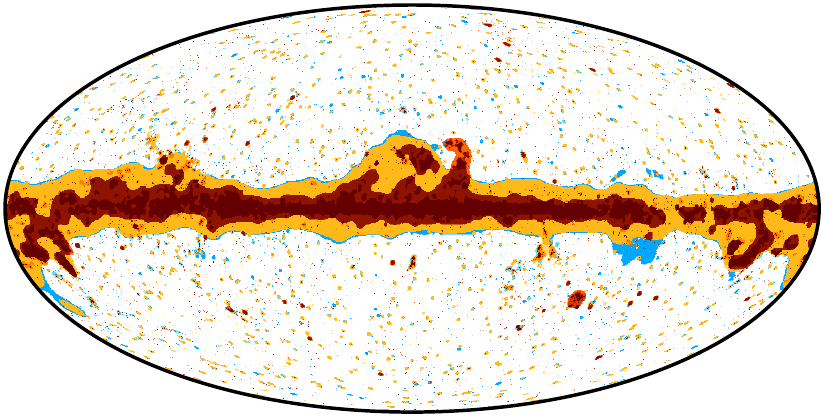}
\end{center}
\caption{Summary of Component Separation (CS) confidence masks.  Each
  pixel is encoded in terms of a sum in which \comrul\ equals 1 (light
  blue), \nilc\ equals 2 (dark red), \sevem\ equals 4 (yellow),
  and \smica\ equals 8 (light red).  The masks are named CS-CR75,
  CS-NILC93, CS-SEVEM76, and CS-SMICA89, respectively, reflecting
  their accepted sky fraction. The union mask (U73), used for
  evaluation purposes in this paper, removes all coloured pixels.}
\label{fig:cmb_mask_union}
\end{figure}

\begin{figure}
\begin{center}
\includegraphics[width=\columnwidth]{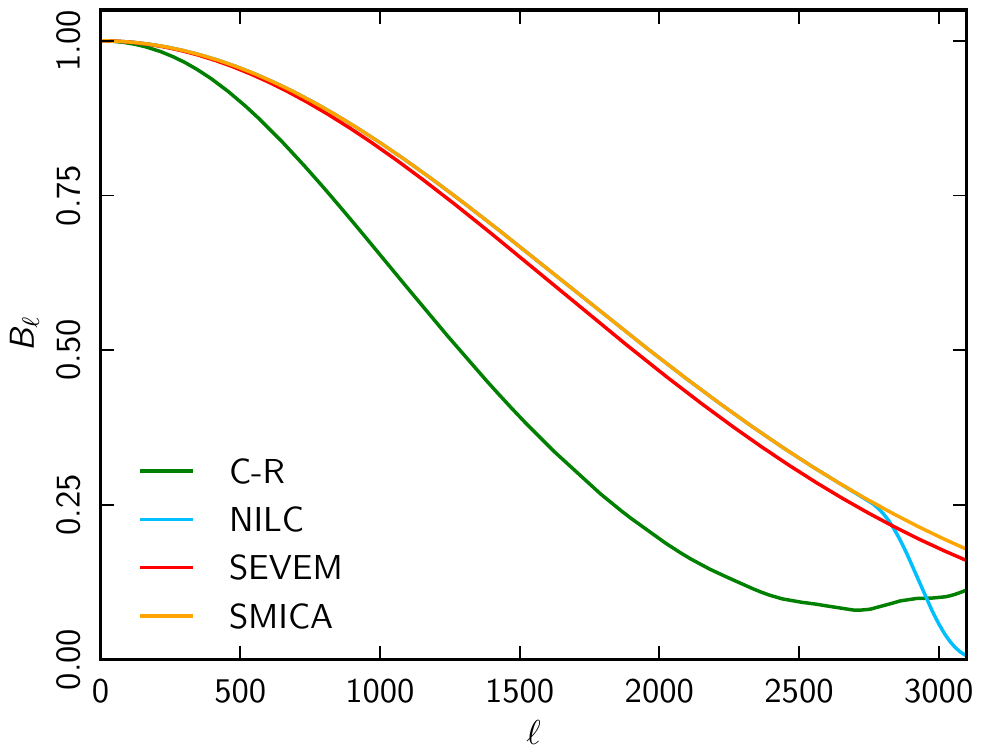}
\end{center}
\caption{Beam transfer functions of the four foreground-cleaned CMB
  maps.}
\label{fig:cmb_beam}
\end{figure}

\begin{figure}
\begin{center}
\includegraphics[width=\columnwidth]{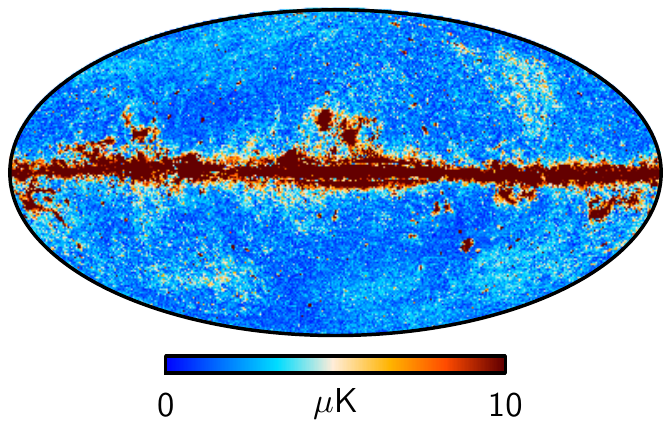}
\end{center}
\caption{Standard deviation between the four foreground-cleaned CMB
  maps. All maps have been downgraded to a \healpix\ resolution of
  $N_{\textrm{side}}=128$. The differences are typically less than
  5\,\microK\ at high Galactic latitudes, demonstrating that the maps
  are consistent over a large part of the sky.}
\label{fig:cmb_rms}
\end{figure}

\begin{figure*}
\begin{center}
\includegraphics[width=\textwidth]{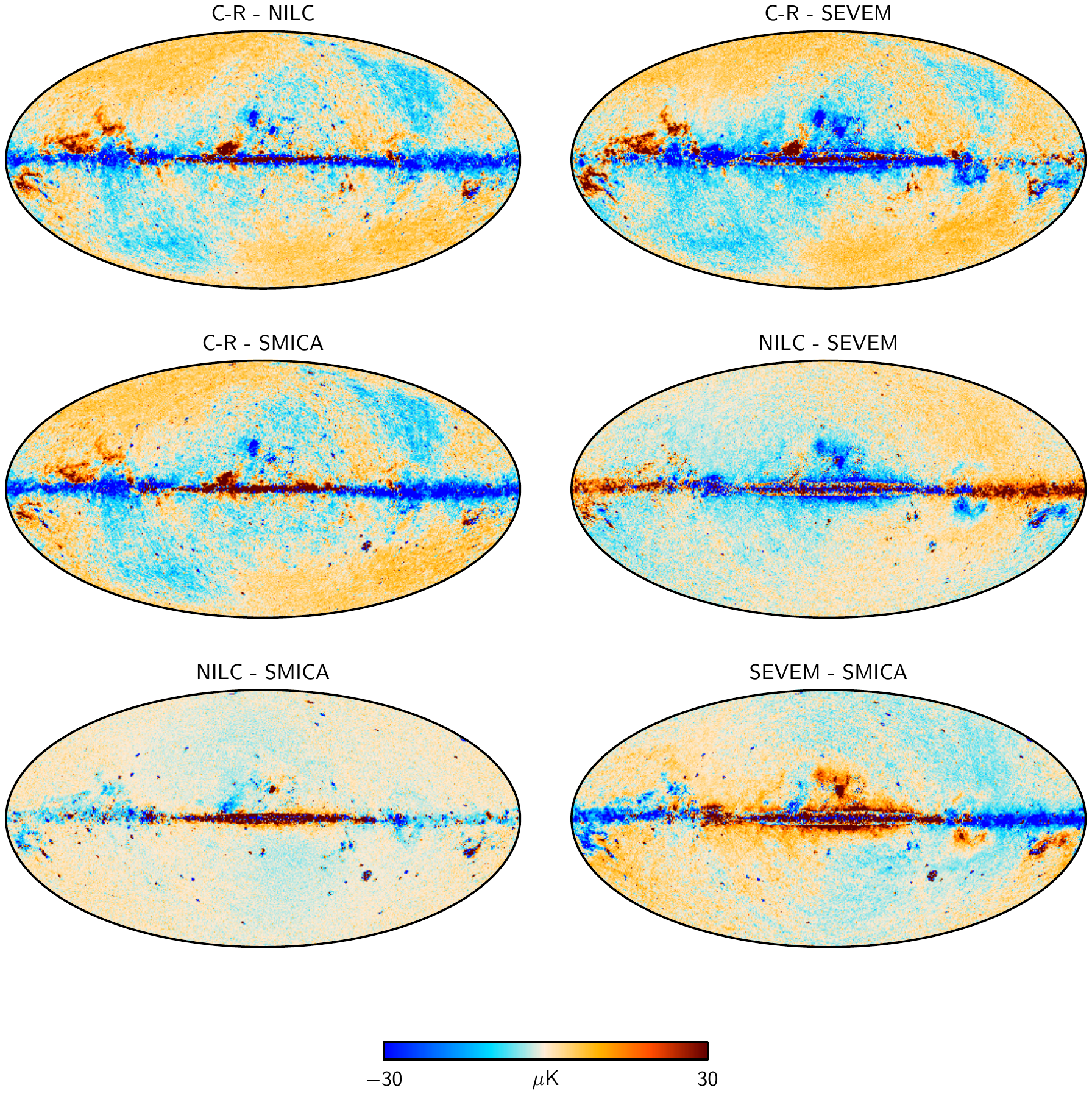}
\end{center}
\caption{Pairwise differences between foreground-cleaned CMB maps. All
  maps have been downgraded to a \healpix\ resolution of
  $N_{\textrm{side}}=128$ to show the large-scale differences.  The
  line-like discontinuities in the differences involving \sevem\ is
  due to the two different regions used in this algorithm to clean the
  sky (see Appendix~\ref{sec:sevem_appendix} for details).}
\label{fig:cmb_diff}
\end{figure*}

Prior to processing the data through each component separation
pipeline, we define masks for the point sources and bright Galactic
regions.  Point source masking is based on the source catalogues
obtained by filtering the input sky maps with the Mexican Hat Wavelet
2 (MHW2) filter and applying a $4\,\sigma$ threshold for the LFI bands
and a $5\,\sigma$ threshold for the HFI
bands \citep{planck2011-6.6,planck2013-p05}.  The mask radius of each
source is different for the LFI and HFI.  Due to the large beam size
of LFI channels, we define a variable masking radius for each source
according to its signal-to-noise ratio (S/N) as $r=(2
\log (A/m))^{1/2}/(2\sqrt{2\log 2}) \times $FWHM, where $r$ is the
radius, $A$ is the S/N, and $m$ is the maximum amplitude (given in
units of the background noise level) allowed for the tail of unmasked
point sources; we set $m = 0.1$, which is a compromise between masking
the source tails and minimizing the number of masked pixels.  For HFI,
the mask radius around each source is $1.27\times$FWHM, using the
average FWHM obtained from the effective beams.

A basic set of Galactic masks is defined as follows. We subtract a
CMB estimate from the 30 and 353\GHz\ maps, mask point sources, and
smooth the resulting maps by a Gaussian with FWHM of 5\deg. We then
threshold and combine them, generating a series of masks with
different amounts of available sky.  The resulting combined Galactic
(CG) masks, shown in Fig.~\ref{fig:masks}, correspond to sky fractions
of 20, 40, 60, 70, 75, 80, 90, 97, and 99\,\%, and are named CG20,
etc.

\section{CMB Maps}
\label{sec:cmb_maps}

\begin{figure*}
  \begin{center}
    \includegraphics[width=\textwidth]{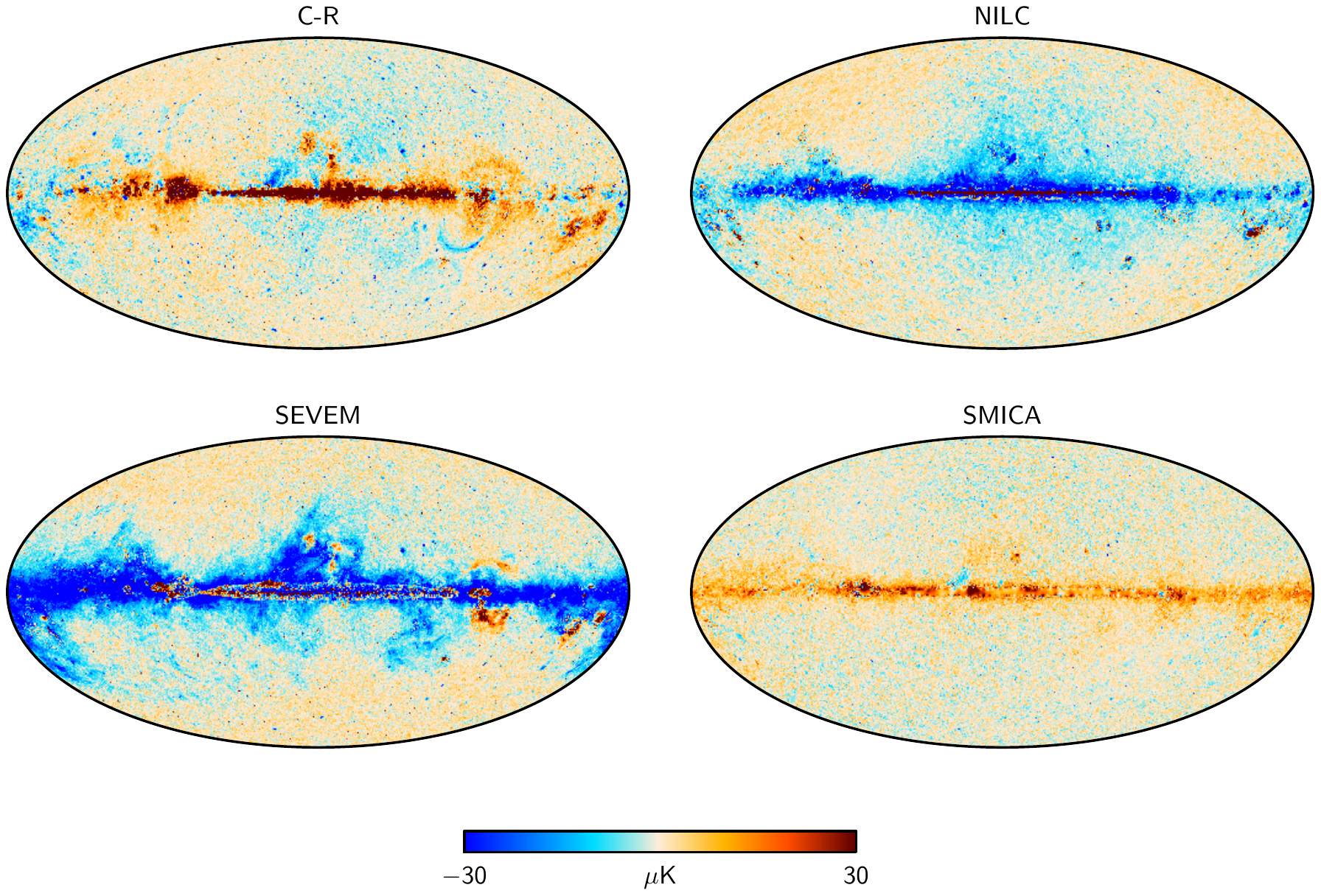}
  \end{center}
  \caption{CMB residual maps from the \ffp\ simulation. A monopole
    determined at high Galactic latitude has been subtracted from the
    maps, and they have been downgraded to a \healpix\ resolution of
    $\nside = 128$ to show the large-scale features. The residuals
    presented here provide a conservative estimate of those expected
    in the data (see text for details).}
  \label{fig:cmb_sim_res}
\end{figure*}

We begin the discussion of our results by presenting the
foreground-cleaned CMB maps. These maps are shown in
Fig.~\ref{fig:cmb_map} for each of the four component separation
algorithms. Already from this figure it is clear that the wide
frequency coverage and high angular resolution of \Planck\ allow a
faithful reconstruction of the CMB field over most of the sky. The
fluctuations appear visually consistent with the theoretical
expectation of a Gaussian and isotropic signal everywhere except
inside a small band very close to the Galactic plane\footnote{Note
  that \smica, being defined in harmonic space, employs a smooth
  filling process inside a small Galactic mask to prevent foreground
  residuals from leaking from low to high Galactic latitudes, and
  therefore appears visually different from the other three solutions
  in this respect; see Appendix \ref{sec:smica_appendix}.}

Each CMB map is accompanied by its own confidence mask outside which
the corresponding solution is considered statistically robust, shown
in Fig.~\ref{fig:cmb_mask_union}; for a definition of each mask, see
Appendices~\ref{sec:commander_appendix}--\ref{sec:smica_appendix}.
Accepted sky fractions are 75, 93, 76, and 89\,\%, respectively, for
\comrul, \nilc, \sevem, and \smica. These masks are denoted CS-CR75,
CS-NILC93, CS-SEVEM76, CS-SMICA89, respectively. The union of the
confidence masks accepts 73\,\% of the sky and is denoted \unionmask.
It is adopted as the default mask for evaluation purposes in this
paper.

In addition to the CMB maps from the full data set, the half-ring
frequency maps have been processed by each algorithm to provide
half-ring CMB maps. They are used to provide estimates of the
instrumental noise contribution to the foreground-cleaned maps in the
power spectrum analysis (see Sect.~\ref{sec:powspec_params}).  The
algorithms were also used to process Monte Carlo simulations: 1000
realizations of the CMB and 1000 realizations of noise.  They are not
used in the analyses presented in this paper, but are used by
\citet{planck2013-p09} and \citet{planck2013-p09a}.

The beam transfer functions of the foreground-cleaned CMB maps have
been estimated for each algorithm, as shown in
Fig.~\ref{fig:cmb_beam}.  The angular resolution of the \nilc, \sevem,
and \smica\ maps corresponds to a Gaussian beam with FWHM of 5\arcm.
The difference between \sevem\ and \nilc/\smica\ is due to their
different treatment of the
\healpix\footnote{\url{http://healpix.sourceforge.net}} pixel window
function \citep{gorski2005}.  The deviation of \nilc\ beam from a
Gaussian shape at $\ell > 2800$ is caused by the last needlet window
(see Appendix~\ref{sec:nilc_appendix}).  \comrul\ has a larger beam,
because it is defined explicitly as a weighted average of frequency
maps in pixel space.  Its resolution is equivalent to a Gaussian beam
with FWHM of approximately $7\parcm 4$.  The beam transfer functions
have been computed assuming the best-fit beam transfer function for
each frequency channel, and the uncertainties in the latter have not
been propagated to these estimates.

In Fig.~\ref{fig:cmb_rms} we show the standard deviation per pixel
among the four foreground-cleaned CMB maps downgraded to $\nside=128$,
and in Fig.~\ref{fig:cmb_diff} we show all pairwise difference
maps. Typical differences at high Galactic latitudes are smaller than
$5\,\mu\textrm{K}$. Considering the difference maps in more detail, it
is clear that the \comrul\ map is the most different from the other
three, whereas \nilc\ and \smica\ are the most similar.  This is not
completely unexpected, because while \comrul\ uses only frequencies
between 30 and 353\GHz\ in its solution, the other three codes
additionally include the dust-dominated 545 and 857\GHz\ maps.

This difference in data selection may explain some of the coherent
structures seen in Fig.\ \ref{fig:cmb_diff}.  In particular, the most
striking large-scale feature in the difference maps involving
\comrul\ is a large negative band roughly following the ecliptic
plane. This is where the ZLE \citep{planck2013-p03} is brightest.
Since the ZLE is also stronger at high frequencies, having a spectrum
close to that of thermal dust, it is possible that this pattern may be
an imprint of residual ZLE either in the \comrul\ map, or in all of
the other three maps. Both cases are plausible.  The \comrul\ solution
may not have enough high-frequency information to distinguish between
ZLE and normal thermal dust emission, and, by assuming a thermal dust
spectrum for the entire high-frequency signal at 353\GHz,
over-subtracts the ZLE at lower frequencies.  It is also possible that
the other three CMB solutions have positive ZLE residuals from
extrapolating the high-frequency signal model from 857\GHz\ to the CMB
frequencies.  Without an accurate and detailed ZLE model, it is
difficult to distinguish between these two possibilities.  It is of
course also possible that the true explanation is in fact unrelated to
ZLE, and the correlation with the ecliptic plane is accidental.  In
either case, it is clear that the residuals are small in amplitude,
with peak-to-peak values typically smaller than $10\,\mu\textrm{K}$,
of which by far the most is contained in a quadrupole aligned with the
ecliptic.  This provides additional evidence that residual ZLE is not
important for the CMB power spectrum and cosmological parameter
estimation, although some care is warranted when using these maps to
study the statistics of the very largest angular scales
\citep[e.g.,][]{planck2013-p09}; checking consistency among all four
maps for a given application alleviates much of this concern.

We end this section by showing in Fig.~\ref{fig:cmb_sim_res} a set of
residual maps derived by analysing the \ffp\ simulation with exactly
the same analysis approaches as applied to the data.  It is evident
that \smica\ produces the map with lowest level of residuals.
Considering the morphology in each case, we see that the main
contaminant for \comrul\ is under-subtracted free-free emission, while
for both \nilc\ and \sevem\ it is over-subtracted thermal dust
emission, and for \smica\ it is under-subtracted thermal dust
emission. However, at high latitudes and outside the confidence masks,
the residuals are generally below a few $\mu$K in amplitude. It is
also worth noting that each algorithm has been optimized (in terms of
model definition, localization parameters, etc.) for the data, and the
same configuration was subsequently used for the \ffp\ simulations
without further tuning.  The simulations presented here therefore
provide a conservative estimate of the residuals in the data.  This is
also reflected in the fact that the differences between CMB
reconstructions for the \ffp\ simulations are larger than those found
in the data. See Appendix~\ref{sec:ffp6_appendix} for further details.

\section{Power spectrum and cosmological parameters}
\label{sec:powspec_params}

\begin{figure}
  \begin{center}
    \includegraphics[width=\columnwidth]{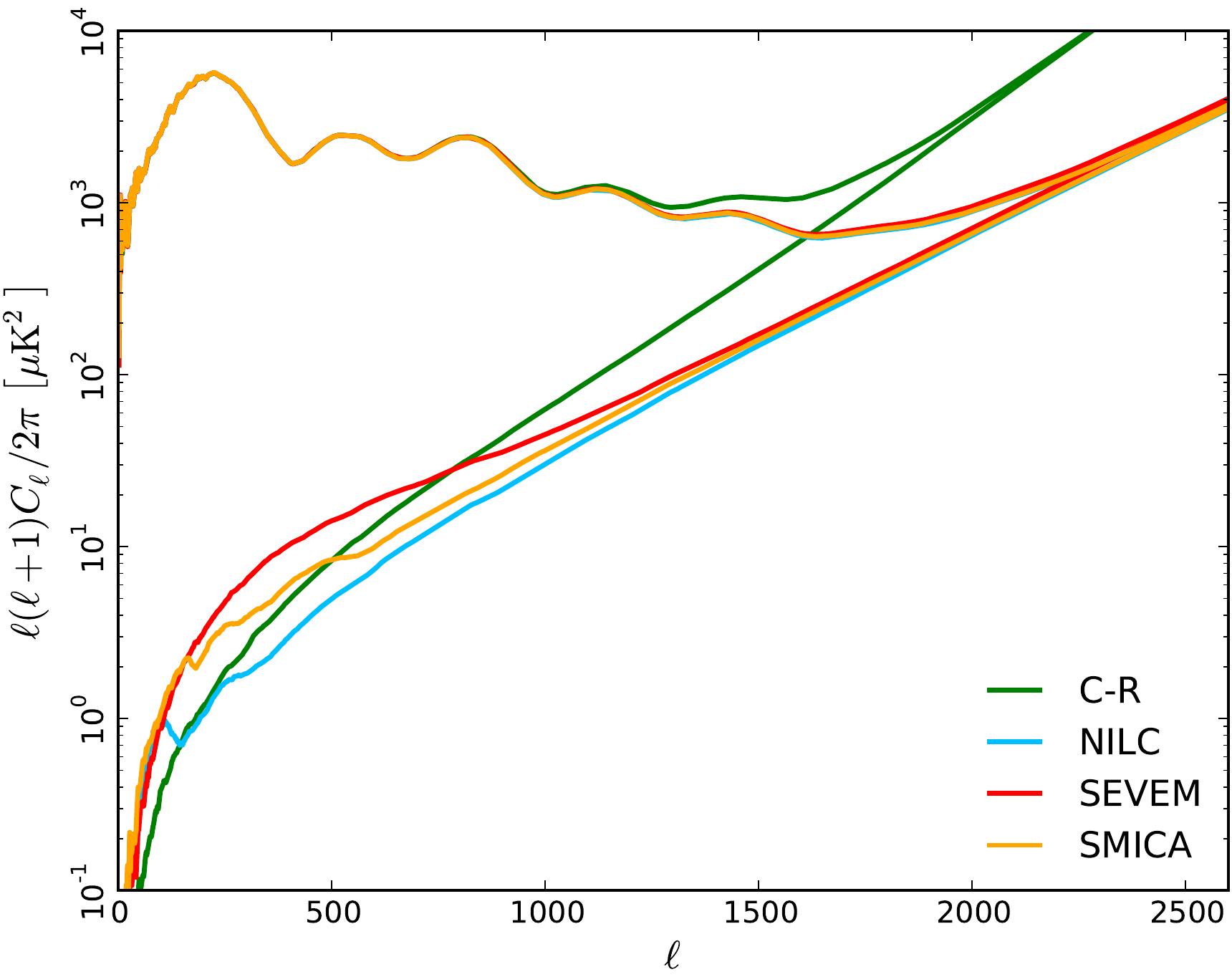}
  \end{center}
  \caption{Angular power spectra of the foreground-cleaned CMB maps
    and half-ring half-difference (HRHD) maps. The spectra have been
    evaluated using the \unionmask\ mask apodized with a
    30\arcm\ cosine function.}
  \label{fig:cmb_spectra}
\end{figure}

\begin{figure}
  \begin{center}
    \includegraphics[width=\columnwidth]{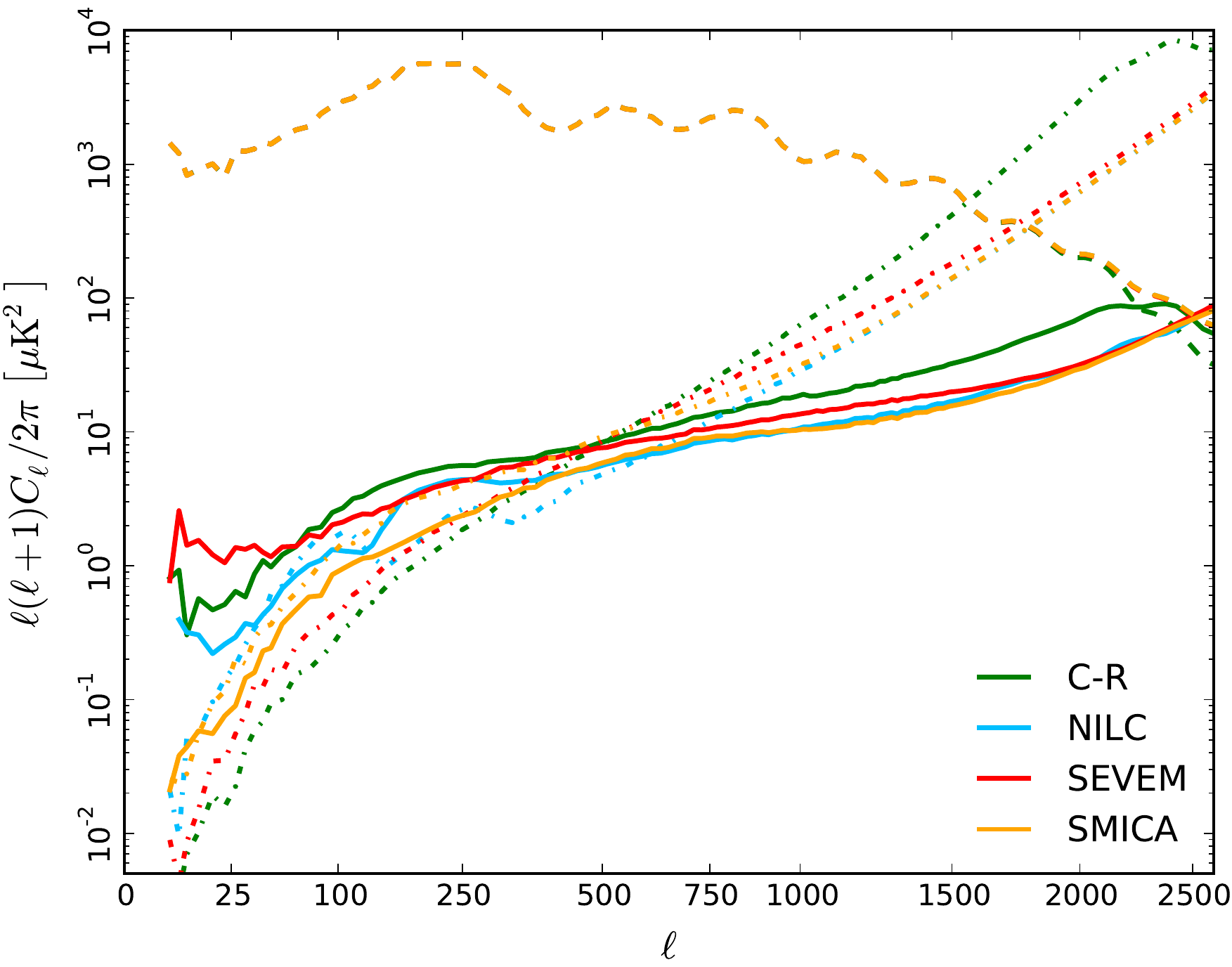}\\ 
    \includegraphics[width=\columnwidth]{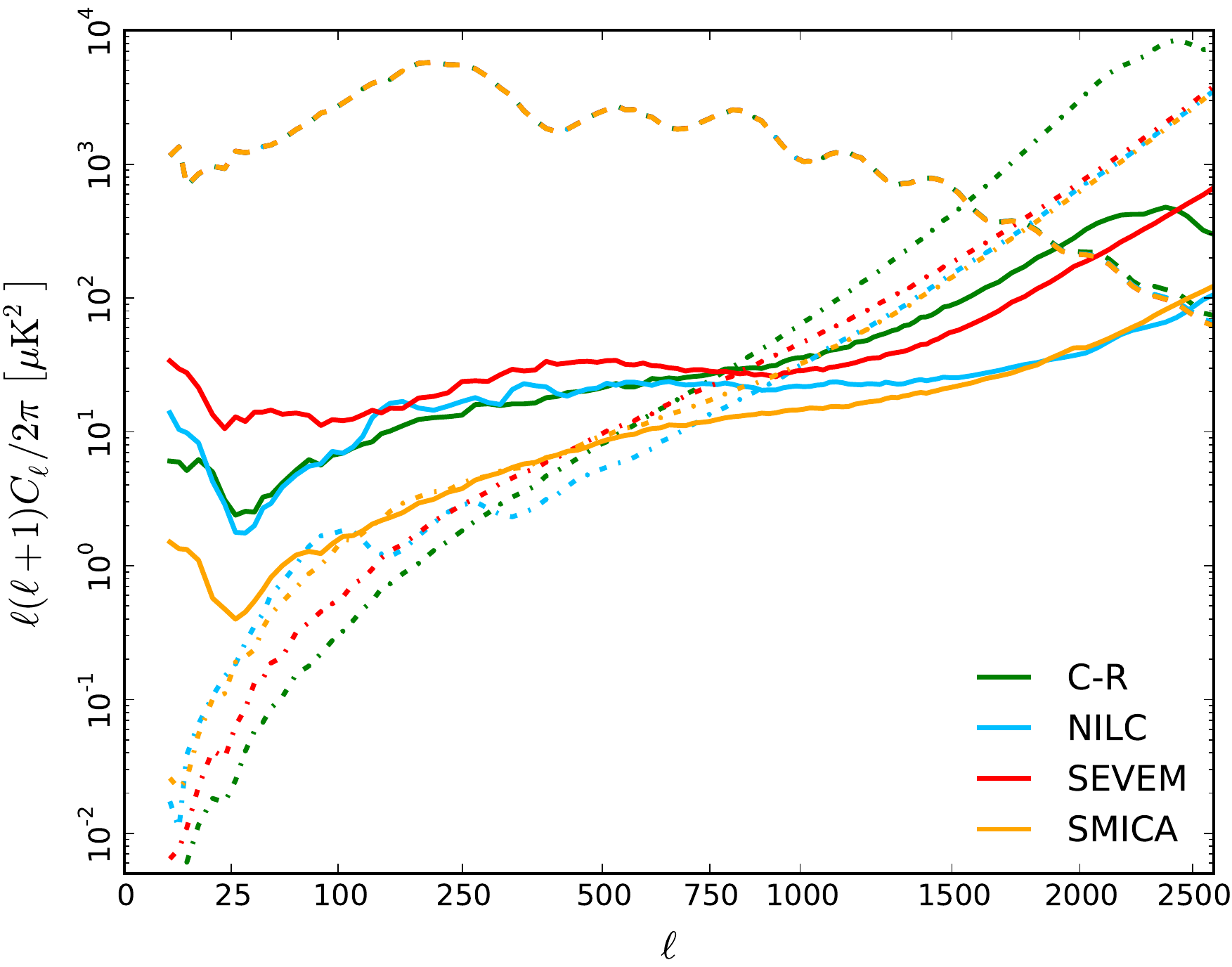}
  \end{center}
  \caption{Angular power spectra of \ffp\ simulated components
    evaluated over the common mask (top) and the common point source
    mask (bottom), both apodized with a 30\arcm\ cosine function.
    Three components are shown: the CMB (dashed line); noise
    (dot-dashed line); and the sum of all foregrounds (solid line). A
    nonlinear scale is used on the horizontal axis to show all the
    features of the spectra.}
  \label{fig:ffp6comp_spectra}
\end{figure}

\begin{figure}
  \begin{center}
    \includegraphics[width=\columnwidth]{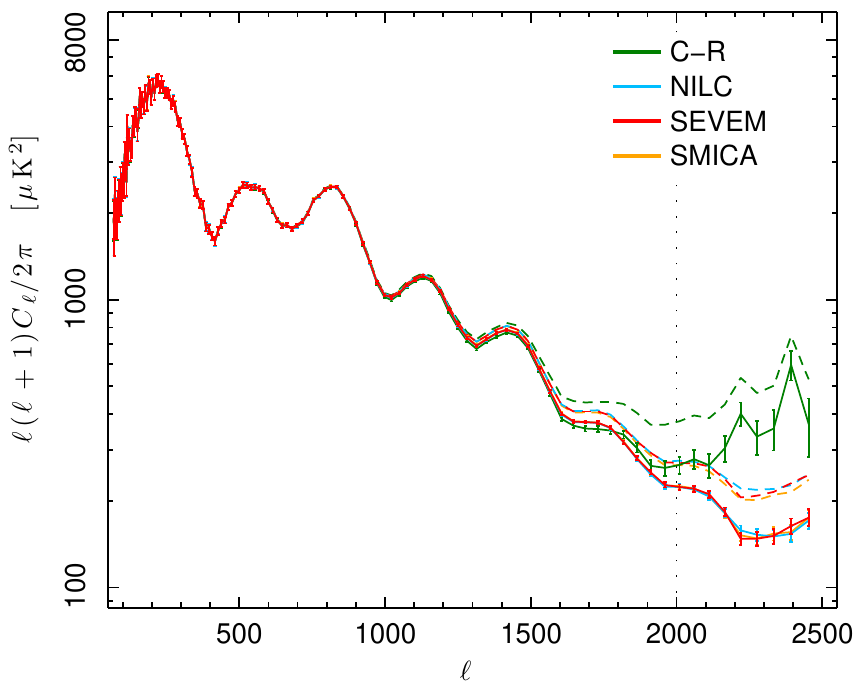}
  \end{center}
  \caption{Estimates of the CMB power spectra from the
    foreground-cleaned maps, computed by \XFaster.  The solid lines
    show the spectra after subtracting the best-fit model of residual
    foregrounds. The vertical dotted line shows the maximum multipole
    ($\ell = 2000$) used in the likelihood for fitting the foreground
    model and cosmological parameters (see
    Sect.~\ref{sec:xfaster_results} for further details).  The dashed
    lines show the spectra before residual foreground subtraction.}
  \label{fig:cb-ddx9}
\end{figure}

\begin{figure*}
  \begin{center}
    \includegraphics[width=\textwidth]{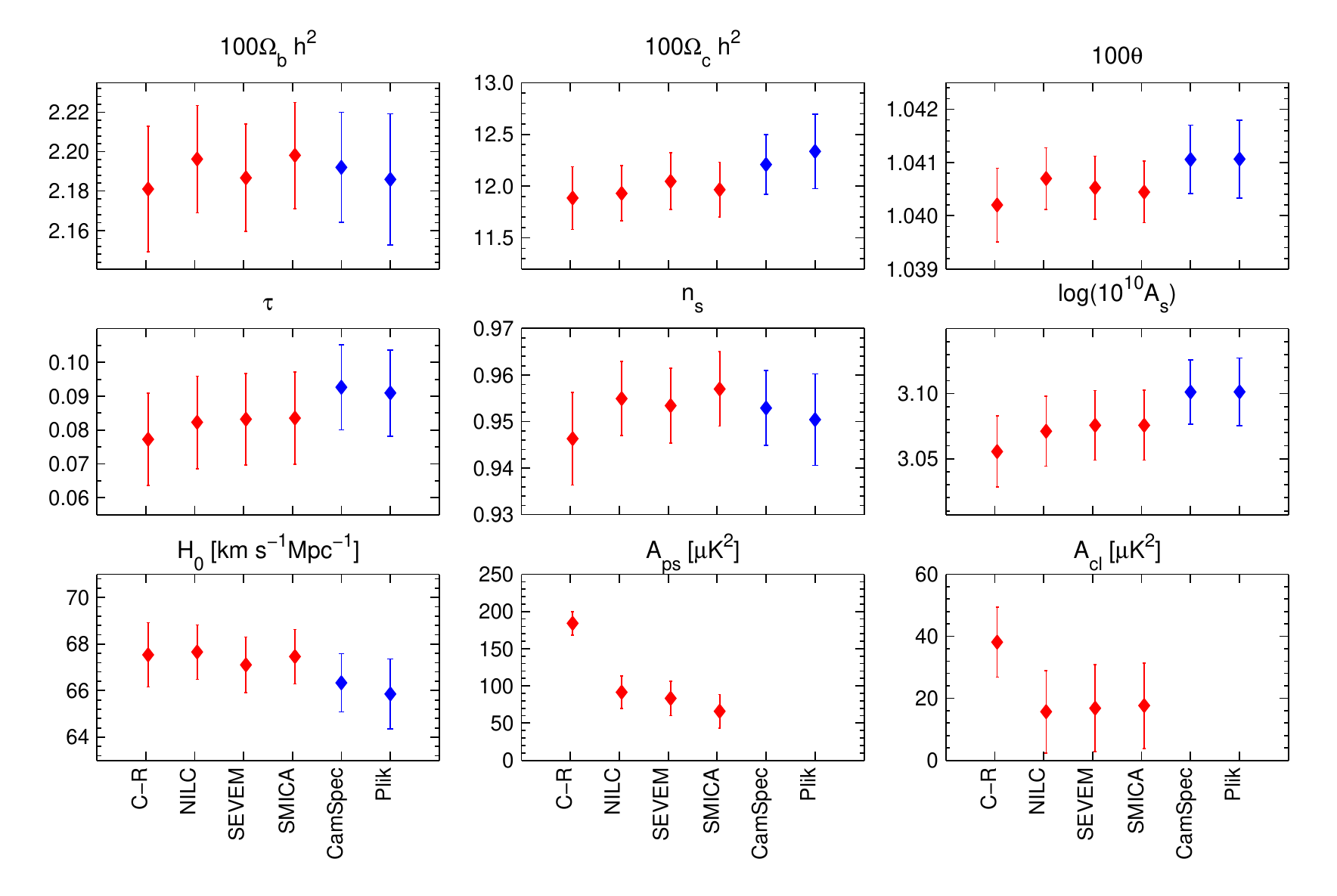}
    \caption{Comparison of cosmological and foreground parameter
      values estimated from the foreground-cleaned CMB maps for
      $\ell_{\mathrm{max}} = 2000$ (in red) and those obtained with
      \CamSpec\ and \Plik\ likelihoods (in blue). The values of the
      foreground parameters are not shown for \CamSpec\ and \Plik,
      since they use a different foreground model.}
    \label{fig:par-summ-ddx9}
  \end{center}
\end{figure*}

\begin{figure}
  \begin{center}
    \includegraphics[width=\columnwidth]{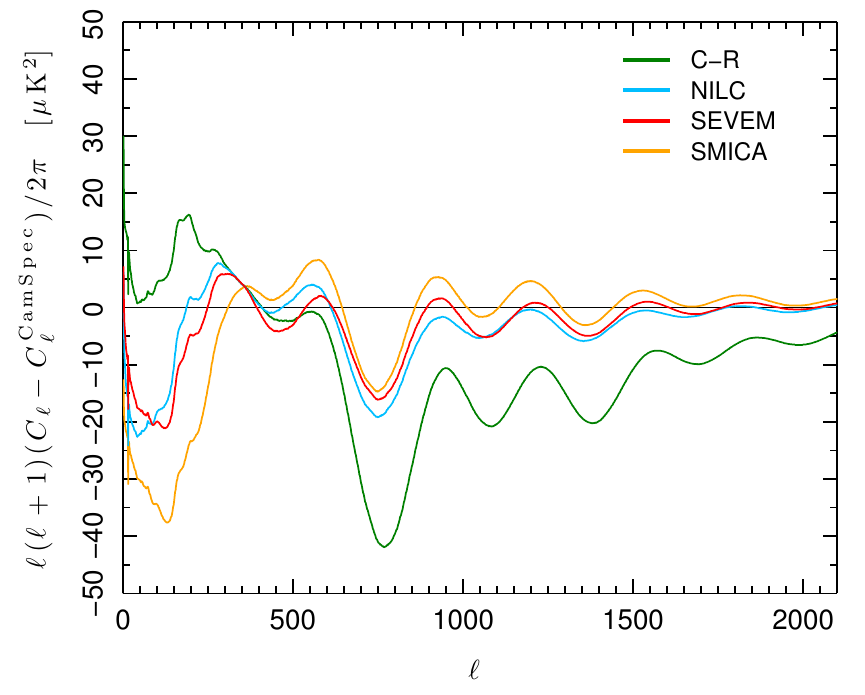}
    \caption{Residuals of all map-based best-fit models relative to
      \CamSpec\ best-fit model (assuming a prior on $\tau$) for
      $\ell_{\textrm{max}}=2000$.}
    \label{fig:bfm-ddx9}
  \end{center}
\end{figure}

\begin{figure}
  \begin{center}
    \includegraphics[width=\columnwidth]{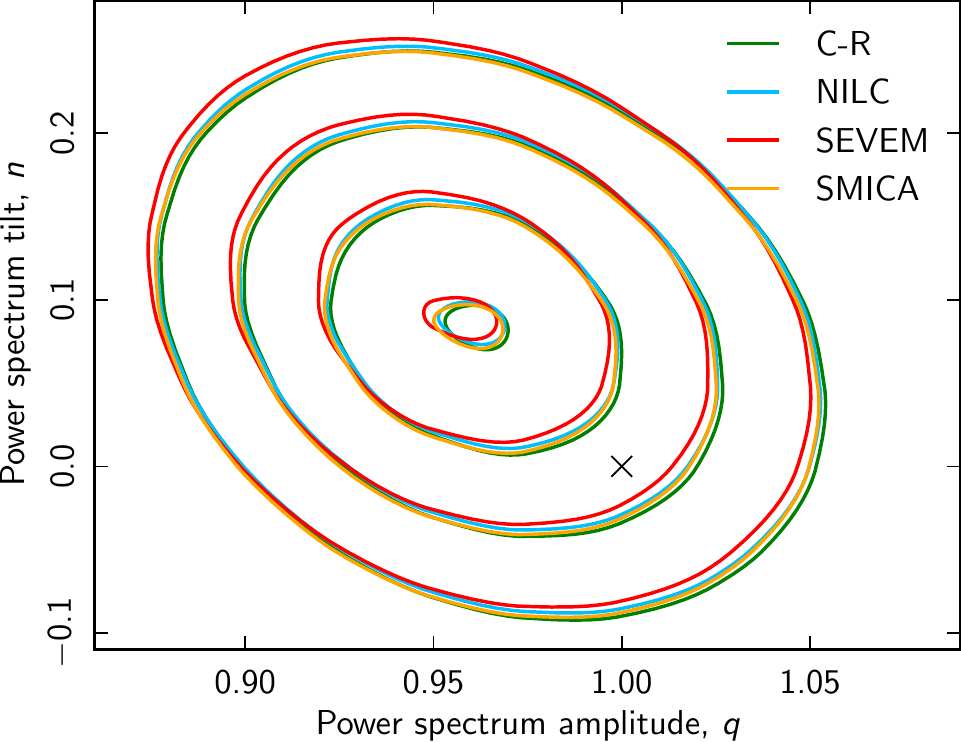}
  \end{center}
  \caption{Low-$\ell$ power spectrum amplitude and tilt constraints
    measured relative to the best-fit \Planck\ \LCDM\ model derived
    from foreground-cleaned CMB maps smoothed to $6^{\circ}$ FWHM. The
    cross shows the best-fit model $(q, n) = (1, 0).$}
  \label{fig:lowl_spectra}
\end{figure}

In this section we evaluate the foreground-cleaned maps in terms of
CMB power spectra and cosmological parameters.  Our purpose in doing
this is to show that the maps are consistent with the high-$\ell$
likelihood obtained from the cross-spectrum analysis of detector set and
frequency maps in \citet{planck2013-p08}, and with the cosmological
parameters derived from them in \citet{planck2013-p11}.  This also establishes
the consistency between \Planck's cosmological constraints and studies of the 
large-scale structure and higher order statistics of the CMB.

\subsection{Power spectra}

Figure~\ref{fig:cmb_spectra} shows the power spectra of the
foreground-cleaned CMB maps and the corresponding HRHD maps, evaluated
using the \unionmask\ mask with a 30\arcm\ cosine apodization.  The
spectra have been corrected for the effect of the mask and the beam
transfer function of each algorithm has been deconvolved.  The spectra
of the HRHD maps give an estimate of the instrumental noise
contribution to the power spectrum of the cleaned map.  The
correlations between the HFI half-ring frequency maps are inherited by
the half-ring CMB maps that use them as input.  At small angular
scales, the CMB solution comes almost entirely from data in the HFI
channels, and therefore the spectrum of the CMB HRHD maps is also
biased low.

At small angular scales, the effective noise levels of \nilc, \sevem,
and \smica\ are very similar, and lower than that of \comrul.  The
last has larger noise because it operates entirely in pixel space and
therefore applies the same weights to all multipoles. It cannot take
advantage of the changing signal-to-noise ratio of the frequency
channels with angular scale.

We can estimate the contribution of residual foregrounds to the
foreground-cleaned CMB maps by making use of the \ffp\ simulations.
In addition to processing the simulated frequency maps, the maps of
the individual input sky components were processed by the algorithms
after fixing their parameters or weights to the values obtained from
the ``observed'' maps.  Figure~\ref{fig:ffp6comp_spectra} shows the
power spectra of the simulated \ffp\ components, in this case CMB,
noise and the sum of the foreground components.  The top panel shows
the spectra computed using the union mask derived from the simulation
with a 30\arcm\ cosine apodization.  The total foreground contribution
becomes comparable to the CMB signal at $\ell \approx 2000$.  The
bottom panel shows the same computed with an apodized point source
mask applied to the maps (i.e., no diffuse masking, although this mask
does removes a large part of the Galactic plane).  The residual
foreground contribution is larger at all angular scales, but still it
only becomes comparable to the CMB signal at $\ell \approx 1800$ in
the worst case. For both masks, \smica\ has the smallest residual
foreground contamination at large angular scales, which is also
demonstrated in Fig.~\ref{fig:cmb_sim_res}.  A more detailed
examination of the contribution of the individual foreground
components to the power spectrum is in
Appendix~\ref{sec:ffp6_appendix}.

\subsection{Likelihood and cosmological parameters}

We estimate the binned power spectra with
\XFaster\ \citep{rocha2009,rocha2010b,XFaster-like2} and determine
cosmological parameter constraints using a correlated Gaussian
likelihood.  Parameter constraints are derived using a
Metropolis-Hastings Markov Chain Monte Carlo sampler.  To speed up
this process, we additionally use \texttt{PICO} \citep[Parameters for
  the Impatient COsmologist,][]{pico}, a tool which interpolates the
CMB power spectra and matter power spectra as a function of
cosmological parameters.
 
\subsubsection{Model and methods} 

We compute the power spectrum for each foreground-cleaned map over the
multipole range $2 \leq \ell \leq 2500$, while parameter constraints
are derived using only $70\leq \ell \leq 2000$; as shown in
Appendix~\ref{sec:ffp6_appendix} through simulations, modelling errors
become non-negligible between $\ell=2000$ and 2500.  For parameter
estimation, we adopt a standard six-parameter \LCDM\ model, and impose
an informative Gaussian prior of $\tau=0.0851 \pm 0.014$, since
polarization data are not included in this analysis.

While the foreground-cleaned maps should have minimal contamination
from diffuse Galactic emission, they do contain significant
contamination from unresolved extragalactic sources.  These
contributions are most easily modelled in terms of residual power
spectra, therefore we marginalize over the corresponding parameters at
the power spectrum level.  To the six \LCDM\ parameters, describing
the standard cosmology, we add two foreground parameters,
$A_{\mathrm{ps}}$, the amplitude of a Poisson component (and hence
constant, $C_{\ell} = A_\mathrm{ps}$), and $A_{\mathrm{cl}}$, the
amplitude of a clustered component with shape $D_{\ell} =
\ell(\ell+1)C_{\ell}/2\pi \propto \ell^{\,0.8}$.  Both are expressed in
terms of $D_{\ell}$ at $\ell = 3000$ in units of $\microK^2$.

The power spectrum calculation is based on the half-ring half-sum
(HRHS) and HRHD CMB maps (see Sect.~\ref{sec:cmb_maps}); the latter is
used to estimate the noise bias in the power spectra extracted from
the HRHS maps.  From these, we calculate the pseudo-spectra,
$\tilde{C}_{\ell}$ and $\tilde{N}_{\ell}$ \citep{master02},
respectively, after applying the \unionmask\ mask.  These are used as
inputs to \XFaster\ together with the beam transfer functions provided
by each method (see Fig.~\ref{fig:cmb_beam}).

To avoid aliasing of power from large to small scales, which would add
an offset between the signal-plus-noise and noise pseudo-spectra at
high $\ell$, we use the apodized version of the \unionmask\ mask.  The
known mismatch in the noise level between the spectra due to the
correlation between the half-ring maps is not explicitly corrected.
It is left to be absorbed into the two foreground parameters.

Using the pseudo-spectra and \XFaster, we then reconstruct an estimate
of the power spectrum of each foreground-cleaned HRHS map, removing
the noise bias as estimated from the corresponding HRHD map.  To this
end we apply an iterative scheme starting from a flat spectrum model.
The result is a binned power spectrum and the associated Fisher matrix,
which are then used to construct the likelihood, approximated here by
a correlated Gaussian distribution.

To study consistency in the low-$\ell$ range, we fit a two-parameter
$q$--$n$ (amplitude-tilt) model relative to the \Planck\ best-fit
\LCDM\ model on the form, $C_{\ell} = q(\ell/\ell_{\textrm{pivot}})^n
C_{\ell}^{\textrm{bf}}$, using a pixel-space likelihood for maps
smoothed to $6^{\circ}$ FWHM; see \citet{planck2013-p08} for further
algorithmic details.

\subsubsection{Results}
\label{sec:xfaster_results}

We perform the power spectrum and parameter estimation analysis for
both the data and the \ffp\ simulations described in
Sect.~\ref{sec:data}.  The results for the latter are given in
Appendix~\ref{sec:ffp6_appendix}.

Figure~\ref{fig:cb-ddx9} shows estimates of the angular power spectrum
for each foreground-cleaned map, with the uncertainties given by the
Fisher matrix.  The parameter summary given in
Fig.~\ref{fig:par-summ-ddx9} shows the parameter constraints derived
using multipoles between $\ell=70$ and 2000, and compares these to
results obtained with the \CamSpec\ and \Plik\ likelihoods
\citep{planck2013-p08}.

Differences in the power spectra at high $\ell$ are mostly absorbed by
the two-parameter foreground model, rendering consistent cosmological
parameters.  For example, the high-$\ell$ power excess seen in the
\comrul\ map is well-fitted in terms of residual point sources, which
makes intuitive sense, considering the lower angular resolution of
this map (see Sect.~\ref{sec:cmb_maps}).  However, the
\LCDM\ parameter uncertainties derived from the four codes are very
consistent. This indicates that most of the cosmological information
content above $\ell\ge 1500$ is degenerate with the extragalactic
foreground model, and a more sophisticated foreground treatment is
required in order to recover significant cosmological parameter
constraints from these scales. Beyond this, deviations among
cosmological parameters are small and within $1\,\sigma$ for all
methods and most of the parameters. Further, the parameters derived
from the four foreground-cleaned CMB maps are in good agreement with
those obtained by \CamSpec\ and \Plik\ using cross-spectra; departures
are well within $1\,\sigma$ for most parameters.

Inspecting the differences between the best-fit models derived from
the four foreground-cleaned maps and from \CamSpec\ plotted in
Fig.~\ref{fig:bfm-ddx9}, we find that the relative residuals are
within $40\,\mu$K$^2$ for all multipole ranges, and smaller than
$20\,\mu$K$^2$ at high $\ell$. This can be compared to the
corresponding residuals for the \ffp\ simulation shown in
Appendix~\ref{sec:ffp6_appendix}.

The likelihood used for this analysis does not take into account some
systematic effects that will affect our foreground-cleaned CMB maps,
such as relative calibration uncertainties between the frequency
channel maps used to construct them, or their beam uncertainties.
These effects are accounted for in the likelihoods in
\citet{planck2013-p08}.  We have also adopted a very simple
two-parameter model for the residual extragalactic foregrounds.
Despite these limitations, the four CMB maps yield cosmological
parameters in agreement with the cross-spectrum based likelihoods for
a basic six-parameter \LCDM\ model.  Thus we can be confident that the
CMB maps are consistent with the power spectrum analysis.

Before concluding this section, we show in Fig.~\ref{fig:lowl_spectra}
the results from a two-parameter fit of an amplitude-tilt model to
each of the four foreground-cleaned maps, downgraded to $6^{\circ}$
and repixelized at an $\nside=32$ grid.  Clearly, the maps are
virtually identical on large angular scales measured relative to
cosmic variance, with any differences being smaller than $0.1\,\sigma$
in terms of cosmological parameters. However, it is worth noting that
the best-fit model, $(q,n)=(1,0)$, is in some tension with the
low-$\ell$ spectrum, at about $1.7\,\sigma$ in this plot. The same
tension between large and small angular scales is observed in
\citet{planck2013-p08} and \citet{planck2013-p11} with higher
statistical significance using the full \Planck\ likelihood.
Irrespective of physical interpretation, the calculations presented
here demonstrate that these low-$\ell$ features are robust with
respect to component separation techniques.

\section{Higher-order statistics}
\label{sec:ng}

The foreground-cleaned CMB maps presented in this paper are used as
inputs for most \Planck\ analyses of higher-order statistics,
including non-Gaussianity studies \citep{planck2013-p09a}, studies of
statistical isotropy \citep{planck2013-p09}, gravitational lensing by
large-scale structure \citep{planck2013-p12}, and of the integrated
Sachs-Wolfe effect \citep{planck2013-p09a}. In this section we provide
a summary of the non-Gaussianity and gravitational lensing results.

\subsection{Non-Gaussianity}

Primordial non-Gaussianity is typically constrained in terms of the
amplitude, $f_{\textrm{NL}}^{\textrm{local}}$, of the quadratic
corrections to the gravitational potential, as well as by means of the
three-point correlation function based on different triangle
configurations. The results from these calculations for the
foreground-cleaned CMB maps are presented in
\citet{planck2013-p09a}. After subtraction of the lensing-ISW
correlation contribution, the final result is
$f_{\textrm{NL}}^{\textrm{local}}=2.7\pm 5.8$, as estimated from the
\smica\ map using the KSW bispectrum estimator \citep{komatsu2005},
consistent within $1\,\sigma$ with results from other methods and
foreground-cleaned maps.

Uncertainties are evaluated by means of the \ffp\ simulations, and
potential biases are studied using both Gaussian and non-Gaussian CMB
realizations. In particular, when a detectable level of primordial
non-Gaussianity ($f_{\textrm{NL}}^{\textrm{local}}=20.4075$) is
injected into the \ffp\ simulations, each foreground-cleaned map
yielded a positive detection within $2\,\sigma$ of the expected value,
recovering values of $f_{\textrm{NL}}^{\textrm{local}}=8.8\pm 8.6$,
$19.0\pm 7.5$, $11.1\pm 7.6$ and $19.7\pm 7.4$ for \comrul, \nilc,
\sevem, \smica, respectively.  We see that \nilc\ and
\smica\ demonstrate the best recovery of the injected non-Gaussianity,
and we favoured the latter for non-Gaussian studies for its faster
performance over \nilc.  The foreground-cleaned CMB maps presented in
this paper do not provide significant evidence of a non-zero value of
$f_{\textrm{NL}}^{\textrm{local}}$, and realistic simulations show
that the component separation methods do not suppress real
non-Gaussian signatures within expected uncertainties.  The
implications of these results in terms of early Universe physics are
discussed in the relevant papers
\citep{planck2013-p09a,planck2013-p17}.

\subsection{Gravitational lensing by large-scale structure}

\begin{figure}
  \begin{center}
    \includegraphics[width=\columnwidth]{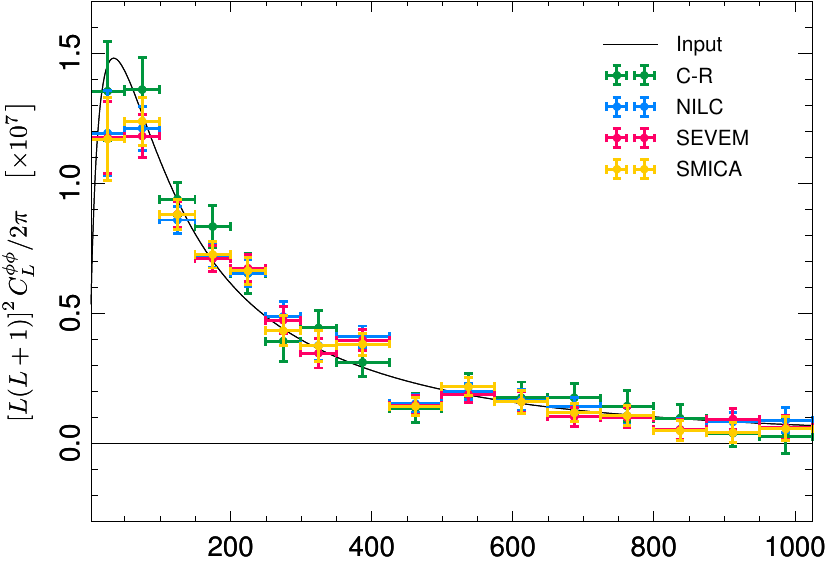}\\
    \includegraphics[width=\columnwidth]{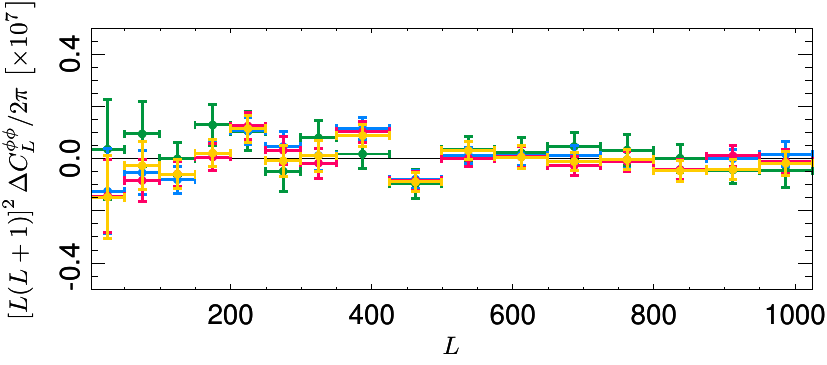}
  \end{center}
  \caption{Lensing power spectrum estimates from \ffp\ simulations
    using an apodized mask covering $f_{\rm{sky},2} \simeq 0.70$ of
    the sky.}
  \label{fig:compsep_0100_gal10}
\end{figure}

Gravitational lensing by the intervening matter imprints a
non-Gaussian signature in the CMB, which allows the reconstruction of
the gravitational potential integrated along the line of sight to the
last scattering surface.  In \citet{planck2013-p12}, this effect has
been detected at a high significance level (greater than $25\,\sigma$)
using the \Planck\ temperature maps. Specifically, the lensing induced
correlations between the total intensity and its gradients have been
used to reconstruct a nearly full sky map of the lensing potential
$\phi$, which has been used for further studies on \Planck\ data,
including the detection of a non-zero correlation with the ISW
\citep{planck2013-p09a,planck2013-p14} and other tracers of
large-scale structure (notably, significant correlation with the CIB
is reported in \citealt{planck2013-p13}), as well as the estimate of
the power spectrum of the lensing potential and the associated
likelihood. The latter was constructed using a simple minimum variance
combination of the 143 and 217\GHz\ maps on about 70\,\% of the sky,
as well as subtracting dust contamination using the
857\GHz\ \Planck\ channel as a template \citep{planck2013-p12}. These
lensing results have improved the cosmological constraints from
\Planck\ \citep{planck2013-p11}.

The foreground-cleaned CMB maps described in Sect.~\ref{sec:cmb_maps}
were used to perform a lensing extraction on a larger sky fraction,
reaching about $87\,\%$ of the sky.  We found the lensing power
spectrum to be in good agreement with the one obtained using the
minimum variance combination, i.e., the signal agrees within
$1\,\sigma$ in the majority of the angular domain bins, and is
characterized by an equivalent uncertainty. The foreground-cleaned
maps were further exploited on the baseline $70\,\%$ sky fraction for
assessing the robustness of the main reconstruction against the
foreground contamination \citep{planck2013-p12}.

We show that the component separation algorithms presented in this
paper do not bias the lensing reconstruction in the case of the large
sky fraction considered here. We consider \ffp\ simulations including
noise and lensed CMB signal, propagated through each of the component
separation algorithms described in Sect.~\ref{sec:comp_sep_approach}.
We perform a lensing potential reconstruction in the pixel domain
based on the CMB maps processed by the four component separation
methods using the \texttt{metis} algorithm described in
\citet{planck2013-p12}. This method uses the quadratic estimator
presented in \citet{Okamoto:2003zw}, which corrects for the mean-field
bias caused by extra sources of statistical anisotropy in addition to
the CMB.

For each method, we combine the masks of CO regions, nearby galaxies
and compact objects as defined in \citet{planck2013-p12}, with the
CG90 mask described in Sect.~\ref{sec:data}. This procedure results in
masks with sky fractions $f_{\rm{sky}} = 0.836, 0.851, 0.850, 0.846$
for \comrul, \nilc\, \sevem, and \smica, respectively.

We estimate the lensing potential power spectrum, $C_L^{\phi\phi}$,
following the methodology described in \citet{planck2013-p12}.  It
consists of a pseudo-$C_\ell$ estimate based on a highly-apodized
version of the lensing potential reconstruction, which has an
effective available sky fraction $f_{\rm{sky},2} = 0.648, 0.690,
0.686, 0.683$ for \comrul, \nilc, \sevem\ and \smica,
respectively. The band-power reconstructions in 17 bins in the range
$2 \le \ell \le 1025$ are plotted in
Fig.~\ref{fig:compsep_0100_gal10}, as well as the residuals relative
to the theoretical lens power spectrum. All algorithms achieved an
unbiased estimation of the underlying lensing power spectrum, with
$\chi^2 = 10.58, 17.34, 18.54, 15.30$, for \comrul, \nilc, \sevem, and
\smica\ respectively, with 17 degrees of freedom.  The associated
probability-to-exceed (PTE) values are $83\,\%, 36\,\%, 29\,\%,
50\,\%$.

The power spectrum estimates are in remarkable agreement with each
other.  However, the \comrul\ solution has significantly larger
uncertainties, as expected from its lower signal-to-noise ratio to
lensing due to its larger beam.  These results on simulated
foreground-cleaned CMB maps demonstrate that the component separation
algorithms do not alter the lensing signal, and this provides a
strategy for achieving a robust lensing reconstruction on the largest
possible sky coverage. The foreground-cleaned maps have been used in
\citet{planck2013-p12} to obtain lensing potential estimates on
$87\,\%$ of the sky.

\section{Foreground components}
\label{sec:foreground_components}

In this section we consider the diffuse Galactic components, and
present full-sky maps of thermal dust and CO emission, as well as a
single low-frequency component map representing the sum of
synchrotron, AME, and free-free emission.  Our all-sky CO map is a
``type 3'' product as presented in \citet{planck2013-p03a}.  To assess
the accuracy of these maps, we once again take advantage of the
\ffp\ simulation. The \comrul\ method used in the following is
described in Appendix~\ref{sec:commander_appendix} and consists of a
standard parametric Bayesian MCMC analysis at low angular resolution,
followed by a generalized least-squares solution for component
amplitudes at high resolution.

\subsection{Data selection and processing}
\label{sec:fg_data}

\begin{figure}
  \center{ 
\begin{subfigure}{\columnwidth} 
   \includegraphics[width=\columnwidth]{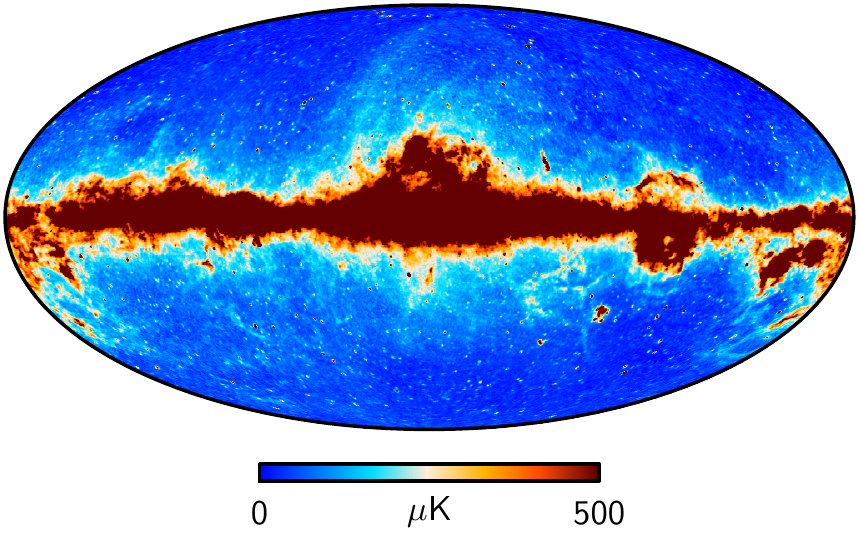}
   \caption{Low-frequency component amplitude at 30\GHz}
\vspace*{2mm}
   \end{subfigure}      
\begin{subfigure}{\columnwidth} 
   \includegraphics[width=\columnwidth]{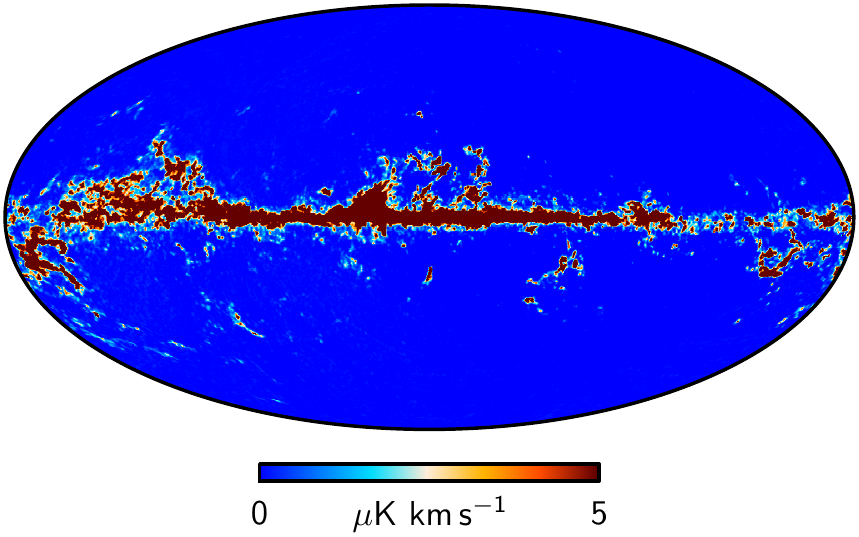}
   \caption{CO amplitude at 100\GHz}
\vspace*{2mm}
   \end{subfigure}      
\begin{subfigure}{\columnwidth} 
   \includegraphics[width=\columnwidth]{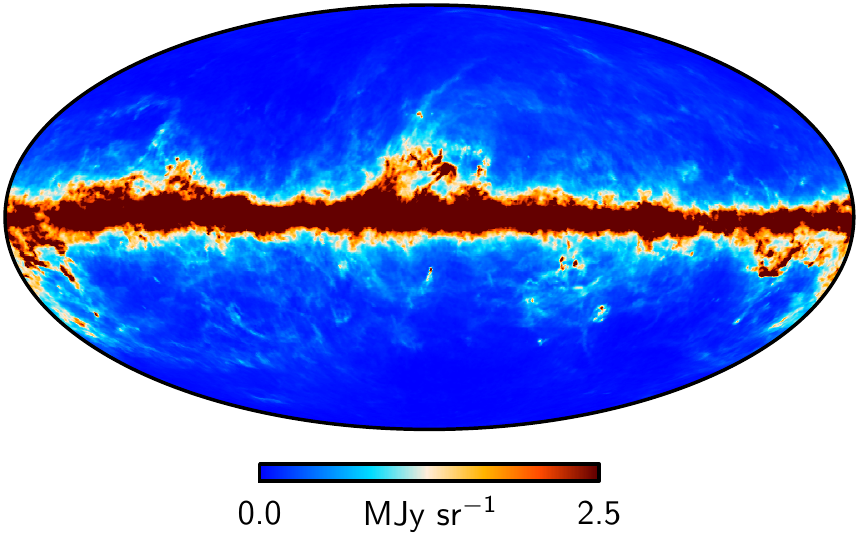}
   \caption{Thermal dust amplitude at 353\GHz}
   \end{subfigure}      
   }
   \caption{Posterior mean foreground amplitude maps derived from the
  low-resolution analysis. From top to bottom are shown the
  low-frequency, CO and thermal dust emission
  maps.}  
  \label{fig:planck_lowres_fg_amps}
\end{figure}

\begin{figure}
  \center{ 
\begin{subfigure}{\columnwidth}
\includegraphics[width=\columnwidth]{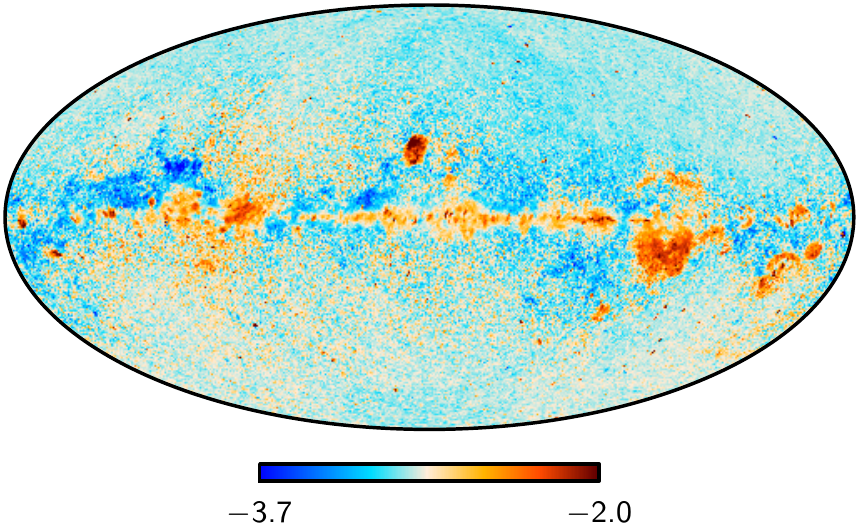} 
\caption{Low-frequency index}
\vspace*{2mm}
\end{subfigure}

\begin{subfigure}{\columnwidth}
\includegraphics[width=\columnwidth]{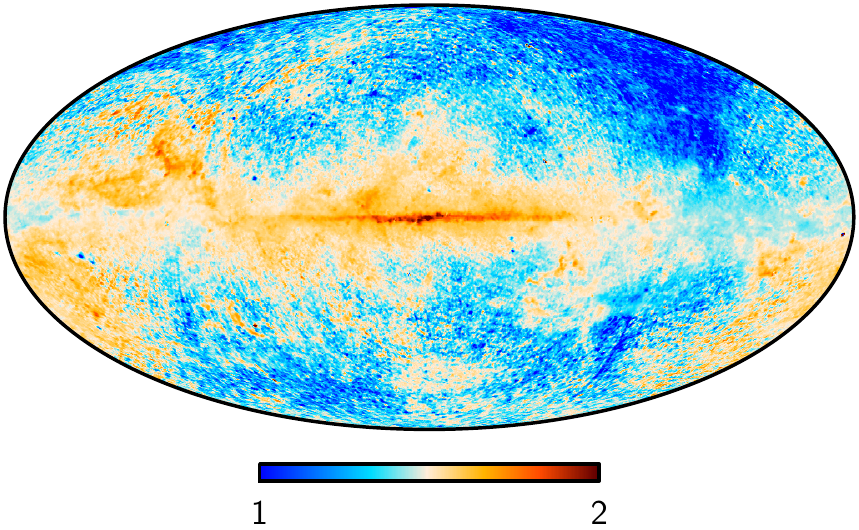}
\caption{Dust emissivity}
\end{subfigure}
  } \caption{Posterior mean spectral parameter maps derived from the
  low-resolution analysis. The top panel shows the power law index of
  the low-frequency component, and the bottom panel shows the
  emissivity index of the one-component thermal dust model. Note that
  the systematic error due to monopole and dipole uncertainties is
  significant for the dust emissivity in regions with a low thermal
  dust amplitude.  } \label{fig:planck_lowres_fg_inds}
\end{figure}

\begin{figure}
  \center{ \includegraphics[width=\columnwidth]{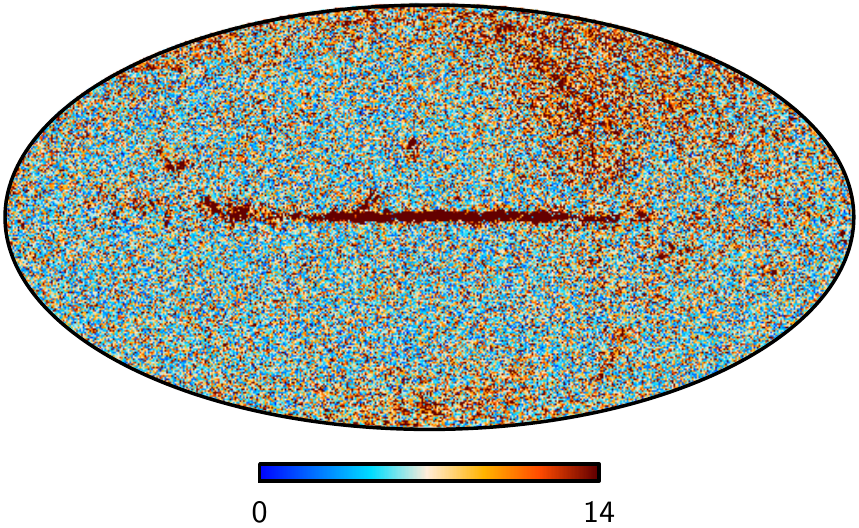}
  } \caption{$\chi^2$ per pixel for the joint CMB and foreground
  analysis. The expected value for an acceptable fit is 7,
  corresponding to the number of frequency bands used in this
  analysis. The pixels with high values can be classified into two
  types, due to either modelling errors (i.e., high residuals in the
  Galactic plane) or to un-modelled correlated noise (i.e., stripes
  crossing through low dust emission
  regions).}  \label{fig:lowres_chisq}
\end{figure}

\begin{figure*}[t]
  \center{
  \includegraphics[width=0.33\textwidth]{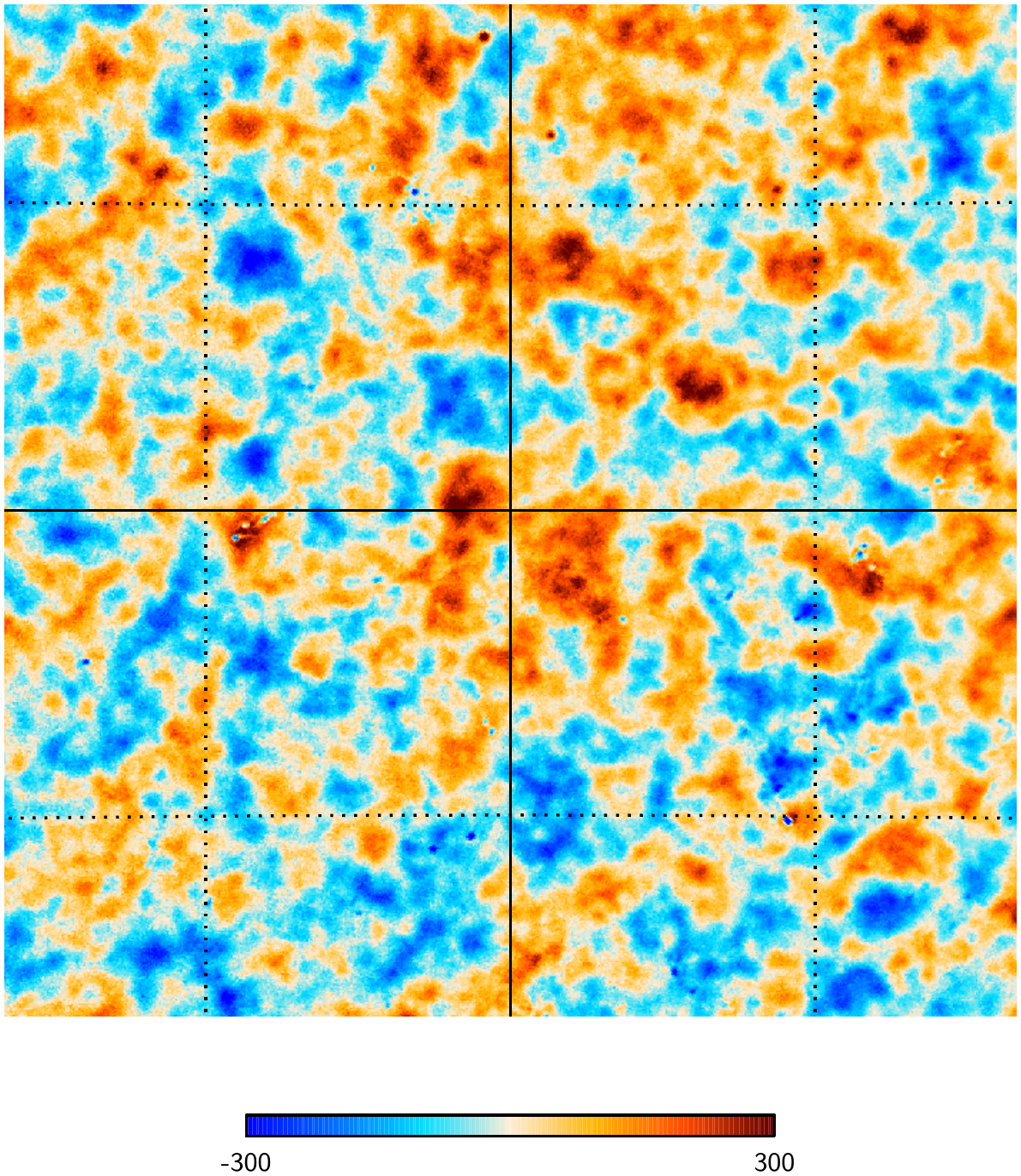}
  \includegraphics[width=0.33\textwidth]{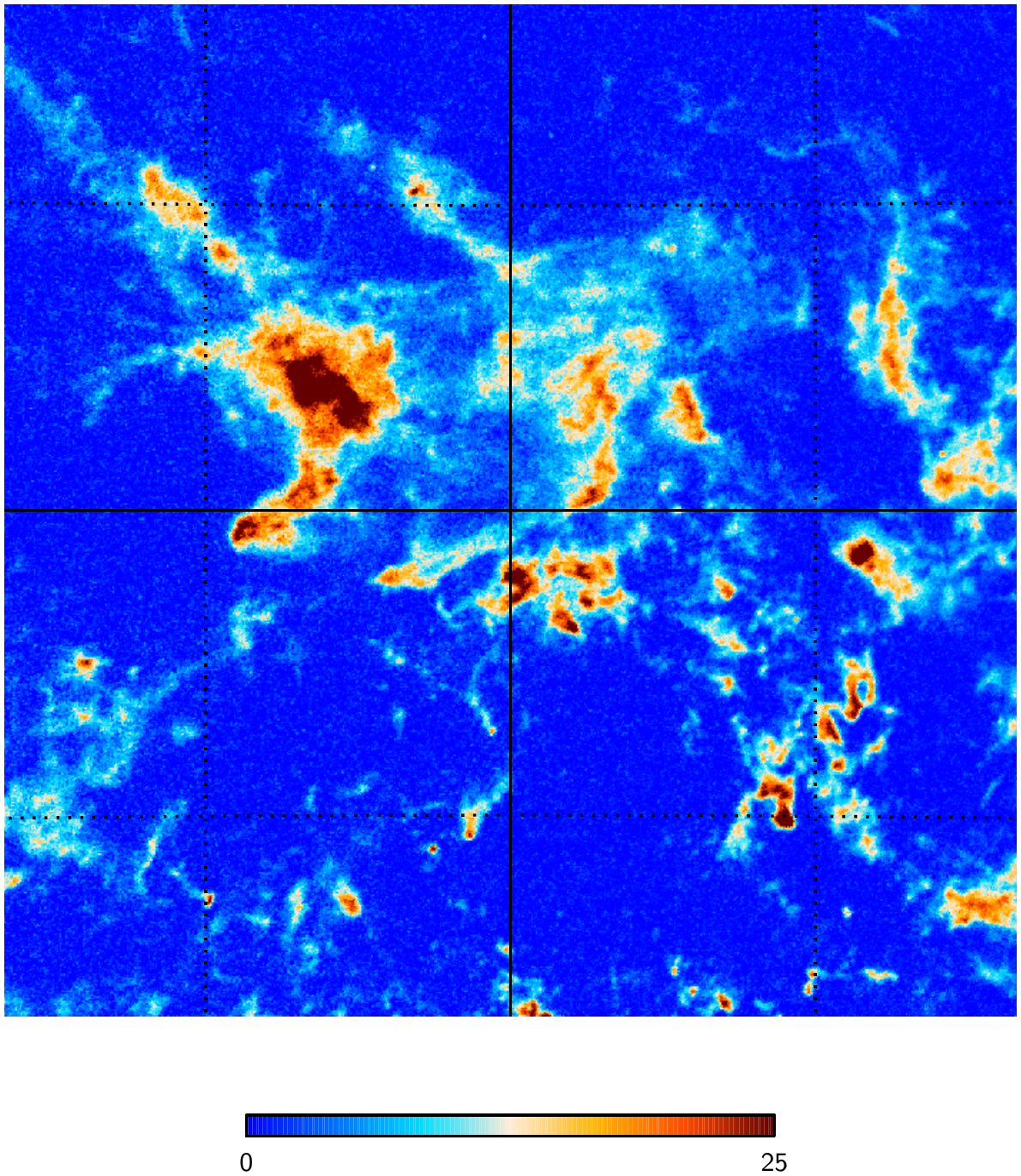}
  \includegraphics[width=0.33\textwidth]{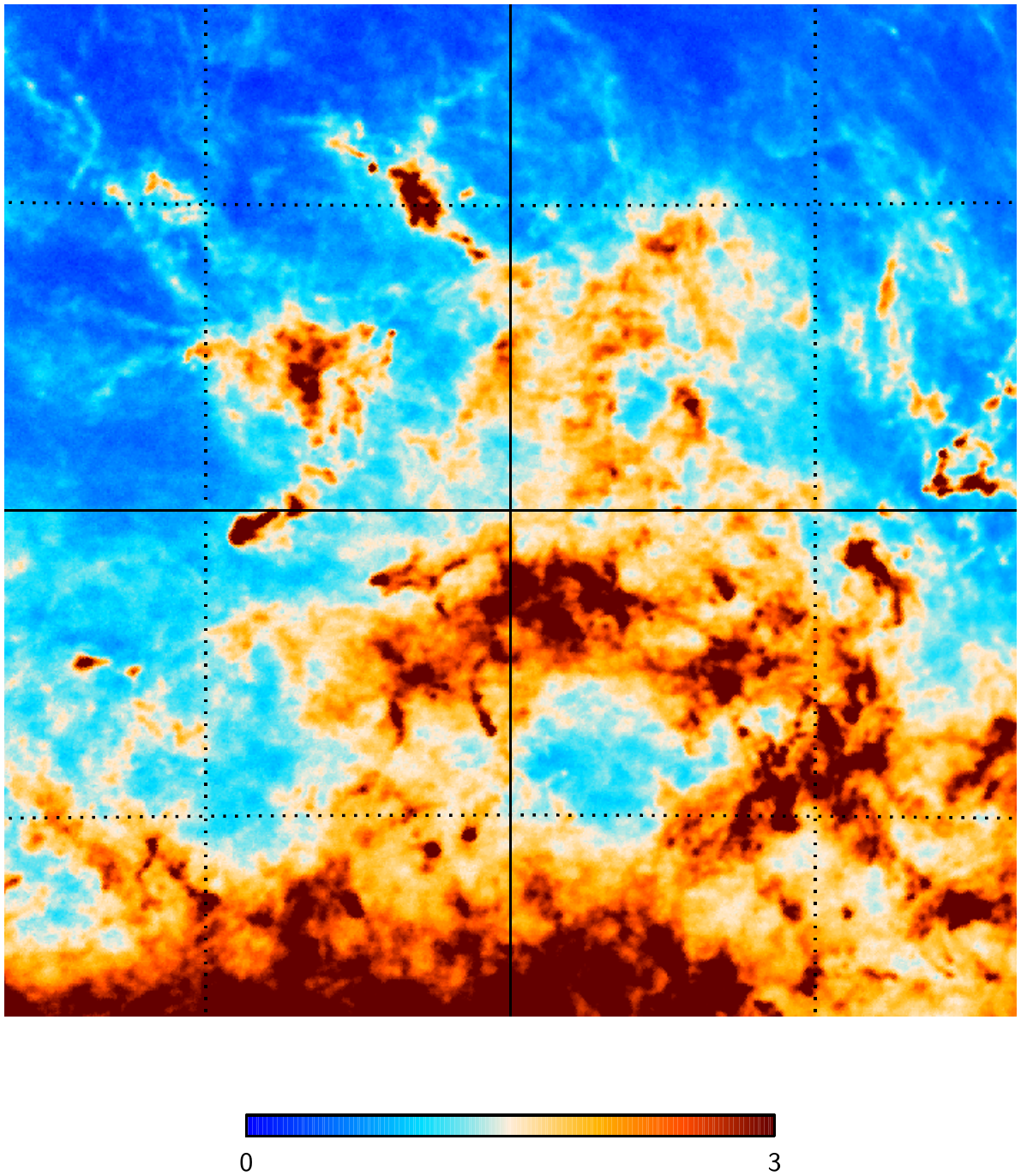}
  }
\vspace*{-15mm} 
  \center{
\begin{subfigure}{0.33\textwidth} 
  \includegraphics[width=\textwidth]{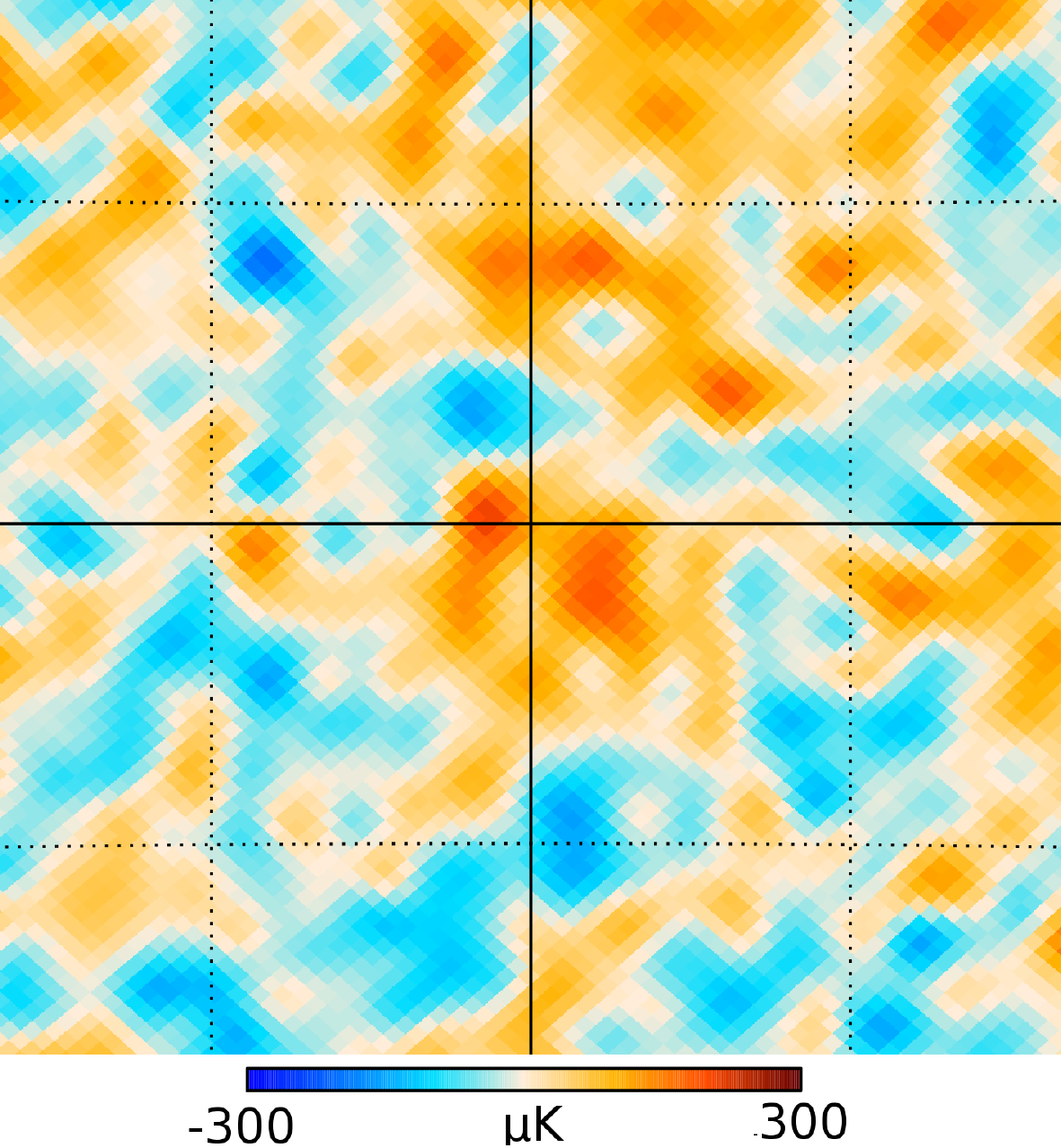}
   \caption{CMB amplitude}
   \end{subfigure}      
\begin{subfigure}{0.33\textwidth} 
  \includegraphics[width=\textwidth]{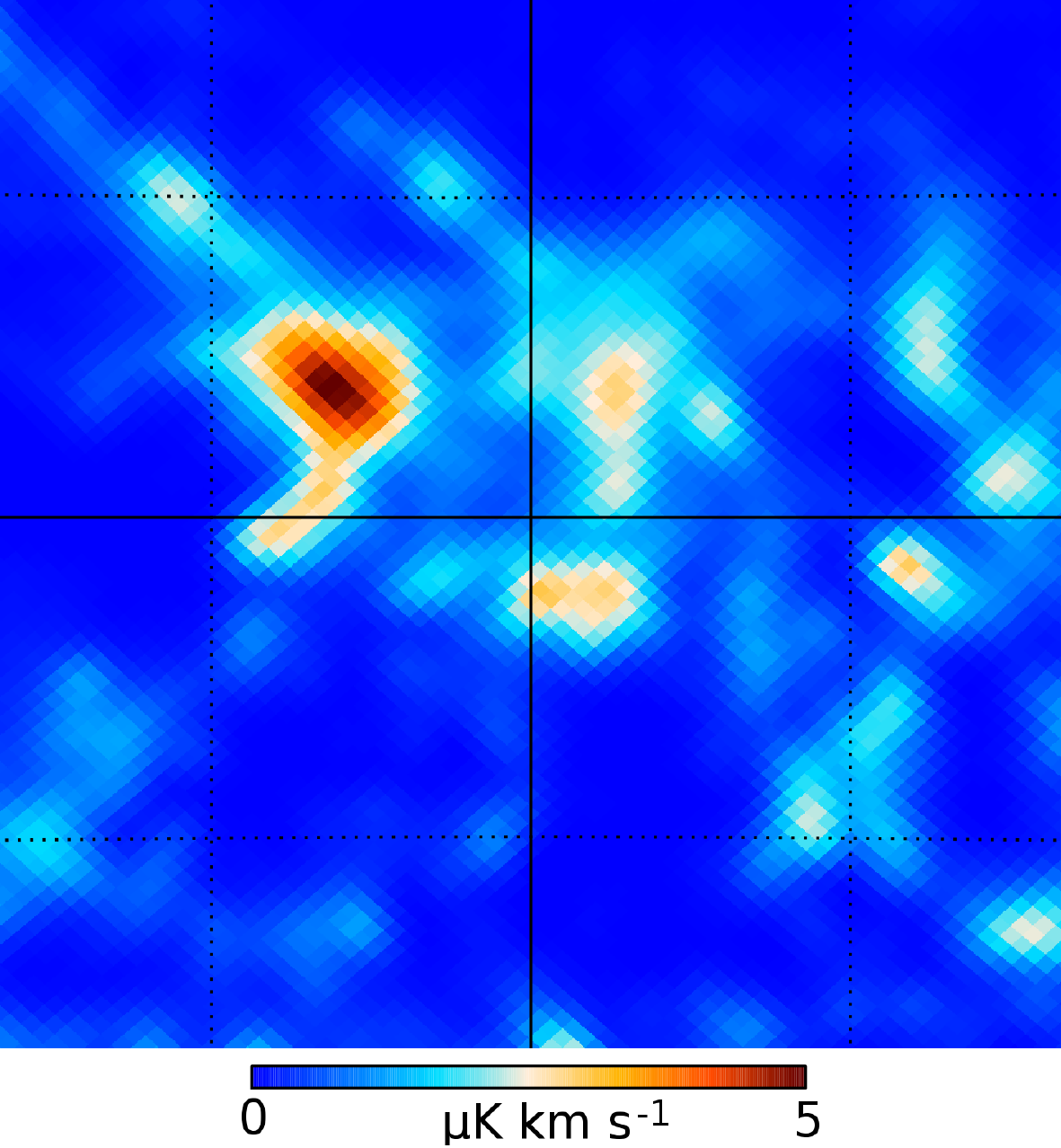}
   \caption{CO amplitude at 100\GHz}
   \end{subfigure}      
\begin{subfigure}{0.33\textwidth} 
  \includegraphics[width=\textwidth]{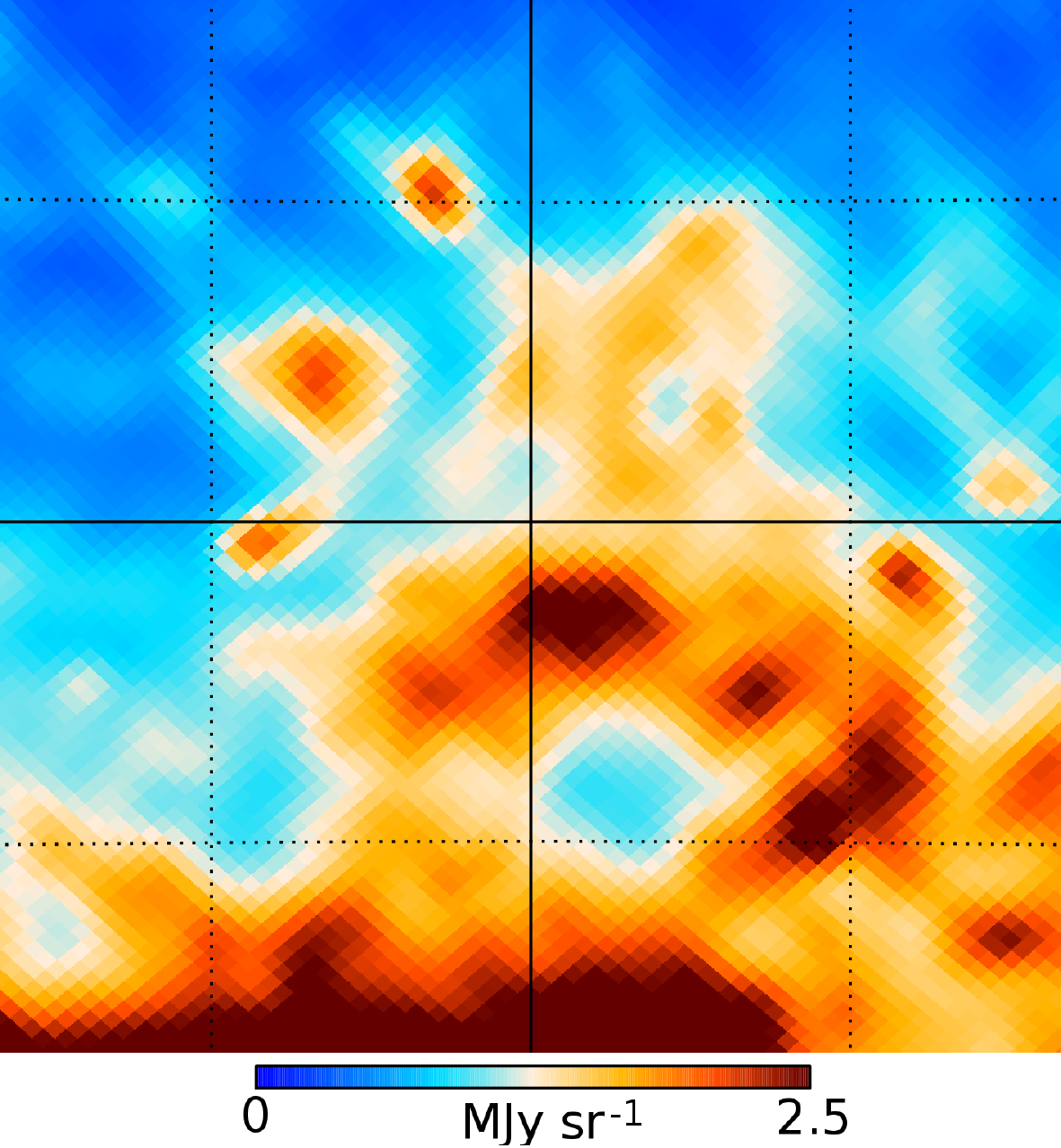}
   \caption{Thermal dust amplitude at 353\GHz}
   \end{subfigure}      
  }
  \caption{Comparison of the high-resolution \ruler\ (top) and
    low-resolution \commander\ (bottom) amplitude maps for a
    particularly strong CO complex near the Fan region; the maps are
    centred on Galactic coordinates $(l,b) = (110^{\circ},
    15^{\circ})$, and the grid spacing is $5^{\circ}$. Columns show,
    left to right: the CMB amplitude; the CO amplitude at 100\GHz\;
    and the thermal dust amplitude at 353\GHz.}
  \label{fig:ruler_vs_commander}
\end{figure*}

We only use the seven lowest \Planck\ frequencies, from 30 to
353\GHz. The two highest channels have significantly different
systematic properties than the lower frequency bands, for instance
concerning calibration, ZLE, and noise correlations, and they are more
relevant to thermal dust and CIB studies than to the present CMB
analysis.  Studies of specific foregrounds (CO, thermal dust, CIB
etc.) using all \Planck\ frequencies as well as ancillary data are
discussed in companion papers \citet{planck2013-p03},
\citet{planck2013-p05b}, \citet{planck2013-p06b}, and additional
future publications will consider extensions to AME, synchrotron and
free-free emission.

In order to obtain unbiased estimates of the spectral parameters
across all frequency bands, each map is downgraded from its native
resolution to a common angular resolution of 40\arcm and repixelized
at $N_{\textrm{side}}=256$, a limit imposed by the LFI
30\GHz\ channel.  Once the spectral indices have been determined, we
re-estimate the component amplitudes at native \Planck\ resolution
(see Appendix \ref{sec:commander_appendix}).

Although the smoothing operation introduces noise correlations between
pixels, we model the noise of the smoothed maps as uncorrelated white
noise with an effective standard deviation, $\sigma(p)$, for each
pixel $p$. This approximation does not bias the final solution,
because the analysis is performed independently for each
pixel. However, it is important to note that correlations between
pixels are not taken into account in this analysis. The effective
noise uncertainty, $\sigma(p)$, is estimated using realistic noise
simulations downgraded in the same way as the data. The measured
instrumental bandpasses are taken into account by integrating the
emission laws over the bandpass for each component at each Monte Carlo
step in the analysis.

The monopole (zero-point) of each frequency map is not constrained by
\Planck, but is rather determined by post-processing, and associated
with a non-negligible uncertainty (see Table 5 of
\citealp{planck2013-p01}). In addition, each frequency map includes a
significant monopole contribution from isotropic extragalactic sources
and CIB fluctuations not traced by local Galactic structure, ranging
from less than about 10--20\microK\ at 70\GHz\ to several hundreds of
\microK\ at 353\GHz. Finally, the effective dipole in each map is
associated with significant uncertainty due to the large kinematic CMB
dipole. In order to prevent these effects from introducing modelling
errors during component separation, they must be fit either prior to
or jointly with the Galactic parameters. Unfortunately, when allowing
free spectral parameters per pixel, there is a near-perfect degeneracy
among the offsets, the foreground amplitudes and the spectral indices,
and in order to break this degeneracy, it is necessary to reduce the
number of spectral degrees of freedom.

We adopt the method described by \citet{wehus2013} for this purpose,
which has the additional advantage of making minimal assumptions about
the foreground spectra. In short, this method uses linear regression
between data from CMB-subtracted maps evaluated on pixels falling
within each large $N_{\textrm{side}}=8$ pixel to estimate the relative
offsets, $m_1$ and $m_2$, between any two maps at each position on the
sky.  Each regression provides a constraint of the form $m_1 = a m_2 +
b$, where $a$ and $b$ are the slope and offset, respectively, and
where each value of $m_i$ consists of the sum of both a monopole and a
dipole term evaluated at that position.  The individual monopoles and
dipoles can then be reconstructed by measuring $a$ and $b$ in
different regions of the sky, exploiting spatial variations in
spectral indices, and solving jointly for two monopoles and dipoles,
including constraints from all positions. To minimize degeneracies, a
positivity prior is imposed on the fit, such that statistically
significant negative pixels are heavily penalized. For 44 and 70\GHz,
we retain the dipole values determined during the mapmaking process,
and do not attempt to fit them.

The resulting complete set of monopole and dipole values is listed in
Table~\ref{tab:offsets}. As a cross-check, we performed a dedicated
\commander\ run in which we fitted for the dipole at 353\GHz, together
with the foreground amplitudes and spectral indices, and only found
sub-\microK\ differences. This channel is by far the most problematic
in our data set in terms of offset determination, because of the very
bright dust emission at this frequency. As a result, there is a large
relative uncertainty between the zero-level of the dust amplitude map
and the 353\GHz\ channel offset not accounted for in the following
analyses. However, the sum of the two terms is well determined, and a
potential error in either therefore does not compromise the quality of
the other signal components (e.g., CMB and low-frequency
components). A similar comment applies between the offset at
30\GHz\ and the zero-level of the low-frequency component, although at
a significantly lower level.

\begin{table*}[tmb]
\begingroup
\newdimen\tblskip \tblskip=5pt
\caption{Estimated monopoles and dipoles in Galactic coordinates, all
  measured in thermodynamic \microK. Errors are estimated by
  bootstrapping, and do not account for correlated errors across
  frequencies. In particular, the 353\GHz\ monopole uncertainty is
  dominated by systematic errors not included in these estimates.
  Note that the dipoles at 44 and 70\GHz\ are fixed at the values
  determined in the mapmaking.}
\label{tab:offsets}
\vskip -3mm
\footnotesize
\setbox\tablebox=\vbox{
\newdimen\digitwidth
\setbox0=\hbox{\rm 0}
\digitwidth=\wd0
\catcode`*=\active
\def*{\kern\digitwidth}
\newdimen\signwidth
\setbox0=\hbox{+}
\signwidth=\wd0
\catcode`!=\active
\def!{\kern\signwidth}
\newdimen\decimalwidth
\setbox0=\hbox{.}
\decimalwidth=\wd0
\catcode`@=\active
\def@{\kern\signwidth}
\halign{ \hbox to 1.0in{#\leaderfil}\tabskip=3em& 
    \hfil#\hfil\tabskip=2em& 
    \hfil#\hfil& 
    \hfil#\hfil& 
    \hfil#\hfil \tabskip=0pt\cr
\noalign{\doubleline}
\omit  \hfill Frequency \hfill & Monopole& X dipole&  Y dipole& Z dipole\cr
\omit  \hfill [\GHz] \hfill & [$\mu$K]& [$\mu\textrm{K}$]&  $[\mu\textrm{K}$]& [$\mu\textrm{K}$]\cr
\noalign{\vskip 2pt\hrule\vskip 2pt} 
 *30&   **$8\pm2$&   *$-4\pm3$*&            *$-6\pm2$*&           !*$6\pm1$*\cr
 *44&   **$2\pm1$&   !*$0\pm0*$&            *!$0\pm0$*&           !*$0\pm0$*\cr
 *70&   *$15\pm1$&   !*$0\pm0*$&           *!$0\pm0$*&            !*$0\pm0$*\cr
 100&   *$15\pm1$&   !*$2\pm1$*&          *!$5\pm1$*&           *$-5\pm1$*\cr
 143&   *$33\pm1$&   !*$2\pm1$*&          *!$7\pm1$*&           *$-6\pm1$*\cr   
 217&   *$86\pm1$&   !*$2\pm1$*&          !$11\pm2$*&           $-10\pm2$*\cr
 353&   $414\pm4$&   !$11\pm10$&        !$52\pm12$&           $-37\pm8$*\cr
\noalign{\vskip 3pt\hrule\vskip 2pt}
}}
\endPlancktablewide
\endgroup
\end{table*}

\subsection{Component models and priors}
\label{sec:fg_models}

Our model for the low-resolution CMB analysis includes four
independent physical components: CMB; ``low-frequency'' emission; CO
emission; and thermal dust emission.  It can be written schematically
in the form
\begin{align}
s_{\nu}(p) &= A_{\textrm{CMB}}(p) +
A_{\textrm{lf}}(p) \left(\frac{\nu}{\nu_{0,\textrm{lf}}}\right)^{\beta_{\textrm{lf}}(p)}
+ \notag\\  &\quad\quad+ A_{\textrm{CO}}(p) f_{\nu,\textrm{CO}} +
A_{\textrm{d}}(p) \frac{e^{\frac{h\nu_{0,\textrm{d}}}{kT_{\textrm{d}}(p)}}-1}{e^{\frac{h\nu}{kT_{\textrm{d}}(p)}}-1} \left(\frac{\nu}{\nu_{0,\textrm{d}}}\right)^{\beta_{\textrm{d}}(p)+1},
\label{eq:fg_model}
\end{align}
where $A_{i}(p)$ denotes the signal amplitude for component $i$ at
pixel $p$, $\nu_{0,i}$ is the reference frequency for each component,
and $\nu$ refers to frequency. (Note that for readability, integration
over bandpass, as well as unit conversions between antenna, flux
density and thermodynamic units, is suppressed in this expression.)
Thus, each component is modelled with a simple frequency spectrum
parameterized in terms of an amplitude and a small set of free
spectral parameters (a power-law index for the low-frequency
component, and an emissivity index and temperature for the thermal
dust component); no spatial priors are imposed. One goal of the
present analysis is to understand how well this simple model captures
the sky signal in terms of effective components over the considered
frequency range, and we exploit the \ffp\ simulation (see
Sect.~\ref{sec:data}) for this purpose.

In order to take into account the effect of bandpass integration, each
term in the above model is evaluated as an integral over the bandpass
as described in Sect.\ 3 of \citet{planck2013-p03d}, and converted
internally to thermodynamic units. Accordingly, the reference
frequencies in Eq.~\ref{eq:fg_model} are computed as effective
integrals over the bandpass, such that the amplitude map, $A_{i}(p)$,
corresponds to the foreground map observed by the reference detector,
i.e., after taking into account the bandpass. In order to minimize
degeneracies between the different signal components, the reference
band for a given component is set to the frequency at which its
relative signal-to-noise ratio is maximized.

The foreground model defined in Eq.~\ref{eq:fg_model} is motivated by
prior knowledge about the foreground composition over the CMB
frequencies as outlined in Sect.~\ref{sec:sky_emissions}, as is our
choice of priors. In addition to the Jeffreys prior\footnote{The
  purpose of the Jeffreys prior is to normalize the parameter volume
  relative to the likelihood, such that the likelihood becomes
  so-called ``data-translated'', i.e., invariant under
  re-parameterizations.} \citep{Eriksen2008ApJ676}, we adopt Gaussian
priors on all spectral parameters with centre values and widths
attempting to strike a balance between prior knowledge and allowing
the data to find the optimal solution. Where needed, we have also run
dedicated analyses, either including particular high signal-to-noise ratio
subsets of the data or using a lower resolution parameterization to
increase the effective signal-to-noise in order to inform our prior
choices. We now consider each foreground component in turn, and note
in passing that the CMB component, by virtue of being a blackbody
signal, is given by a constant in thermodynamic temperature units.

We approximate the low-frequency component by a straight power law in
antenna temperature with a free spectral index per pixel, and adopt a
prior of $\beta=-3 \pm 0.3$ (this is the index in terms of brightness
temperature).  This choice is determined by noting that the prior is
in practice only relevant at high Galactic latitudes where the
signal-to-noise ratio is low and the dominant foreground component is
expected to be synchrotron emission; in the signal-dominated and
low-latitude AME and free-free regions, the data are sufficiently
strong to render the prior irrelevant. For validation purposes, we
have also considered minor variations around this prior, such as
$\beta=-2.9\pm0.3$ and $\beta=-3.05\pm0.2$, finding only small
differences in the final solutions. The reference band for the
low-frequency component is set to 30\GHz, where the low-frequency
foreground signal peaks. The final low-frequency amplitude map is
provided in units of thermodynamic microkelvin.

The CO emission is modelled in terms of a single line ratio for each
frequency. Specifically, the CO amplitude is normalized to the
100\GHz\ band, and defined in units of \microK\,km\,s$^{-1}$
\citep{planck2013-p03a}. The amplitude at other frequencies is
determined by a single multiplicative factor relative to this, with a
numerical value of 0.595 at 217\GHz\ and 0.297 at 353\GHz; all other
frequencies are set to zero. These values are obtained from a
dedicated CO analysis that includes only high signal-to-noise ratio CO
regions covering a total of 0.5\% of the sky. The derived values are
in good agreement with those presented by \citet{planck2013-p03a}.

Thermal dust emission is modelled by a one-component modified
blackbody emission law with a free emissivity spectral index,
$\beta_\textrm{d}$, and dust temperature, $T_{\textrm{d}}$, per pixel.
However, since we only include frequencies below 353\GHz, the dust
temperature is largely unconstrained in our fits, and we therefore
impose a tight prior around the commonly accepted mean value of
$T_{\textrm{d}} = 18\pm0.05$\,K.  The only reason we do not fix it
completely to 18\,K is to allow for modelling errors near the Galactic
centre. The dust emissivity prior is set to $\beta_{\textrm{d}} =
1.5\pm0.3$, where the mean is determined by a dedicated run fitting
for a single best-fit value for the high-latitude sky, where the prior
is relevant. The reference band for the thermal dust component is
353\GHz, and the final map is provided in units of megajansky per
steradian.

\subsection{Results and validation}
\label{sec:fg_results}

The output of the Bayesian component separation algorithm is a set of
samples drawn from the joint posterior distribution of the model
parameters, as opposed to a single well-defined value for each.  For
convenience, we summarize this distribution in terms of posterior mean
and standard deviation maps, computed over the sample set, after
rejecting a short burn-in phase. The goodness-of-fit is monitored in
terms of the $\chi^2$ per pixel.  Although convenient, it is, however,
important to note that this description does not provide a
comprehensive statistical representation of the full posterior
distribution, which is intrinsically non-Gaussian.  One should be
careful about making inferences in the low signal-to-noise regime
based on this simplified description.

The low-resolution \commander\ posterior mean amplitude maps are shown
in Fig.~\ref{fig:planck_lowres_fg_amps} for the low-frequency, CO, and
thermal dust components, and the spectral index maps in
Fig.~\ref{fig:planck_lowres_fg_inds}.  The associated $\chi^2$ map is
plotted in Fig.~\ref{fig:lowres_chisq}. Note that because we are
sampling from the posterior instead of searching for the
maximum-likelihood point, the expected number of degrees of freedom is
equal to $N_{\textrm{band}}=7$ in this plot, not
$N_{\textrm{band}}-N_{\textrm{par}}$.

Figure~\ref{fig:ruler_vs_commander} compares the high-resolution
\ruler\ solution to the low-resolution \commander\ solution for CMB,
CO and thermal dust on a particularly strong CO complex near the Fan
region, centred on Galactic coordinates $(l,b) = (110^{\circ},
15^{\circ})$.

\begin{figure}
  \center{ 
\begin{subfigure}{\columnwidth} 
  \includegraphics[width=\columnwidth]{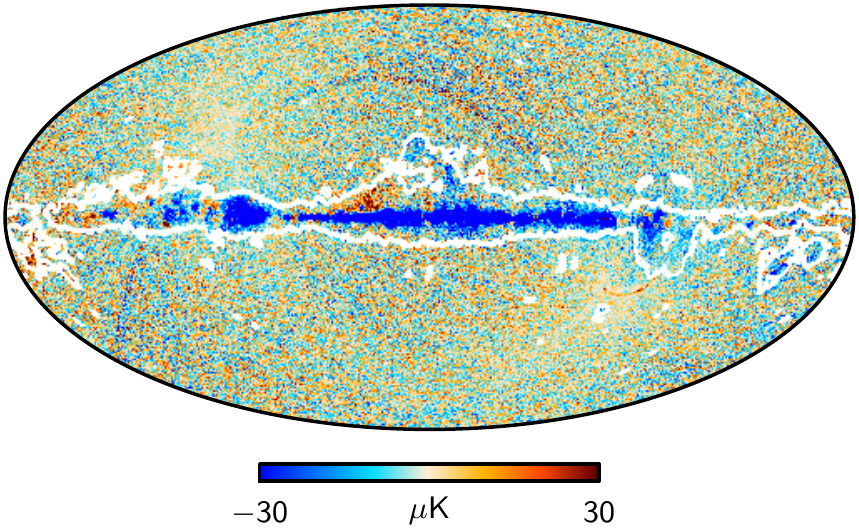} 
   \caption{Low-frequency component residual at 30\GHz}
\vspace*{2mm}
   \end{subfigure}      
\begin{subfigure}{\columnwidth} 
  \includegraphics[width=\columnwidth]{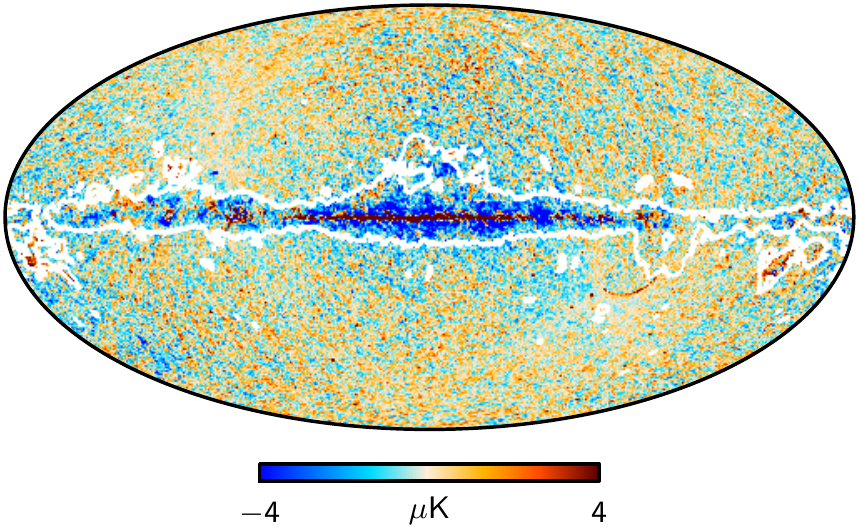} 
   \caption{CO residual at 100\GHz}
\vspace*{2mm}
   \end{subfigure}      
\begin{subfigure}{\columnwidth} 
  \includegraphics[width=\columnwidth]{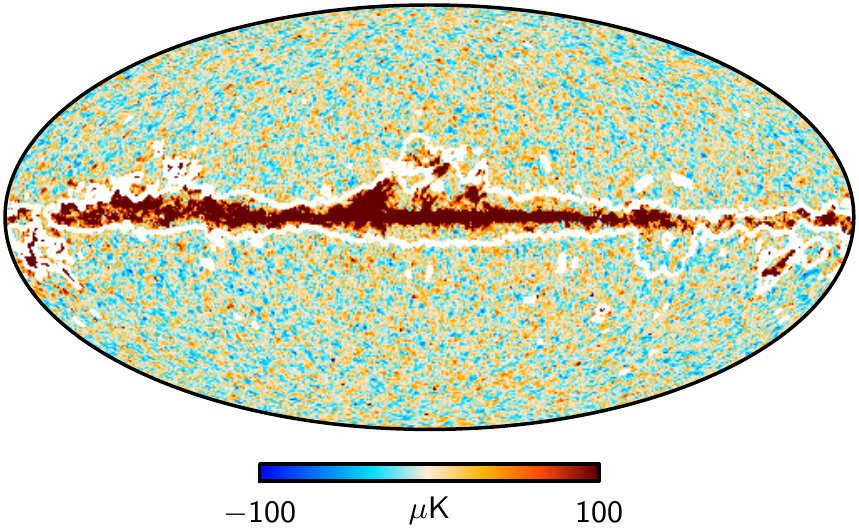}
   \caption{Thermal dust residual at 353\GHz}
   \end{subfigure}      
  }  
  \caption{Amplitude residual maps,
    $A_{\textrm{out}}-A_{\textrm{in}}$, computed blindly from the
    \ffp\ simulation. The panels show (from top to bottom) the
    low-frequency residual at 30\GHz, the CO residual at 100\GHz\ and
    the thermal dust residual at 353\GHz. All units are thermodynamic
    $\mu\textrm{K}$. The white lines indicate the boundary of the
    \commander\ likelihood analysis mask, removing 13\% of the
    sky.}  \label{fig:ffp6_amp_diff}
\end{figure}

Several features can be seen here, foremost of which is that the
Galactic plane is strikingly obvious, with $\chi^2$ values exceeding
$10^4$ for seven degrees-of-freedom in a few pixels. This is not
surprising, given the very simplified model at low frequencies (i.e.,
a single power law accounting for AME, synchrotron, and free-free
emission), as well as the assumption of a nearly constant dust
temperature of 18\,K.  Second, there is an extended region of
moderately high $\chi^2$ roughly aligned with a great circle going
through the Ecliptic South Pole, indicating the presence of correlated
noise in the scanning rings not accounted for in our white noise
model.

Based on these, and other considerations, it is clear that parts of
the sky must be masked before proceeding to CMB power spectrum and
likelihood analyses.  This masking process is discussed at greater
length in \citet{planck2013-p08}, and results in different masks for
specific applications.  The goal of our present discussion is to
evaluate the adequacy of the mask adopted for low-$\ell$ likelihood
analysis (L87), which is based on the fits presented here. This mask
removes 13\% of the sky, and is derived from a combination of $\chi^2$
and component amplitude thresholding.

\begin{figure}
  \begin{center}
    \includegraphics[width=\columnwidth]{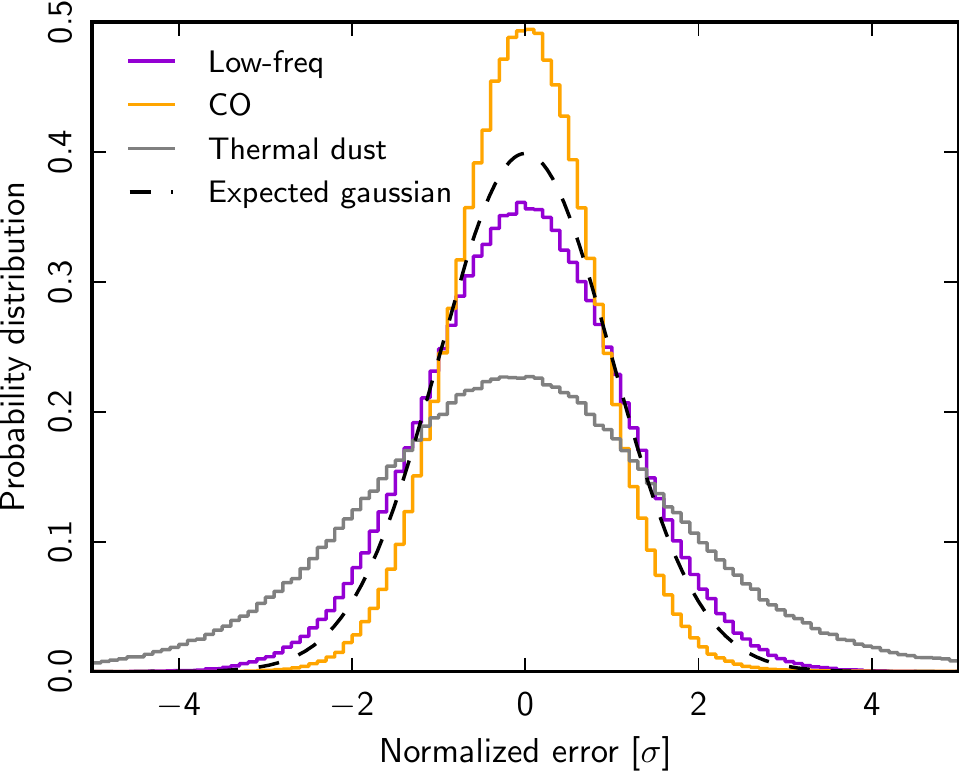}
    \vspace*{2mm}
    \includegraphics[width=\columnwidth]{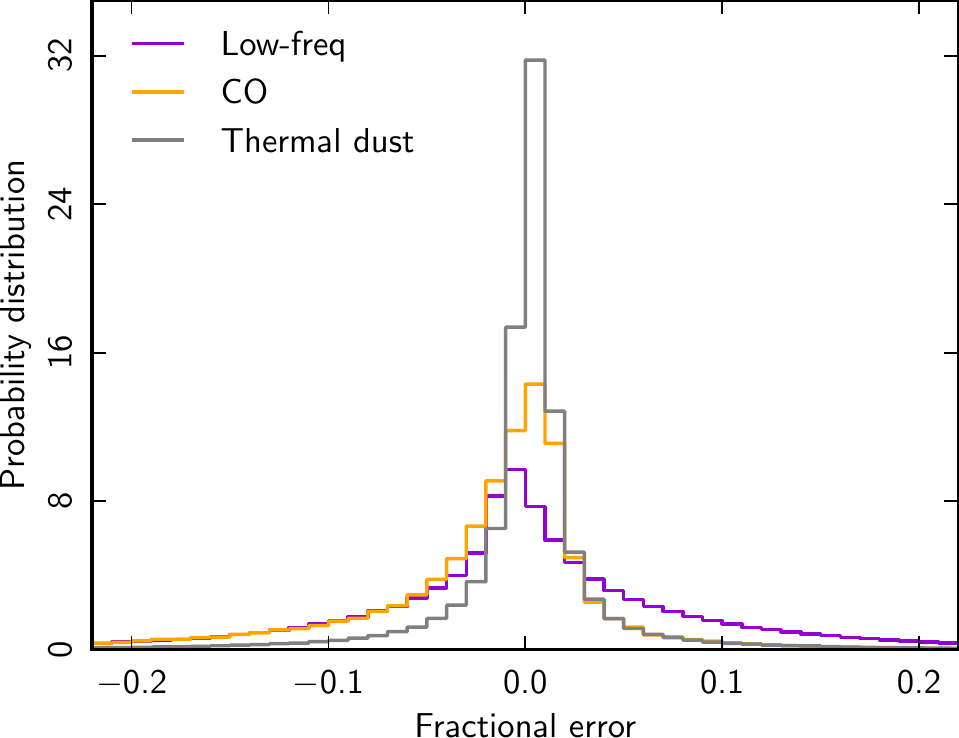}
  \end{center}
  \caption{Error validation for component amplitudes, evaluated from
    the \ffp\ simulation.  The upper panel shows histograms of the
    normalized errors $\delta =
    (A_{\textrm{out}}-A_{\textrm{in}})/\sigma_{\textrm{out}}$ for the
    three foreground components and including all pixels outside the
    \commander\ likelihood analysis mask. The lower panel shows
    histograms of the fractional error $f \equiv
    (A_{\textrm{out}}-A_{\textrm{in}})/A_{\textrm{in}}$ for pixels
    with a foreground detection level above $5\,\sigma$. No evidence
    of significant bias is observed for any component, and the
    uncertainty estimates for the low-frequency and CO components are
    accurate to about 12\,\%; the thermal dust uncertainty is
    underestimated by a factor of 2 due to the presence of unmodelled
    fluctuations.}
  \label{fig:fg_errors}
\end{figure}

\begin{figure}
  \center{
  \includegraphics[width=\columnwidth]{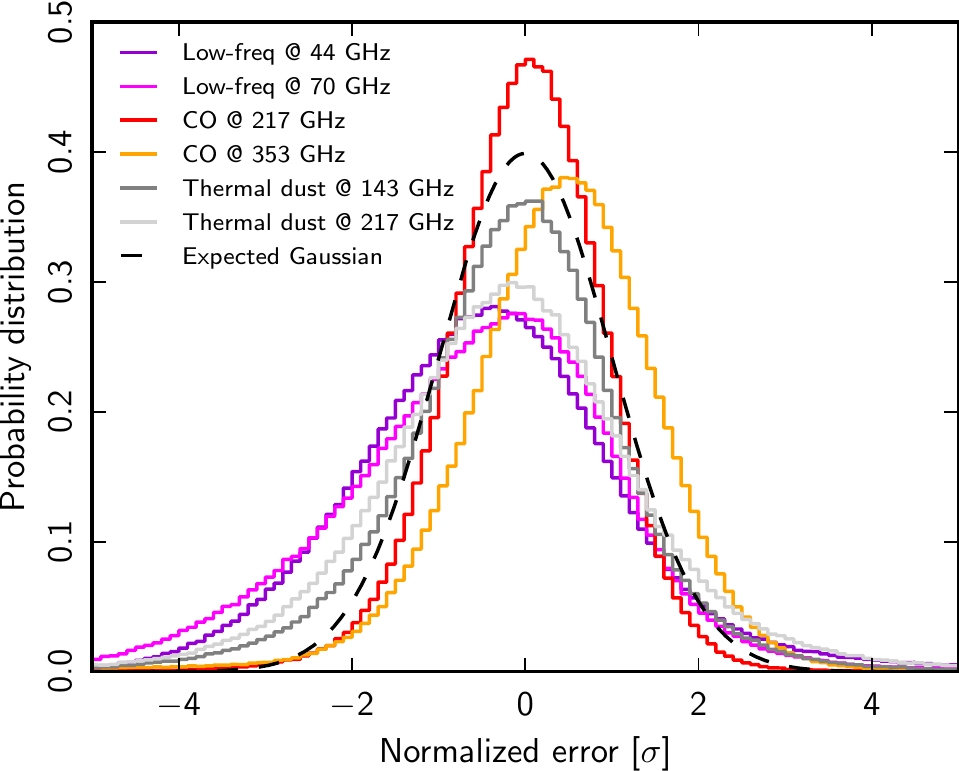}
  }
  \caption{Validation of spectral parameters for low-frequency
    foregrounds, thermal dust, and CO emission, evaluated from the
    \ffp\ simulation. Each histogram shows the error distribution at
    the two leading sub-dominant frequencies in the form of the
    normalized errors $\delta =
    (A_{\textrm{out}}(\nu)-A_{\textrm{in}}(\nu))/\sigma_{\textrm{out}}(\nu)$
    for all pixels outside the \commander\ likelihood analysis mask,
    where $A_{\textrm{out}}(\nu)$ is the predicted foreground
    amplitude at frequency $\nu$ given the estimated amplitude and
    spectral parameters, and $\sigma_{\textrm{out}}(\nu)$ is the
    corresponding standard deviation computed over the sample set.}
  \label{fig:mixmat_errors}
\end{figure}

For validation purposes, we analyse the simulations described in
Section \ref{sec:data} in the same way as the real data, including
monopole and dipole determination, CO line ratio estimation and
spectral index estimation.  Individual component maps at each observed
frequency are available from the simulation process, and used for
direct comparison with the reconstructed products.

In Fig.~\ref{fig:ffp6_amp_diff} we show the differences between the
recovered and input component maps at their respective reference
frequencies. The boundary of the 13\% \commander\ mask is traced by
the white contours, and a best-fit monopole and dipole have been
subtracted from each difference map.  All difference maps are shown in
units of thermodynamic \microK. The top panel of
Fig.~\ref{fig:fg_errors} gives the error histograms outside the masked
region for each component, normalized to the respective estimated
standard deviation; if the recovered solution has both correct mean
and standard deviation, these histograms should match a Gaussian
distribution with zero mean and unit variance, indicated by the dashed
black line. Conversely, a significant bias would be visible as a
horizontal shift in this plot, while under-estimation of the errors
would result in too wide a distribution and vice versa.  The bottom
panel shows the fractional error (i.e., the error divided by the true
input value) for all pixels with signal above $5\,\sigma$; the
fractional error is not a useful quantity for noisier signals.

The difference maps in Fig.~\ref{fig:ffp6_amp_diff} display
significant errors in the Galactic plane.  For the low-frequency
component, the residuals are dominated by free-free emission, while
for thermal dust the dominant contaminant is CO emission. However,
outside the mask the residuals are small, and, at least for the
low-frequency and CO components, the spatial characteristics appear
similar to instrumental noise. This is more clear in the histograms
shown in the top panel of Fig.~\ref{fig:fg_errors}; the mean and
standard deviations are $\delta_{\textrm{lf}} = 0.01\pm1.12$,
$\delta_{\textrm{CO}} = 0.00\pm0.87$, and $\delta_{\textrm{td}} =
0.00\pm2.01$, respectively, for the low-frequency, CO and thermal dust
components. There is no evidence of bias outside the mask in any
component, and the error estimates are accurate to 12 and 13\,\% for
the low-frequency and CO components. Note, though, that the estimated
error for the CO component is actually larger than the true
uncertainty, suggesting that the white noise approximation for the
100\GHz\ channel overestimates the true noise. This can occur if the
correlated instrumental noise is important in regions where there is
no significant CO emission.  Locally re-scaling the white noise to
account for spatially varying correlated noise would correct this
effect.

\begin{figure}
  \center{
  \vspace*{3mm}
  \includegraphics[width=\columnwidth]{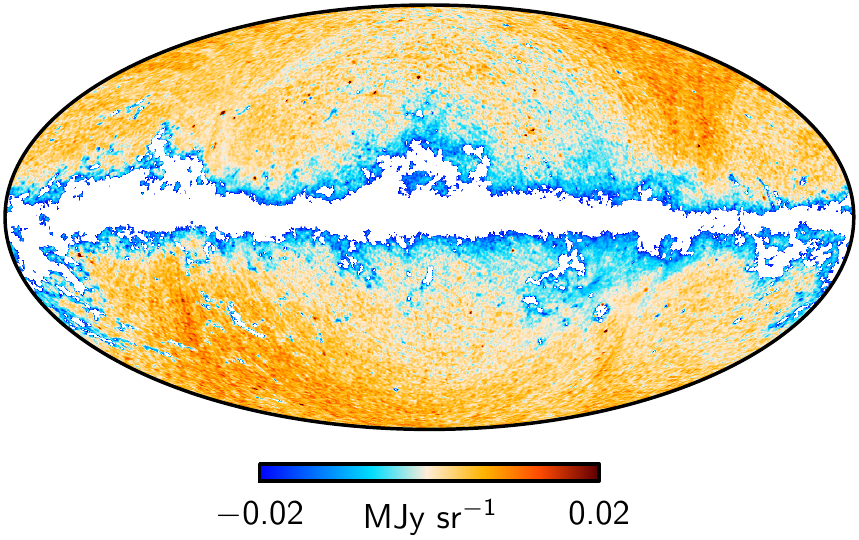}
  \includegraphics[width=\columnwidth]{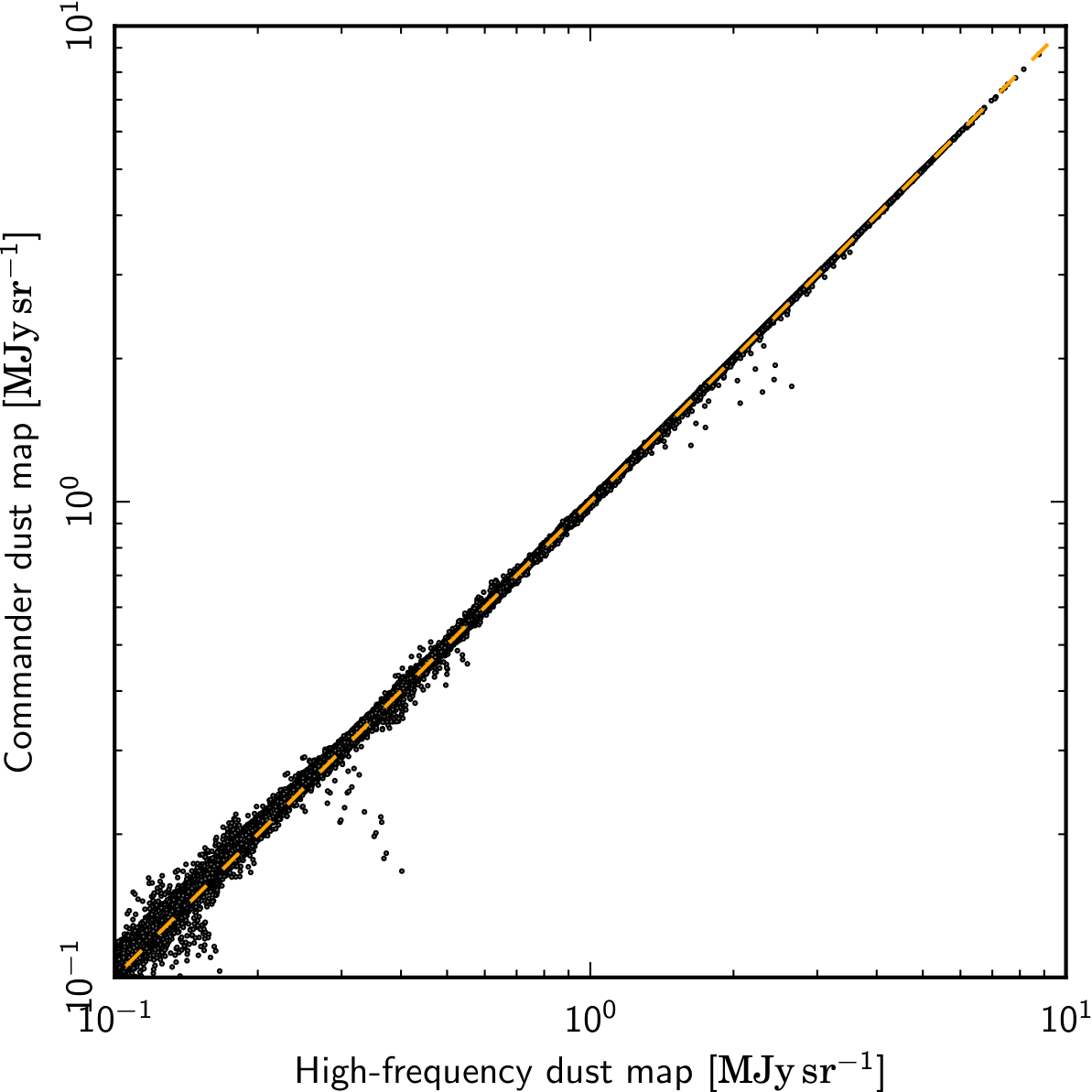}
  }
  \caption{Top: Difference map between the estimated thermal dust
    amplitude at 353\GHz\ derived by \citet{planck2013-p06b} and the
    low-resolution dust map presented here, both smoothed to
    40\arcm. A monopole and dipole term outside the \commander\ mask
    has been removed. The former includes only high-frequency
    observations (the \Planck\ 353, 545 and 857\GHz\ channels and
    observations at 100\,\micron), while the one derived here only
    uses low-frequency data (\Planck\ 30--353\GHz). Note that the
    colour scale ranges between $-0.02$ and
    $0.02\,\mathrm{MJy}\,\mathrm{sr}^{-1}$ in this plot, whereas it
    ranges from 0 to $2.5\,\mathrm{MJy}\,\mathrm{sr}^{-1}$ in the
    bottom panel of Fig.~\ref{fig:planck_lowres_fg_amps}. Masked
    regions indicate pixels with a CO amplitude at 100\GHz\ larger
    than 1$\,\mu\textrm{K}\,\textrm{km}\,\textrm{s}^{-1}$.  Bottom:
    $T$--$T$ plot between the same two maps.}
  \label{fig:diff_dust_MAMD_commander}
\end{figure}

For the thermal dust component, on the other hand, the error is
underestimated by a factor of 2. The explanation for this is most
easily seen from the lower panel of Fig.~\ref{fig:ffp6_amp_diff}.
This map is dominated by isotropic CIB fluctuations, rather than
instrumental noise. Because these fluctuations have a slightly
different spectrum than the dominant Galactic dust emission, and the
model does not account for a separate CIB component, the error on the
Galactic component is underestimated.  When using the Galactic map
presented here for detailed analysis near the noise limit, taking into
account these residual fluctuations is essential, and the effective
noise per pixel should be increased by a factor of 2.

As clearly seen in Fig.~\ref{fig:ffp6_amp_diff}, the residuals inside
the mask are highly significant in a strict statistical
sense. However, as seen in the bottom panel of
Fig.~\ref{fig:fg_errors}, they are relatively small in terms of
fractional errors. Specifically, the three histograms have means and
standard deviations of $f_{\textrm{lf}} = 0.00\pm0.10$,
$f_{\textrm{CO}} = -0.03\pm0.10$, and $f_{\textrm{td}} = 0.00\pm0.06$,
respectively, for the low-frequency, CO and thermal dust components.
The largest bias is observed for the CO component, for which the
absolute amplitude is biased by 3\,\%. The bias in the low-frequency
and thermal dust components is negligible, and the fractional
uncertainties are 10 and 6\,\%, respectively. This confirms that
approximating the sum of the three low-frequency components by a
single power-law over the \Planck\ frequency bands is reasonable; if
modelling errors dominated, one would expect to see a significant bias
in the resulting amplitude.

In order to validate the spectral parameters, we show in
Fig.~\ref{fig:mixmat_errors} histograms of the normalized residuals
for each foreground component evaluated at its two leading
sub-dominant frequencies (i.e., at 44 and 70\GHz\ for the
low-frequency component; at 217 and 353\GHz\ for the CO component; and
at 143 and 217\GHz\ for the thermal dust component). The means and
standard deviations of these distributions are:
$\delta_{\textrm{lf}}(44\,\textrm{GHz}) = -0.41\pm1.98$ and
$\delta_{\textrm{lf}}(70\,\textrm{GHz}) = -0.34\pm2.04$ for the
low-frequency component; $\delta_{\textrm{CO}}(217\,\textrm{GHz}) =
0.10\pm0.84$ and $\delta_{\textrm{CO}}(353\,\textrm{GHz}) =
0.51\pm1.00$ for the CO component; and
$\delta_{\textrm{td}}(143\,\textrm{GHz}) = -0.02\pm1.53$ and
$\delta_{\textrm{td}}(217\,\textrm{GHz}) = -0.13\pm1.87$ for the
thermal dust component. As expected, the effect of modelling errors is
more significant at the sub-dominant frequencies than at the pivot
frequencies, when measured in terms of statistical uncertainties,
since the foreground signal is weaker and the confusion with the other
components relatively larger.  Nevertheless, we see that the absolute
bias is at most $0.5\,\sigma$ for the CO component at 353\GHz, while
the thermal dust bias is negligible even at 143\GHz. The estimated
uncertainties are generally underestimated by up to a factor of two
due to these modelling errors.

Finally, the efficiency of the adopted foreground model for CMB
analysis is quantified in Appendix \ref{sec:ffp6_appendix} in terms of
power spectrum residuals and cosmological parameter estimation.

To summarize, we find that the simplified model, defined by
Eq.~\ref{eq:fg_model}, provides a good fit to the realistic
\ffp\ simulation for most of the sky. Absolute residuals are small,
and the amplitude uncertainty estimates are accurate to around 12\,\%,
except for the thermal dust component for which unmodelled CIB
fluctuations are important. Further, we find that the real
\Planck\ data behave both qualitatively and quantitatively very
similarly to the \ffp\ simulation, suggesting that this approach also
performs well on the real sky.

\subsection{Interpretation and comparison with other results}

The maps shown in Figs.~\ref{fig:planck_lowres_fg_amps} and
\ref{fig:planck_lowres_fg_inds} provide a succinct summary of the
average foreground properties over the \Planck\ frequency range.  We
now consider their physical interpretation and compare them to
products from alternative methods.

First, the top panel of Fig.~\ref{fig:diff_dust_MAMD_commander} shows
a difference map between the dust map at 353\GHz\ derived in the
present paper and one determined from only the three highest
\Planck\ frequencies and the 100\,\micron\ IRIS map by
\citet{planck2013-p06b}, shown in flux density units, after removing a
small monopole and dipole difference. This map is to be compared to
the corresponding dust amplitude map in the bottom panel of Figure
\ref{fig:planck_lowres_fg_amps}; note that the colour range varies
from $-$0.02 to 0.02\,MJy\,sr$^{-1}$ in the difference plot, and
between 0 and 2.5\,MJy\,sr$^{-1}$ in the amplitude plot. Thus, despite
the very different data sets and methods, we see that the two
reconstructions agree to very high accuracy outside the Galactic
plane.  Inside the Galactic plane, the differences are dominated by
residuals due to different CO modelling, seen as solid blue colours in
Fig.~\ref{fig:diff_dust_MAMD_commander}; however, even in this region
the differences are smaller than 5\,\% of the amplitude. The bottom
panel shows a corresponding $T$--$T$ plot between the two maps,
excluding any pixel for which the \commander\ CO amplitude at
100\GHz\ is larger than
1$\,\mu\textrm{K}\,\textrm{km}\,\textrm{s}^{-1}$. The two maps agree
to 0.2\% in terms of best-fit amplitude.

From Fig.~\ref{fig:planck_lowres_fg_inds}, we see that the dust
emissivity ranges between 1.3 and 1.7 for most of the sky; considering
only the pixels with a posterior distribution width that is a third of
the prior width (i.e., $\sigma(\beta_{\textrm{d}}) < 0.1$), we find a
mean value of 1.49. The two exceptions are a large region of shallow
indices northeast of the Galactic centre, and steep indices near the
Galactic plane. The former region corresponds to a part of the sky
with low dust emission, where we expect the spectral index to be
sensitive to both monopole and dipole residuals, as well as
instrumental systematics, such as correlated $1/f$ noise. The latter
appears to be particularly pertinent here because the shallow index
region at least partially traces the \Planck\ scanning strategy; as a
result, the systematic error on the spectral index in this region is
considerable. The main systematic uncertainty connected to the region
of steep indices around the Galactic plane is confusion with CO
emission.

The CO map shown in Figs.~\ref{fig:planck_lowres_fg_amps} and
\ref{fig:ruler_vs_commander} is discussed in greater detail in
\citet{planck2013-p03a}.  A distinct advantage of this particular
solution over available alternatives is its high signal-to-noise ratio
per pixel, which is achieved by reducing all information into a single
value per pixel. Consequently, this map serves as a unique tool for
follow-up CO observations. However, the assumption of a constant line
ratio over the full sky may lead to a significant systematic
uncertainty on CO amplitude per pixel. This possibility is
investigated in a forthcoming work \citep{planck-pip58}.

\begin{figure}
  \center{
\begin{subfigure}{\columnwidth} 
  \includegraphics[width=\columnwidth]{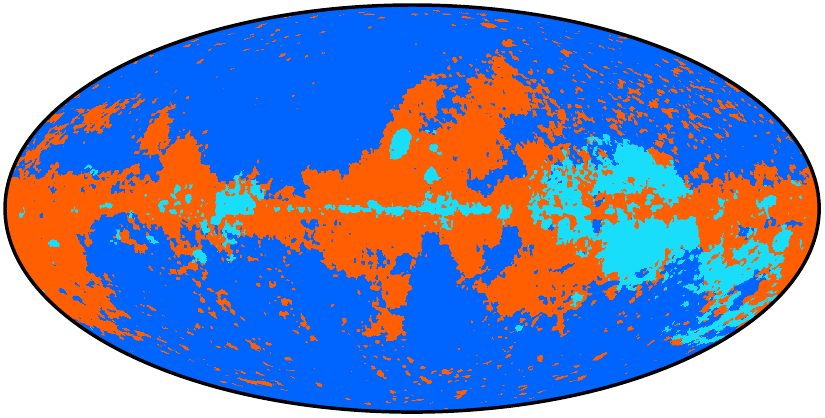}
   \caption{Dominant low-frequency component map}
\vspace*{2mm}
   \end{subfigure}      
\begin{subfigure}{\columnwidth} 
  \includegraphics[width=\columnwidth]{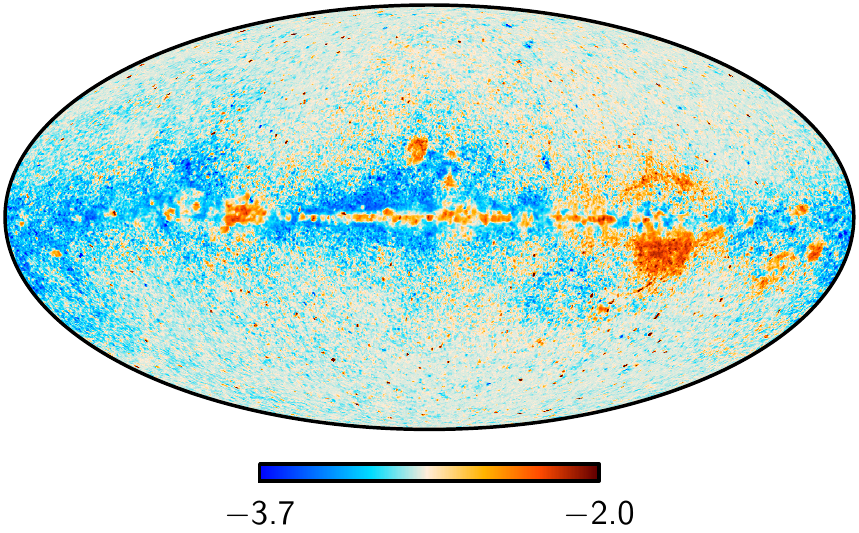}
   \caption{Low-frequency component power-law index}
   \end{subfigure}      
  }
  \caption{Top: Dominant foreground component per pixel at 30\GHz\ in
    the \ffp\ simulation. Dark blue indicates that synchrotron
    emission is the strongest component at 30\GHz, light blue
    indicates that free-free dominates, and orange indicates that
    spinning dust (AME) is the strongest component. Bottom: The
    recovered low-frequency power-law index derived from the same
    simulation.}
  \label{fig:lowfreq_components}
\end{figure}

Finally, the spectral index map for the low-frequency component shown
in Fig.~\ref{fig:planck_lowres_fg_inds} can be used to determine the
dominant low-frequency component (synchrotron, free-free or AME) as a
function of position on the sky.  To illustrate this connection, we
once again take advantage of the \ffp\ simulation for which we know
the amplitude of each low-frequency component per pixel.  In the top
panel of Fig.~\ref{fig:lowfreq_components}, we use this information to
make a ``dominant component map''; dark blue indicates that
synchrotron emission is strongest at a given pixel, light blue that
free-free is strongest, and orange that spinning dust (AME) dominates.
In the bottom panel, we show our derived power-law index map from the
same simulation.  As expected, the correspondence between the
power-law index and the dominant component is very strong, implying
that the spectral index map can be used to trace the individual
components. In particular, we see that an index below about $-3.3$
reflects the presence of a component consistent with a spinning dust
model peaking below 20\GHz\ over the \Planck\ frequency range, while
an index higher than around $-2.3$ signals the importance of free-free
emission. Intermediate values typically indicate synchrotron emission,
although it should be noted that the signal-to-noise ratio at very
high latitudes is low and the results are therefore prior-driven in
these regions.

Returning to the spectral index map shown in
Fig.~\ref{fig:planck_lowres_fg_inds}, we see a good correspondence
between the real data and the simulation.  Features present in the
simulation also appear in the data.  For instance, we see that the
spectral index in the so-called Fan Region (i.e., near Galactic
coordinates $(l,b)=(90^\circ,20^\circ)$) is low in both cases, and
this alone provides strong evidence for the presence of AME.  Further,
the AME spectral index is consistent with the spinning dust
interpretation.  This power-law index map may be used to identify
particular AME regions for follow-up observations. Finally, we note
that regions known for strong free-free emission, such as the Gum
Nebula or Zeta~Ophiuchi, have spectral indices close to $-2.1$ or
$-2.2$, as expected.

\section{Conclusions}
\label{sec:conclusions}

We have tested four component separation algorithms on the \Planck\
frequency maps to produce clean maps of the CMB anisotropies over a
large area of sky.  These CMB maps are used for studies of statistics
and isotropy
\citep{planck2013-p09}, primordial non-Gaussianity
\citep{planck2013-p09a}, gravitational lensing \citep{planck2013-p12},
the ISW effect \citep{planck2013-p14}, cosmic geometry and
topology \citep{planck2013-p19}, searches for cosmic defects from
primordial phase transitions \citep{planck2013-p20}, as well as an
integral part of the low-$\ell$ \Planck\ likelihood
\citep{planck2013-p08}.  Two of the methods, one using internal
foreground templates (\sevem) and the other an ILC in needlet space
(\nilc), are non-parametric, extracting the CMB map by minimizing the
variance of the total contamination.  The other two methods fit models
of the foregrounds to clean the CMB of their emission.  One fits a
parametric model in real space (\CR) and one fits a non-parametric in
the harmonic domain (\smica).

All four methods have been demonstrated to work well both on real and
simulated data, and to yield consistent results. Nevertheless, there
are differences between the methods, making them more or less suitable
for specific applications. For instance, \comrul\ allows a joint
parametric foreground estimation and CMB power spectrum estimation,
with full propagation of foreground uncertainties to cosmological
parameters, but is limited to a lower angular resolution than the
other codes.  This method has therefore been selected for the
low-$\ell$ \Planck\ likelihood
\citep{planck2013-p08} and to produce astrophysical component maps
(Sect.\ \ref{sec:foreground_components}), while it is sub-optimal for
applications requiring full angular resolution, e.g., gravitational
lensing reconstruction or estimation of primordial non-Gaussianity.
For these purposes, we use the three higher-resolution maps.  We
take \smica\ to be the leading method, based on its superior
performance on the \ffp\ simulation, where it has be shown to have the
lowest residual foreground contamination at large scales and to
preserve primordial non-Gaussianity.  When subjecting
foreground-cleaned \Planck\ maps to scientific analysis, we use the
other two or three maps, as appropriate, to assess the uncertainties
inherent in the choice of methods and the assumptions they make.
Indeed, this is the main purpose for presenting four different CMB
solutions to the general community.

The CMB anisotropies are robustly recovered over a large fraction
(73\,\%) of the sky and down to small angular scales, reaching to
multipoles $\ell\approx 2000$.  We characterize the CMB maps with
angular power spectra and cosmological parameter constraints.
Parameter constraints from these maps are consistent with those from
the \Planck\ likelihood function based on cross-spectra and large sky
cuts \citep{planck2013-p08}.  This agreement supports the robustness
of both our component separation methodology and cosmological
parameter constraints.

The real-space parametric fits of \comrul\ enable us to characterize
the diffuse Galactic foregrounds.  We parameterize them with a
low-frequency power-law component, representing the sum of
synchrotron, free-free, and AME emission, a high-frequency modified
blackbody spectrum describing thermal dust emission, and a molecular
CO component.  Using only the \Planck\ data from 30 to 353\GHz, we fit
for the amplitude and spectral parameters of the three foregrounds and
the CMB simultaneously at each pixel of a 40-arcmin resolution map.
The spectral parameters are the low-frequency component power-law
exponent and the modified blackbody emissivity power-law exponent; the
CO line ratios are spatially fixed.  These parameters give us the
source mixing matrix, which we then use in a direct inversion to
deduce the component amplitudes at higher resolution.  Through Gibbs
sampling, we obtain realizations drawn from the full posterior
distribution of possible foreground and CMB solutions, giving us a
powerful ability to statistically characterize our results.
 
Our in-depth analysis of the recovered CMB anisotropies is
unprecedented for component separation studies, concerning both the
accuracy of cosmological parameter constraints, and studies of early
Universe physics and structure formation through gravitational
lensing.  On the other hand, the complex nature of the foreground
emission over such a large frequency range limits us to the use of
relatively simple methods when analysing \Planck\ data alone.  An
extensive study in combination with other probes of Galactic
foregrounds will be presented in forthcoming papers.  In particular,
the separation of individual components at low frequencies requires
the use of ancillary data, for example, from the Wilkinson Microwave
Anisotropy Probe (\textit{WMAP}) and radio surveys.

\begin{acknowledgements}

The development of Planck has been supported by: ESA; CNES and
CNRS/INSU-IN2P3-INP (France); ASI, CNR, and INAF (Italy); NASA and DoE
(USA); STFC and UKSA (UK); CSIC, MICINN, JA and RES (Spain); Tekes,
AoF and CSC (Finland); DLR and MPG (Germany); CSA (Canada); DTU Space
(Denmark); SER/SSO (Switzerland); RCN (Norway); SFI (Ireland);
FCT/MCTES (Portugal); PRACE (EU). A description of the Planck
Collaboration and a list of its members, including the technical or
scientific activities in which they have been involved, can be found
at
\url{http://www.sciops.esa.int/index.php?project=planck&page=Planck_Collaboration}. The
authors acknowledge the support provided by the Advanced Computing and
e-Science team at IFCA.  This work made use of the COSMOS
supercomputer, part of the STFC DiRAC HPC Facility.  Some of the
results in this paper have been derived using the \healpix\ package.

\end{acknowledgements}

\bibliographystyle{aa}
\bibliography{Planck_bib,component_separation}

\begin{thebibliography}{84}
\expandafter\ifx\csname natexlab\endcsname\relax\def\natexlab#1{#1}\fi

\bibitem[{{Ali-Ha{\"i}moud} {et~al.}(2009){Ali-Ha{\"i}moud}, {Hirata}, \&
  {Dickinson}}]{spdust1}
{Ali-Ha{\"i}moud}, Y., {Hirata}, C.~M., \& {Dickinson}, C. 2009, \mnras, 395,
  1055

\bibitem[{{Banday} {et~al.}(2003){Banday}, {Dickinson}, {Davies}, {Davis}, \&
  {G{\'o}rski}}]{banday2003}
{Banday}, A.~J., {Dickinson}, C., {Davies}, R.~D., {Davis}, R.~J., \&
  {G{\'o}rski}, K.~M. 2003, \mnras, 345, 897

\bibitem[{{Bennett} {et~al.}(2003){Bennett}, {Hill}, {Hinshaw}, {Nolta},
  {Odegard}, {Page}, {Spergel}, {Weiland}, {Wright}, {Halpern}, {Jarosik},
  {Kogut}, {Limon}, {Meyer}, {Tucker}, \& {Wollack}}]{bennett2003b}
{Bennett}, C.~L., {Hill}, R.~S., {Hinshaw}, G., {et~al.} 2003, \apjs, 148, 97

\bibitem[{{Benoit-L{\'e}vy} {et~al.}(2013){Benoit-L{\'e}vy}, {D{\'e}chelette},
  {Benabed}, {Cardoso}, {Hanson}, \& {Prunet}}]{BenoitLevy13}
{Benoit-L{\'e}vy}, A., {D{\'e}chelette}, T., {Benabed}, K., {et~al.} 2013,
  \aap, accepted

\bibitem[{{Betoule} {et~al.}(2009){Betoule}, {Pierpaoli}, {Delabrouille}, {Le
  Jeune}, \& {Cardoso}}]{2009A&A...503..691B}
{Betoule}, M., {Pierpaoli}, E., {Delabrouille}, J., {Le Jeune}, M., \&
  {Cardoso}, J.-F. 2009, \aap, 503, 691

\bibitem[{{Bonaldi} {et~al.}(2007){Bonaldi}, {Ricciardi}, {Leach}, {Stivoli},
  {Baccigalupi}, \& {de Zotti}}]{bonaldi2007}
{Bonaldi}, A., {Ricciardi}, S., {Leach}, S., {et~al.} 2007, \mnras, 382, 1791

\bibitem[{{Broadbent} {et~al.}(1989){Broadbent}, {Osborne}, \&
  {Haslam}}]{broad1989}
{Broadbent}, A., {Osborne}, J.~L., \& {Haslam}, C.~G.~T. 1989, \mnras, 237, 381

\bibitem[{{Cardoso} {et~al.}(2008){Cardoso}, {Martin}, {Delabrouille},
  {Betoule}, \& {Patanchon}}]{smicaIEEE}
{Cardoso}, J.-F., {Martin}, M., {Delabrouille}, J., {Betoule}, M., \&
  {Patanchon}, G. 2008, IEEE Journal of Selected Topics in Signal Processing,
  2, 735

\bibitem[{{Casaponsa} {et~al.}(2011){Casaponsa}, {Barreiro}, {Curto},
  {Mart{\'{\i}}nez-Gonz{\'a}lez}, \& {Vielva}}]{Casaponsa2012}
{Casaponsa}, B., {Barreiro}, R.~B., {Curto}, A.,
  {Mart{\'{\i}}nez-Gonz{\'a}lez}, E., \& {Vielva}, P. 2011, \mnras, 411, 2019

\bibitem[{{Davies} {et~al.}(2006){Davies}, {Dickinson}, {Banday}, {Jaffe},
  {G{\'o}rski}, \& {Davis}}]{davies2006}
{Davies}, R.~D., {Dickinson}, C., {Banday}, A.~J., {et~al.} 2006, \mnras, 370,
  1125

\bibitem[{{Davies} {et~al.}(1996){Davies}, {Watson}, \&
  {Gutierrez}}]{davies1996}
{Davies}, R.~D., {Watson}, R.~A., \& {Gutierrez}, C.~M. 1996, \mnras, 278, 925

\bibitem[{{de Oliveira-Costa} {et~al.}(2004){de Oliveira-Costa}, {Tegmark},
  {Davies}, {Guti{\'e}rrez}, {Lasenby}, {Rebolo}, \& {Watson}}]{oliveira2004}
{de Oliveira-Costa}, A., {Tegmark}, M., {Davies}, R.~D., {et~al.} 2004, \apjl,
  606, L89

\bibitem[{{Delabrouille} {et~al.}(2012){Delabrouille}, {Betoule}, {Melin},
  {Miville-Desch{\^e}nes}, {Gonzalez-Nuevo}, {Le Jeune}, {Castex}, {de Zotti},
  {Basak}, {Ashdown}, {Aumont}, {Baccigalupi}, {Banday}, {Bernard}, {Bouchet},
  {Clements}, {da Silva}, {Dickinson}, {Dodu}, {Dolag}, {Elsner}, {Fauvet},
  {Fa{\"y}}, {Giardino}, {Leach}, {Lesgourgues}, {Liguori}, {Macias-Perez},
  {Massardi}, {Matarrese}, {Mazzotta}, {Montier}, {Mottet}, {Paladini},
  {Partridge}, {Piffaretti}, {Prezeau}, {Prunet}, {Ricciardi}, {Roman},
  {Schaefer}, \& {Toffolatti}}]{delabrouille2012}
{Delabrouille}, J., {Betoule}, M., {Melin}, J.-B., {et~al.} 2012, ArXiv
  e-prints

\bibitem[{Delabrouille {et~al.}(2009)Delabrouille, Cardoso, {Le Jeune}, M.,
  Fay, \& Guilloux}]{nilc2009}
Delabrouille, J., Cardoso, J.-F., {Le Jeune}, M., {et~al.} 2009, A\&A, 493, 835

\bibitem[{{Dickinson} {et~al.}(2003){Dickinson}, {Davies}, \&
  {Davis}}]{dickinson2003}
{Dickinson}, C., {Davies}, R.~D., \& {Davis}, R.~J. 2003, \mnras, 341, 369

\bibitem[{{Dobler} \& {Finkbeiner}(2008)}]{dobler2008}
{Dobler}, G. \& {Finkbeiner}, D.~P. 2008, \apj, 680, 1222

\bibitem[{{Draine} \& {Lazarian}(1998)}]{DL1998}
{Draine}, B.~T. \& {Lazarian}, A. 1998, \apj, 508, 157

\bibitem[{{Eriksen} {et~al.}(2006){Eriksen}, Dickinson, Lawrence, Baccigalupi,
  Banday, Górski, Hansen, Lilje, Pierpaoli, Seiffert, Smith, \&
  Vanderlinde}]{Eriksen2006ApJ641}
{Eriksen}, H.~K., Dickinson, C., Lawrence, C.~R., {et~al.} 2006, \apj, 641, 665

\bibitem[{{Eriksen} {et~al.}(2007){Eriksen}, {Huey}, {Saha}, {Hansen}, {Dick},
  {Banday}, {G{\'o}rski}, {Jain}, {Jewell}, {Knox}, {Larson}, {O'Dwyer},
  {Souradeep}, \& {Wandelt}}]{Eriksen2007ApJ656}
{Eriksen}, H.~K., {Huey}, G., {Saha}, R., {et~al.} 2007, \apj, 656, 641

\bibitem[{{Eriksen} {et~al.}(2008){Eriksen}, {Jewell}, {Dickinson}, {Banday},
  {G{\'o}rski}, \& {Lawrence}}]{Eriksen2008ApJ676}
{Eriksen}, H.~K., {Jewell}, J.~B., {Dickinson}, C., {et~al.} 2008, \apj, 676,
  10

\bibitem[{{Eriksen} {et~al.}(2004){Eriksen}, {O'Dwyer}, {Jewell}, {Wandelt},
  {Larson}, {G{\'o}rski}, {Levin}, {Banday}, \& {Lilje}}]{Eriksen2004ApJS155}
{Eriksen}, H.~K., {O'Dwyer}, I.~J., {Jewell}, J.~B., {et~al.} 2004, \apjs, 155,
  227

\bibitem[{{Fendt} \& {Wandelt}(2008)}]{pico}
{Fendt}, W. \& {Wandelt}, B. 2008, in APS April Meeting Abstracts, J8004

\bibitem[{{Fern{\'a}ndez-Cobos} {et~al.}(2012){Fern{\'a}ndez-Cobos}, {Vielva},
  {Barreiro}, \& {Mart{\'{\i}}nez-Gonz{\'a}lez}}]{Fernandezcobos2012}
{Fern{\'a}ndez-Cobos}, R., {Vielva}, P., {Barreiro}, R.~B., \&
  {Mart{\'{\i}}nez-Gonz{\'a}lez}, E. 2012, \mnras, 420, 2162

\bibitem[{{Finkbeiner}(2004)}]{finkbeiner2004}
{Finkbeiner}, D.~P. 2004, \apj, 614, 186

\bibitem[{{Finkbeiner} {et~al.}(1999){Finkbeiner}, {Davis}, \&
  {Schlegel}}]{finkbeiner1999}
{Finkbeiner}, D.~P., {Davis}, M., \& {Schlegel}, D.~J. 1999, \apj, 524, 867

\bibitem[{{Finkbeiner} {et~al.}(2004){Finkbeiner}, {Langston}, \&
  {Minter}}]{fink2004}
{Finkbeiner}, D.~P., {Langston}, G.~I., \& {Minter}, A.~H. 2004, \apj, 617, 350

\bibitem[{{Ghosh} {et~al.}(2012){Ghosh}, {Banday}, {Jaffe}, {Dickinson},
  {Davies}, {Davis}, \& {Gorski}}]{ghosh2011}
{Ghosh}, T., {Banday}, A.~J., {Jaffe}, T., {et~al.} 2012, \mnras, 422, 3617

\bibitem[{{Gold} {et~al.}(2011){Gold}, {Odegard}, {Weiland}, {Hill}, {Kogut},
  {Bennett}, {Hinshaw}, {Chen}, {Dunkley}, {Halpern}, {Jarosik}, {Komatsu},
  {Larson}, {Limon}, {Meyer}, {Nolta}, {Page}, {Smith}, {Spergel}, {Tucker},
  {Wollack}, \& {Wright}}]{gold2010}
{Gold}, B., {Odegard}, N., {Weiland}, J.~L., {et~al.} 2011, \apjs, 192, 15

\bibitem[{{G{\'o}rski} {et~al.}(2005){G{\'o}rski}, {Hivon}, {Banday},
  {Wandelt}, {Hansen}, {Reinecke}, \& {Bartelmann}}]{gorski2005}
{G{\'o}rski}, K.~M., {Hivon}, E., {Banday}, A.~J., {et~al.} 2005, \apj, 622,
  759

\bibitem[{{Haslam} {et~al.}(1982){Haslam}, {Stoffel}, {Salter}, \&
  {Wilson}}]{haslam1982}
{Haslam}, C., {Stoffel}, H., {Salter}, C.~J., \& {Wilson}, W.~E. 1982,
  Astronomy and Astrophysics Supplement Series, 47, 1

\bibitem[{{Hivon} {et~al.}(2002){Hivon}, G{\' o}rski, Netterfield, P., \&
  et~al.}]{master02}
{Hivon}, E., G{\' o}rski, K.~M., Netterfield, C.~B., P., C.~B., \& et~al. 2002,
  Astrophys.J., 567

\bibitem[{{Hoang} \& {Lazarian}(2012)}]{hoang2012}
{Hoang}, T. \& {Lazarian}, A. 2012, Advances in Astronomy, 2012

\bibitem[{{Jewell} {et~al.}(2004){Jewell}, {Levin}, \&
  {Anderson}}]{Jewell2004ApJ609}
{Jewell}, J., {Levin}, S., \& {Anderson}, C.~H. 2004, \apj, 609, 1

\bibitem[{{Kogut}(1996)}]{kogut1996}
{Kogut}, A. 1996, in Bulletin of the American Astronomical Society, Vol.~28,
  American Astronomical Society Meeting Abstracts, 1295

\bibitem[{{Komatsu} {et~al.}(2005){Komatsu}, {Spergel}, \&
  {Wandelt}}]{komatsu2005}
{Komatsu}, E., {Spergel}, D.~N., \& {Wandelt}, B.~D. 2005, \apj, 634, 14

\bibitem[{{Lagache}(2003)}]{lagache2003}
{Lagache}, G. 2003, \aap, 405, 813

\bibitem[{{Leach} {et~al.}(2008){Leach}, {Cardoso}, {Baccigalupi}, {Barreiro},
  {Betoule}, {Bobin}, {Bonaldi}, {Delabrouille}, {de Zotti}, {Dickinson},
  {Eriksen}, {Gonz{\'a}lez-Nuevo}, {Hansen}, {Herranz}, {Le Jeune},
  {L{\'o}pez-Caniego}, {Mart{\'{\i}}nez-Gonz{\'a}lez}, {Massardi}, {Melin},
  {Miville-Desch{\^e}nes}, {Patanchon}, {Prunet}, {Ricciardi}, {Salerno},
  {Sanz}, {Starck}, {Stivoli}, {Stolyarov}, {Stompor}, \& {Vielva}}]{leach2008}
{Leach}, S.~M., {Cardoso}, J., {Baccigalupi}, C., {et~al.} 2008, \aap, 491, 597

\bibitem[{{Leitch} {et~al.}(1997){Leitch}, {Readhead}, {Pearson}, \&
  {Myers}}]{leitch1997}
{Leitch}, E.~M., {Readhead}, A.~C.~S., {Pearson}, T.~J., \& {Myers}, S.~T.
  1997, \apjl, 486, L23

\bibitem[{{Mart{\i}nez-Gonzalez} {et~al.}(2003){Mart{\i}nez-Gonzalez}, Diego,
  Vielva, \& Silk}]{martinez-gonzalez_etal_2003}
{Mart{\i}nez-Gonzalez}, E., Diego, J.~M., Vielva, P., \& Silk, J. 2003, MNRAS,
  345

\bibitem[{{Miville-Desch{\^e}nes} {et~al.}(2008){Miville-Desch{\^e}nes},
  {Ysard}, {Lavabre}, {Ponthieu}, {Mac{\'{\i}}as-P{\'e}rez}, {Aumont}, \&
  {Bernard}}]{mamd}
{Miville-Desch{\^e}nes}, M.-A., {Ysard}, N., {Lavabre}, A., {et~al.} 2008,
  \aap, 490, 1093

\bibitem[{Okamoto \& Hu(2003)}]{Okamoto:2003zw}
Okamoto, T. \& Hu, W. 2003, Phys. Rev., D67, 083002

\bibitem[{{Patanchon} {et~al.}(2005){Patanchon}, {Cardoso}, {Delabrouille}, \&
  {Vielva}}]{2005MNRAS.364.1185P}
{Patanchon}, G., {Cardoso}, J.-F., {Delabrouille}, J., \& {Vielva}, P. 2005,
  \mnras, 364, 1185

\bibitem[{{Pietrobon} {et~al.}(2012){Pietrobon}, {G{\'o}rski}, {Bartlett},
  {Banday}, {Dobler}, {Colombo}, {Hildebrandt}, {Jewell}, {Pagano}, {Rocha},
  {Eriksen}, {Saha}, \& {Lawrence}}]{pietrobon_et_al_2011}
{Pietrobon}, D., {G{\'o}rski}, K.~M., {Bartlett}, J., {et~al.} 2012, \apj, 755,
  69

\bibitem[{{Planck Collaboration}(2012)}]{marta}
{Planck Collaboration}. 2012, {Planck intermediate results: Dust emission at
  millimetre wavelenghts in the Galactic Plane} ({in preparation})

\bibitem[{{Planck Collaboration}(2013{\natexlab{a}})}]{planck-pip58}
{Planck Collaboration}. 2013{\natexlab{a}}, In preparation

\bibitem[{{Planck Collaboration}(2013{\natexlab{b}})}]{planck2013-p06b}
{Planck Collaboration}. 2013{\natexlab{b}}, Submitted to \aap

\bibitem[{{Planck Collaboration ES}(2013)}]{planck2013-p28}
{Planck Collaboration ES}. 2013, {The Explanatory Supplement to the Planck 2013
  results} ({ESA})

\bibitem[{{Planck Collaboration I}(2013)}]{planck2013-p01}
{Planck Collaboration I}. 2013, Submitted to \aap

\bibitem[{{Planck Collaboration II}(2013)}]{planck2013-p02}
{Planck Collaboration II}. 2013, Submitted to \aap

\bibitem[{{Planck Collaboration Int. IX}(2013)}]{planck2012-IX}
{Planck Collaboration Int. IX}. 2013, Submitted to \aap

\bibitem[{{Planck Collaboration Int. XII}(2013)}]{planck2013-XII}
{Planck Collaboration Int. XII}. 2013, Submitted to \aap

\bibitem[{{Planck Collaboration IX}(2013)}]{planck2013-p03d}
{Planck Collaboration IX}. 2013, Submitted to \aap

\bibitem[{{Planck Collaboration VI}(2013)}]{planck2013-p03}
{Planck Collaboration VI}. 2013, Submitted to \aap

\bibitem[{{Planck Collaboration VIII}(2013)}]{planck2013-p03f}
{Planck Collaboration VIII}. 2013, Submitted to \aap

\bibitem[{{Planck Collaboration X}(2013)}]{planck2013-p03e}
{Planck Collaboration X}. 2013, Submitted to \aap

\bibitem[{{Planck Collaboration XIII}(2011)}]{planck2011-6.1}
{Planck Collaboration XIII}. 2011, \aap, 536, A13

\bibitem[{{Planck Collaboration XIII}(2013)}]{planck2013-p03a}
{Planck Collaboration XIII}. 2013, Submitted to \aap

\bibitem[{{Planck Collaboration XIX}(2013)}]{planck2013-p14}
{Planck Collaboration XIX}. 2013, Submitted to \aap

\bibitem[{{Planck Collaboration XV}(2013)}]{planck2013-p08}
{Planck Collaboration XV}. 2013, Submitted to \aap

\bibitem[{{Planck Collaboration XVI}(2013)}]{planck2013-p11}
{Planck Collaboration XVI}. 2013, Submitted to \aap

\bibitem[{{Planck Collaboration XVII}(2013)}]{planck2013-p12}
{Planck Collaboration XVII}. 2013, Submitted to \aap

\bibitem[{{Planck Collaboration XVIII}(2011)}]{planck2011-6.6}
{Planck Collaboration XVIII}. 2011, \aap, 536, A18

\bibitem[{{Planck Collaboration XVIII}(2013)}]{planck2013-p13}
{Planck Collaboration XVIII}. 2013, Submitted to \aap

\bibitem[{{Planck Collaboration XX}(2011)}]{planck2011-7.2}
{Planck Collaboration XX}. 2011, \aap, 536, A20

\bibitem[{{Planck Collaboration XXI}(2013)}]{planck2013-p05b}
{Planck Collaboration XXI}. 2013, Submitted to \aap

\bibitem[{{Planck Collaboration XXII}(2013)}]{planck2013-p17}
{Planck Collaboration XXII}. 2013, Submitted to \aap

\bibitem[{{Planck Collaboration XXIII}(2013)}]{planck2013-p09}
{Planck Collaboration XXIII}. 2013, Submitted to \aap

\bibitem[{{Planck Collaboration XXIV}(2013)}]{planck2013-p09a}
{Planck Collaboration XXIV}. 2013, Submitted to \aap

\bibitem[{{Planck Collaboration XXIX}(2013)}]{planck2013-p05a}
{Planck Collaboration XXIX}. 2013, Submitted to \aap

\bibitem[{{Planck Collaboration XXV}(2013)}]{planck2013-p20}
{Planck Collaboration XXV}. 2013, Submitted to \aap

\bibitem[{{Planck Collaboration XXVI}(2013)}]{planck2013-p19}
{Planck Collaboration XXVI}. 2013, Submitted to \aap

\bibitem[{{Planck Collaboration XXVIII}(2013)}]{planck2013-p05}
{Planck Collaboration XXVIII}. 2013, Submitted to \aap

\bibitem[{{Platania} {et~al.}(2003){Platania}, {Burigana}, {Maino}, {Caserini},
  {Bersanelli}, {Cappellini}, \& {Mennella}}]{platania2003}
{Platania}, P., {Burigana}, C., {Maino}, D., {et~al.} 2003, \aap, 410, 847

\bibitem[{{Reich} \& {Reich}(1988)}]{reich&reich1988}
{Reich}, P. \& {Reich}, W. 1988, \aaps, 74, 7

\bibitem[{{Rocha} {et~al.}(2009){Rocha}, {Contaldi}, {Bond}, \&
  {Gorski}}]{rocha2009}
{Rocha}, G., {Contaldi}, C.~R., {Bond}, J.~R., \& {Gorski}, K.~M. 2009, ArXiv
  e-prints

\bibitem[{{Rocha} {et~al.}(2011){Rocha}, Contaldi, Bond, \&
  Gorski}]{XFaster-like2}
{Rocha}, G., Contaldi, C.~R., Bond, J.~R., \& Gorski, K.~M. 2011, MNRAS, 414,
  823R

\bibitem[{{Rocha} {et~al.}(2010){Rocha}, {Contaldi}, {Colombo}, {Bond},
  {Gorski}, \& {Lawrence}}]{rocha2010b}
{Rocha}, G., {Contaldi}, C.~R., {Colombo}, L.~P.~L., {et~al.} 2010, ArXiv
  e-prints

\bibitem[{{Stompor} {et~al.}(2009){Stompor}, Leach, Stivoli, \&
  Baccigalupi}]{Stompor2009MNRAS392}
{Stompor}, R., Leach, S., Stivoli, F., \& Baccigalupi, C. 2009, MNRAS, 392, 216

\bibitem[{{Tegmark} {et~al.}(2003){Tegmark}, de~Oliveira-Costa, \&
  Hamilton}]{Tegmark2003}
{Tegmark}, M., de~Oliveira-Costa, A., \& Hamilton, A. 2003, Phys.Rev.D, 68,
  123523

\bibitem[{{Tristram} {et~al.}(2005){Tristram}, {Patanchon},
  {Mac{\'{\i}}as-P{\'e}rez}, {Ade}, {Amblard}, {Ansari}, {Aubourg},
  {Beno{\^i}t}, {Bernard}, {Blanchard}, {Bock}, {Bouchet}, {Bourrachot},
  {Camus}, {Cardoso}, {Couchot}, {de Bernardis}, {Delabrouille}, {D{\'e}sert},
  {Douspis}, {Dumoulin}, {Filliatre}, {Fosalba}, {Giard}, {Giraud-H{\'e}raud},
  {Gispert}, {Guglielmi}, {Hamilton}, {Hanany}, {Henrot-Versill{\'e}},
  {Kaplan}, {Lagache}, {Lamarre}, {Lange}, {Madet}, {Maffei}, {Magneville},
  {Masi}, {Mayet}, {Nati}, {Perdereau}, {Plaszczynski}, {Piat}, {Ponthieu},
  {Prunet}, {Renault}, {Rosset}, {Santos}, {Vibert}, \&
  {Yvon}}]{2005A&A...436..785T}
{Tristram}, M., {Patanchon}, G., {Mac{\'{\i}}as-P{\'e}rez}, J.~F., {et~al.}
  2005, \aap, 436, 785

\bibitem[{{Wandelt} {et~al.}(2004){Wandelt}, {Larson}, \&
  {Lakshminarayanan}}]{Wandelt2004PhRvD70}
{Wandelt}, B.~D., {Larson}, D.~L., \& {Lakshminarayanan}, A. 2004, \prd, 70,
  083511

\bibitem[{{Wehus et al.}(2013)}]{wehus2013}
{Wehus et al.} 2013, in preparation

\bibitem[{{Ysard} {et~al.}(2010){Ysard}, {Miville-Desch{\^e}nes}, \&
  {Verstraete}}]{ysard2010}
{Ysard}, N., {Miville-Desch{\^e}nes}, M.~A., \& {Verstraete}, L. 2010, \aap,
  509, L1

\bibitem[{{Ysard} \& {Verstraete}(2010)}]{ysard-ves2010}
{Ysard}, N. \& {Verstraete}, L. 2010, \aap, 509, A12

\end{thebibliography}

\begin{appendix}

\section{Physical parametrization}
\label{sec:commander_appendix}

The \comrul\ (\CR) approach implements Bayesian component separation
in pixel space, fitting a parametric model to the data by sampling the
posterior distribution for the model parameters.  For computational
reasons, the fit is performed in a two-step procedure: First, both
foreground amplitudes and spectral parameters are found at
low-resolution using Markov Chain Monte Carlo (MCMC)/Gibbs sampling
algorithms
\citep{Jewell2004ApJ609,Wandelt2004PhRvD70,Eriksen2004ApJS155,Eriksen2007ApJ656,Eriksen2008ApJ676}. Second,
the amplitudes are recalculated at high resolution by solving the
generalized least squares system (GLSS) per pixel with the spectral
parameters fixed to the their values from the low-resolution run.

For the CMB-oriented analysis presented in this paper, we only use the
seven lowest \Planck\ frequencies, i.e., from 30 to 353\GHz.  We first
downgrade each frequency map from its native angular resolution to a
common resolution of 40~arcminutes and repixelize at
\healpix\ $\nside=256$.  Second, we set the monopoles and dipoles for
each frequency band as described in Sect.~\ref{sec:fg_data} using a
method that locally records spectral indices \citep{wehus2013}.  We
approximate the effective instrumental noise as white with a root mean
square (RMS) per pixel given by the \Planck\ scanning pattern and an
amplitude calibrated by smoothing simulations of the instrumental
noise including correlations to the same resolution.  For the
high-resolution analysis, the important pre-processing step is the
upgrading of the effective low-resolution mixing matrices to full
\Planck\ resolution: this is done by repixelizing from $\nside=256$ to
2048 in harmonic space, ensuring that potential pixelization effects
from the low-resolution map do not introduce sharp boundaries in the
high-resolution map.

Our model for the data, a map $\mathbf{d}_\nu$ at frequency $\nu$,
consists of a linear combination of $N_{\textrm{c}}$ astrophysical
components, plus instrumental noise,
\begin{equation}
\mathbf{d}_\nu = \sum_{i=1}^{N_\mathrm{c}} \mathbf{F}^i_\nu(\theta)
\cdot \mathbf{A}^i + \mathbf{n}_\nu,
\end{equation}
where $\mathbf{A}^i$ denotes a sky map vector containing the
foreground amplitude map for component $i$ normalized at a reference
frequency, and $\mathbf{F}^{i}_\nu(\theta^i)$ is a diagonal matrix
describing the spectral emission law for component $i$ as a function
of frequency and which depends on a (small) set of spectral
parameters, $\theta$.  The CMB signal is included in the sum and, as a
special case, it may be represented either in harmonic or pixel space,
depending on whether the main goal of the analysis is CMB power
spectrum analysis or component separation. The former representation
is used for the \commander-based low-$\ell$ likelihood presented in
\citet{planck2013-p08}, while the latter is used for the foreground
fits presented in Section \ref{sec:foreground_components} of this
paper.

Bayes theorem specifies the posterior distribution for the model parameters,
\begin{equation}
P(\mathbf{A}^i, \theta|\mathbf{d}) \propto \mathcal{L}(\mathbf{A}^i,
\theta) P(\mathbf{A}^i, \theta),
\end{equation}
where $\mathcal{L}(\mathbf{A}^i, \theta)$ is a Gaussian likelihood,
and the prior $P(\mathbf{A}^i, \theta^i)$ depends on the
application. In this paper, the prior on spectral indices is a product
of a Jeffreys prior and physical priors, as detailed in Section
\ref{sec:fg_models}; no priors are imposed on the foreground
amplitudes.

In the low-resolution \commander\ analysis, we exploit a Gibbs sampler
to map out the posterior distribution \citep{Eriksen2008ApJ676},
adopting the following minimal two-step scheme:
\begin{align}
\label{eq:commander_ampsamp}
\mathbf{A} &\leftarrow P(\mathbf{A} | \theta, \mathbf{d}) \\
\label{eq:commander_indxsamp}
\theta     &\leftarrow P(\theta | \mathbf{A}, \mathbf{d}).
\end{align}
The first conditional distribution is a multivariate Gaussian, while
the second distribution does not have an analytic form and must be
mapped out numerically.

The high-resolution \ruler\ analysis maximizes the foreground
amplitude conditional in Eq.~\ref{eq:commander_ampsamp} numerically by
solving the generalized least squares system
\begin{equation}
\label{eq:cr_glss}
\mathbf{A}^i(\theta) = \Big( \sum_\nu \mathbf{F}_\nu^{i,\rm
  T}\mathbf{N}_\nu^{-1}\mathbf{F}_\nu^i\Big)^{-1} \sum_\mu
\mathbf{F}_\mu^{i,\rm T}\mathbf{N}_\mu^{-1} \mathbf{D}_\mu \equiv
\sum_\mu \mathbf{W}(\theta)_\mu^i \mathbf{D}_\mu,
\end{equation}
were $\mathbf{N}_\nu$ is the noise covariance matrix (assumed to be
diagonal) of the $\nu$th channel, $\mathbf{F}^i \equiv
\mathbf{F}^i(\theta)$ and we have neglected the different angular
resolutions of the channels.  The posterior marginal average for the
high-resolution amplitude maps is then given by
$\langle\mathbf{A}^i\rangle = \int \mathbf{A}^i(\theta)P(\theta) d
\theta \simeq \frac{1}{N_{\rm sample}}\sum_{\theta,\nu}
\mathbf{W}(\theta)_\nu^i \mathbf{D}_\nu \equiv \sum_\nu
\mathbf{W}_\nu^i \mathbf{D}$, a sum over the $N_{\rm sample}$ samples
of the spectral parameters $\theta$.

Once the channel weights, ${\bf W}_\nu$, have been computed,
processing a large number of simulations requires negligible
computational resources. This feature has been extensively used for
computation of the effective beam of \ruler\ maps: \ffp\ CMB
simulations for the 30 to 353\GHz\ channels are combined according to
${\bf W}_\nu$ and the effective beam transfer function is found as
$b_\ell^2 \equiv \langle C_\ell^{\rm out} /(C_\ell^{\rm inp} w_\ell^2)
\rangle$.  Here, $C_\ell^{\rm inp}$ is the power spectrum of the input
simulation before convolution with the instrumental beam, $w_\ell$ is
the \healpix\ pixel window function, and the average is taken over the
set of simulations. Missing pixels are set to $0$ when computing
$C_\ell$ in the above expression. The low number ($\sim 500$) of
missing pixels in the data renders the impact of such a choice
negligible at $\ell < 2000$. A similar procedure is used for defining
the effective beam of the non-CMB components.

The above algorithm produces a set of samples drawn from the posterior
distribution, as opposed to a direct estimate of individual component
amplitudes or spectral parameters.  While this sample set provides a
statistically complete representation of the posterior, it is
non-trivial to visualize or to compare the distribution with external
data.  For convenience, we therefore summarize the distribution in
terms of mean and standard deviation maps for each component. We
emphasize, however, that the distribution is significantly
non-Gaussian, and when searching for features in the maps at low
signal-to-noise levels, one must take into account the exact
distribution.

Finally, the \comrul\ confidence mask (see Sect.~\ref{sec:cmb_maps})
is primarily defined by the product of the CG80 mask and the point
source mask described in Sect.~\ref{sec:data}.  We additionally remove
any pixels excluded by the 13\% \commander\ likelihood mask described
by \citet{planck2013-p08}; however, this is almost entirely included
within the CG80 mask, and this step therefore has very little impact
on the final result. To complete the mask, we remove any pixels for
which the high-resolution \ruler\ CMB map, smoothed to 40 arcminutes,
differs by more than $3\,\sigma$ from the low-resolution
\commander\ CMB map, which can happen due to spatial spectral
variations on pixel scales.

\section{Internal linear combination}
\label{sec:nilc_appendix}

\nilc\ is a method to extract the CMB (or any component
with known spectral behaviour) by applying the ILC technique to
multi-channel observations in needlet space, that is, with weights
that are allowed to vary over the sky and over the full multipole
range.

\begin{figure}[!h]
  \begin{center}
    \includegraphics[width=\columnwidth]{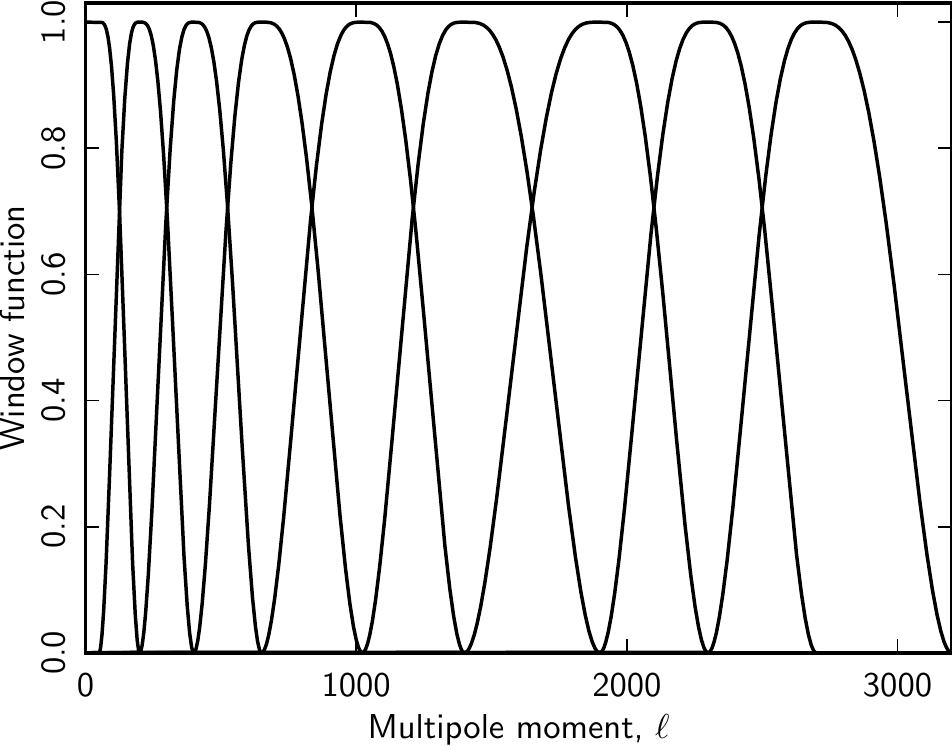}\\
    \includegraphics[width=\columnwidth]{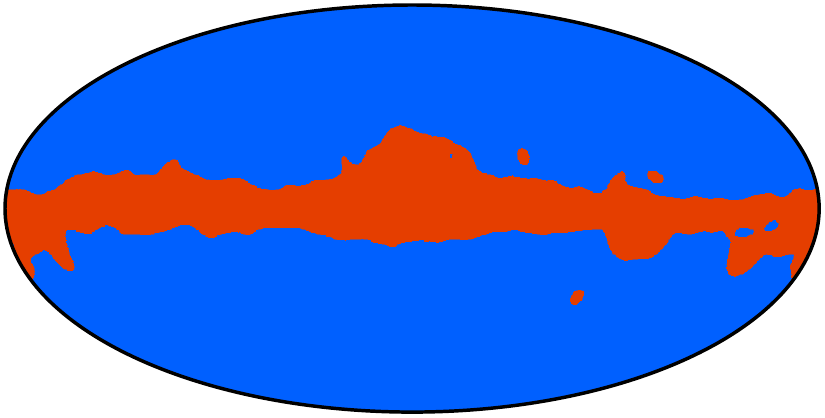}\\
    \includegraphics[width=\columnwidth]{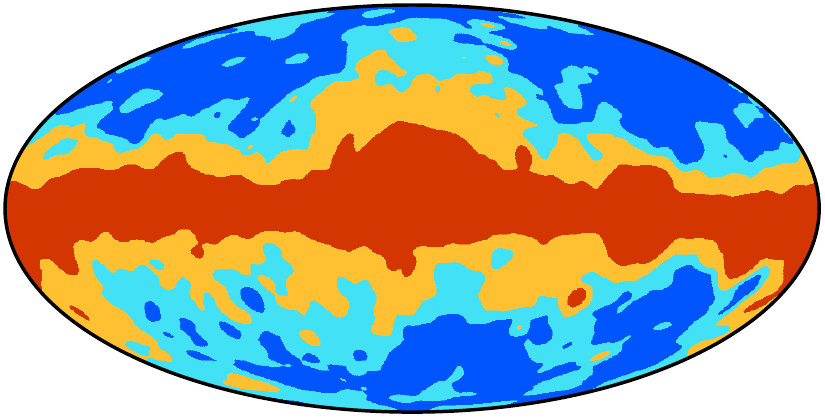}\\
    \includegraphics[width=\columnwidth]{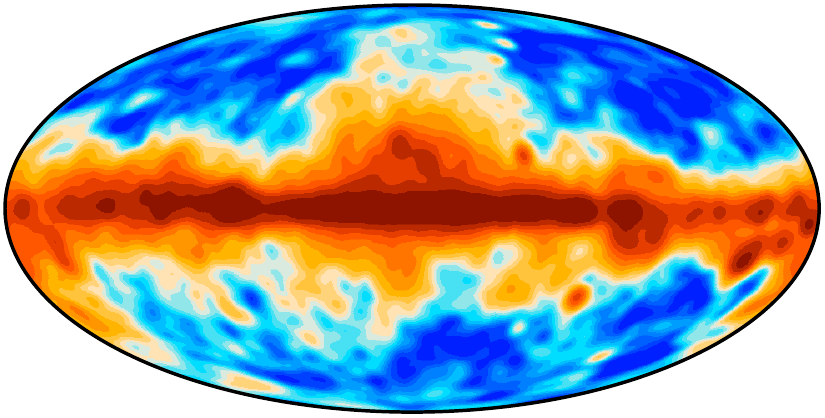}
  \end{center}
  \caption{Spectral localization for \nilc\ using nine spectral window
    functions defining nine needlet scales (\emph{top panel}).  The
    scale-dependent spatial localization partitions the sky in one
    zone (for scale 1), two zones (for scale 2), four zones (for scale
    3), or twenty zones (for scales 5, 6, 7, 8, 9).  The two-zone,
    4-zone and 20-zone partitions are depicted in the lower panels.}
  \label{fig:nilc:localization}
\end{figure}

The ability to linearly combine input maps varying over the sky and
over multipoles is called localization.  In the needlet framework,
harmonic localization is achieved using a set of bandpass filters
defining a series of scales, and spatial localization is achieved at
each scale by defining zones over the sky.  The harmonic localization
adopted here uses nine spectral bands covering multipoles up to
$\ell=3200$ (see Fig.~\ref{fig:nilc:localization}).  The spatial
localization depends on the scale. At the coarsest scale, which
includes the multipoles of lowest degree, we use a single zone (no
localization), while at the finest scales (which include the highest
multipoles) the sky is partitioned into 20 zones (again, see
Fig.~\ref{fig:nilc:localization}).

The \nilc\ method amounts to computing an ILC in each zone of each
scale, allowing the ILC weights to adapt naturally to the varying
strength of the other components as a function of position and
multipole.  A complete description of the basic \nilc\ method can be
found in~\citet{nilc2009}.

In the present work, however, there is an important difference in the
processing of the coarsest scale. Since the coarsest scale of the
\nilc\ filter is not localized, a plain \nilc\ map would be equivalent
to a pixel-based ILC for all the multipoles of that scale.  This
procedure, however, is known to be quite susceptible to the ``ILC
bias'' due to chance correlations between the CMB and foregrounds. In
order to mitigate this effect, the (single) covariance matrix which
determines the ILC coefficients at the coarsest scale is not computed
as a pixel average, but is rather estimated in the harmonic domain as
an average over spherical harmonic coefficients using a spectral
weight which equalizes the power of the CMB modes (based on a fiducial
spectrum).  This can be shown significantly to decrease the large
scale errors.

In practice, our \nilc\ processing depends on several implementation
choices, as follows:

\begin{itemize}
  
\item{Input channels:} In this work, the \nilc\ algorithm is applied to
  all Planck channels from 44 to 857\GHz\ omitting only the 30\GHz\
  channel.

\item{Pre-processing of point sources:} Identical to the
  \smica\ pre-processing (see Appendix~\ref{sec:smica_appendix}).
  
\item{Masking and inpainting:} The \nilc\ CMB map is actually produced
  in a three-step process.  In a first step, the \nilc\ weights are
  computed from covariance matrices evaluated using a Galactic mask
  removing about 2\,\% of the sky (and is apodized at 1\deg).  In a
  second step, those \nilc\ weights are applied to needlet
  coefficients computed over the complete sky (except for point source
  masking/subtraction), yielding a \nilc\ CMB estimate over the full
  sky (except for the point source mask).  In short, the weights are
  computed over a masked sky but are applied to a full sky (up to
  point sources).  In a final step, the pixels masked due to point
  source processing are replaced by the values of a constrained
  Gaussian realization (inpainting).

\item{Spatial localization:} The boundaries of the zones used for
 spatial localisation (shown at Fig.~\ref{fig:nilc:localization}) are
 obtained as iso-level curves of a low resolution map of Galactic
 emission.

\item{Beam control and transfer function:} As in the
  \smica\ processing, the input maps are internally re-beamed to a
  5\arcm resolution, so the resulting CMB map is automatically
  synthesized with an effective Gaussian beam of five arcminutes,
  according to the unbiased nature of the ILC.

\item{Using \smica\ recalibration:} In our current implementation, the
  \nilc\ solution uses the values determined by \smica\ for the CMB
  spectrum, given in Eq.~(\ref{eq:smica:recalibnumbers}).

\end{itemize}

\section{Template fitting}

\begin{table*}[tmb]
\begingroup
\newdimen\tblskip \tblskip=5pt
  \caption{Linear coefficients, $\alpha_j$, and templates used to
    clean individual frequency maps with \sevem. The first column
    lists the templates constructed to produce clean maps. Before
    subtraction, the maps are smoothed to a common resolution. The
    (353-143)\GHz\ template is constructed at the resolution of the 44
    and 70\GHz\ channels, in order to clean the 44 and 70\GHz\ maps,
    respectively. For the rest of the templates, the first map used to
    construct the templates was filtered with the beam corresponding
    to the second map and vice versa.  Note that for 100, 143 and
    217\GHz\ channels, we give the coefficients used for the largest
    of the two regions considered in the cleaning, which covers 97\,\%
    of the sky.}
\label{table:sevem_coef}
\nointerlineskip
\vskip -3mm
\footnotesize
\setbox\tablebox=\vbox{
\newdimen\digitwidth
\setbox0=\hbox{\rm 0}
\digitwidth=\wd0
\catcode`*=\active
\def*{\kern\digitwidth}
\newdimen\signwidth
\setbox0=\hbox{+}
\signwidth=\wd0
\catcode`!=\active
\def!{\kern\signwidth}
\newdimen\expsignwidth
\setbox0=\hbox{$^{-}$}
\expsignwidth=\wd0
\catcode`@=\active
\def@{\kern\expsignwidth}
\halign{\hfil#\hfil\tabskip=2em&\hfil#\hfil&\hfil#\hfil&\hfil#\hfil&\hfil#\hfil&\hfil#\hfil&\hfil#\hfil\tabskip=0pt\cr
\noalign{\doubleline}
      Template&   44\GHz& 70\GHz& 100\GHz& 143\GHz& 217\GHz& 353\GHz\cr
\noalign{\vskip 3pt\hrule\vskip 5pt}
   *30--70*   &$3.65\times 10^{-1}$&&&&&\cr  
   *30--44*   &&$1.25\times 10^{-1}$&$-2.35\times 10^{-2}$&!$2.14\times 10^{-2}$&$-1.03\times 10^{-1}$&\cr  
   *44--70*   &&&!$1.67\times 10^{-1}$&!$1.23\times 10^{-1}$&!$1.76\times 10^{-1}$&\cr  
   217--100 &&&&&&$-0.12\times 10^{1@}$\cr  
   217--143 &&&&&&!$8.99\times 10^{-1}$\cr  
   353--143 &$4.05\times 10^{-3}$&$9.31\times 10^{-3}$&&&&\cr
   545--217 &&&&&&!$9.92\times 10^{-2}$\cr
   545--353 &&&!$5.21\times 10^{-3}$&!$7.52\times 10^{-3}$&!$1.84\times 10^{-2}$&\cr  
   857--545 &&&$-4.66\times 10^{-5}$&$-6.67\times 10^{-5}$&$-1.21\times 10^{-4}$&$-5.02\times 10^{-4}$\cr 
\noalign{\vskip 5pt\hrule\vskip 3pt}}}
\endPlancktablewide
\endgroup
\end{table*} 

The original \sevem\ algorithm produced clean CMB maps at several
frequencies through template fitting, followed by an estimation of the
CMB power spectrum from these clean maps using a method based on the
Expectation Maximization algorithm
\citep{martinez-gonzalez_etal_2003,leach2008,Fernandezcobos2012}. From
this power spectrum, a multifrequency Wiener-filtered CMB map was
produced. For the present work, only the first step of the method,
producing clean CMB maps at different frequencies, is considered. In
addition, two of these clean maps are optimally combined to produce a
final CMB map.

The templates used for cleaning are internal, i.e., they are
constructed from \Planck\ data, avoiding the need for external data
sets, which usually complicate the analyses and may introduce
inconsistencies. In the cleaning process, no assumptions about the
foregrounds or noise levels are needed, rendering the technique very
robust. The fitting can be done in real or wavelet space (using a fast
wavelet adapted to the \healpix\ pixelization;
\citealp{Casaponsa2012}) to properly deal with incomplete sky
coverage.  For expediency, however, we fill in the small number of
unobserved pixels at each channel with the mean value of their
neighbouring pixels before applying \sevem.

We construct our templates by subtracting two close \Planck\ frequency
channel maps, after first smoothing them to a common resolution, and
converting to CMB temperature units if necessary, to ensure that the
CMB signal is properly removed. A linear combination of the templates,
$t_j$, is then subtracted from (hitherto unused) map, $d$, to produce
a clean CMB map at that frequency.  This is done either in real or
wavelet space (i.e., scale by scale) at each position on the sky,
\begin{equation}
T_c(\vec{x},\nu)=d(\vec{x},\nu)- \sum_{j=1}^{n_t} \alpha_j
t_j(\vec{x}),
\end{equation}
where $n_t$ is the number of templates. If the cleaning is performed
in real space, the $\alpha_j$ coefficients are obtained by minimizing
the variance of the clean map, $T_c$, outside a given mask. When
working in wavelet space, the cleaning is performed in the same way at
each wavelet scale independently (i.e., the linear coefficients depend
on the scale). Although we exclude very contaminated regions during
the minimization, the subtraction is performed for all pixels and,
therefore, the cleaned maps cover the full-sky (although foreground
residuals are expected to be present in the excluded areas).

An additional level of flexibility may also be considered: the linear
coefficients can be fixed over the full sky, or in several regions.
The regions are then combined in a smooth way, by weighting the pixels
at the boundaries to reduce possible discontinuities in the clean
maps.

Since the method is linear, we may easily propagate the noise
properties to the final CMB map. Moreover, it is very fast and permits
the generation of thousands of simulations to characterize the
statistical properties of the outputs, a critical need for many
cosmological applications. The final CMB map retains the angular
resolution of the original frequency map.

There are several configurations possible for \sevem, depending the
number of frequency maps to be cleaned or the number of templates used
in the fitting. Note that the production of clean maps at different
frequencies is of great interest in order to test the robustness of
the results. Therefore, to define the best strategy, one needs to find
a compromise between the number of maps that can be cleaned
independently and the number of templates that can be constructed.

In particular, we have cleaned the 143\GHz\ and 217\GHz\ maps using
four templates constructed as the difference of the following
\Planck\ channels (smoothed to a common resolution): (30$-$44),
(44$-$70), (545$-$353) and (857$-$545). For simplicity, the two maps
have been cleaned in real space, since there was no significant
improvement when using wavelets, especially at high latitude.  In
order to take into account the different spectral behaviour of the
foregrounds at low and high Galactic latitudes, we considered two
independent sky regions, using different sets of coefficients (see
Table~\ref{table:sevem_coef} for the values of the linear coefficients
for the main considered region). The first region corresponds to the
brightest 3\,\% of Galactic emission, while the second region is
defined by the remaining 97\,\% of the sky (see Sect.~\ref{sec:data}
for a detailed description of our Galactic mask construction). For the
first region, the coefficients are actually estimated over the
complete sky (we find that this is better than performing the
minimization on only the brightest 3\,\% of the sky, where the CMB is
very sub-dominant), while for the second region, we exclude the bright 3\,\%
sky fraction, point sources detected at any frequency and those
pixels which have not been observed in all channels.

Note that, for consistency, we have used the same configuration (four
templates, cleaning in real space, two regions) for the analysis of
the \ffp\ simulations. However, we find that this simple configuration
produces more contaminated CMB maps than for the data (although the
region outside the confidence mask still has low contamination),
indicating some differences between the foreground level in the data
and in simulations. Therefore, conclusions derived from the
\ffp\ results for \sevem\ should be taken with caution, since they are
expected to provide overestimated residuals.

Our final CMB map was constructed by combining the 143 and 217\GHz\
maps by weighting the maps in harmonic space, taking into account the
noise level, the resolution, and a rough estimation of the foreground
residuals of each map (obtained from realistic simulations). This
final map has a resolution corresponding to a Gaussian beam of
5\arcm\ FWHM.

Moreover, additional clean CMB maps (at frequencies 44, 70, 100 and
353\GHz) were also produced using different combinations of
templates. In particular, to clean the 100\GHz\ map, we used the same
templates and regions as for 143 and 217\GHz. This allows us to
produce three (almost) independent clean maps, in the sense that none
of the three maps to be cleaned is used to construct the templates.
For 44, 70 and 353\GHz, different combinations of templates are used
and the linear coefficients are chosen to be the same over the full
sky. They are obtained by minimizing the variance of the map outside
the same mask as that used to clean the central frequency maps on the
largest region. The templates and the corresponding linear
coefficients used for each of the considered frequencies are given in
Table~\ref{table:sevem_coef}.

The \sevem\ clean frequency maps have been used in analyses of the
isotropy and statistics of the CMB \citep{planck2013-p09} and to
obtain cosmological constraints from the integrated Sachs-Wolfe effect
\citep{planck2013-p14}. In particular, clean maps from 44 to 353\GHz\
were used for the stacking analysis presented in
\cite{planck2013-p14}, while frequencies from 70 to 217\GHz\ were used
for consistency tests in \cite{planck2013-p09}.

The confidence mask provided for \sevem\ is constructed by combining
four types of selected regions. In particular, it excludes zones with
high residuals identified through the subtraction of \sevem\ clean
maps at different frequencies, as well as the sources detected at all
Planck channels using the Mexican Hat Wavelet algorithm
\citep{planck2013-p05}.  These point sources are masked with holes of
varying radius, according to the flux of each source. The pixels that
are not observed by all channels are also masked. Finally, to ensure
that the area left outside the mask is statistically robust, we also
exclude from the analysis the brightest 20\,\% of Galactic emission,
leaving a useful area of around 76\,\%. This provides a conservative
mask for CMB analysis; however, we point out that smaller masks could
also be used in specific applications, such as the lensing potential
reconstruction described in Sect.~\ref{sec:ng}).

\label{sec:sevem_appendix}

\section{Spectral matching}
\label{sec:smica_appendix}

\begin{figure}
  \begin{center}
    \includegraphics[width=\columnwidth]{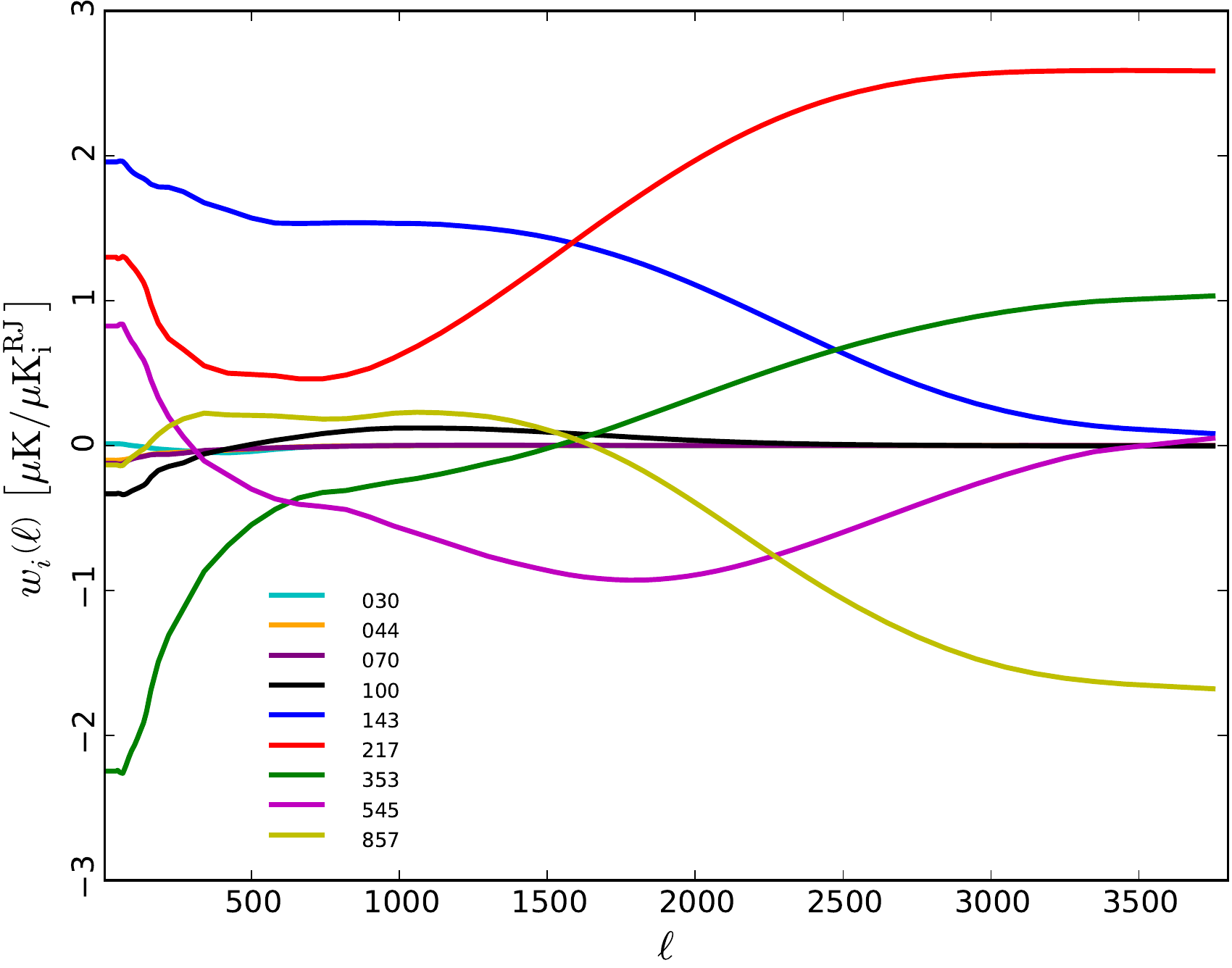} \\
    \includegraphics[width=\columnwidth]{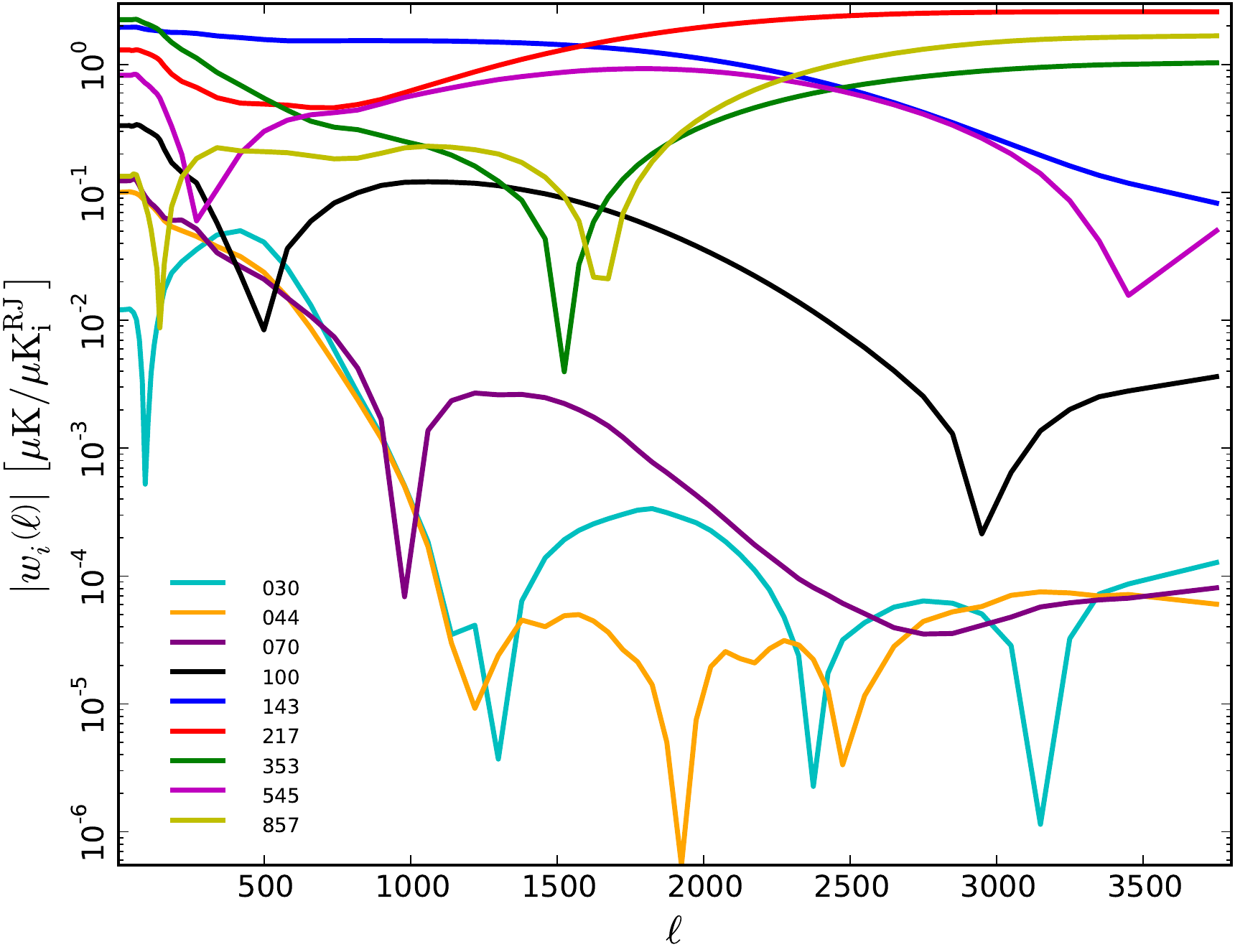}
  \end{center}
  \caption{Weights, $\mathbf{w}_\ell$, given by \smica\ to the input
    maps, after they are re-beamed to 5\arcm and expressed in
    $K_\mathrm{RJ}$, as a function of multipole.  Top panel: linear
    scale; bottom panel: the absolute value of the weights on a
    logarithmic scale.}
  \label{fig:smica:filtersDX9}
\end{figure}

\begin{figure}
  \begin{center}
    \includegraphics[width=\columnwidth]{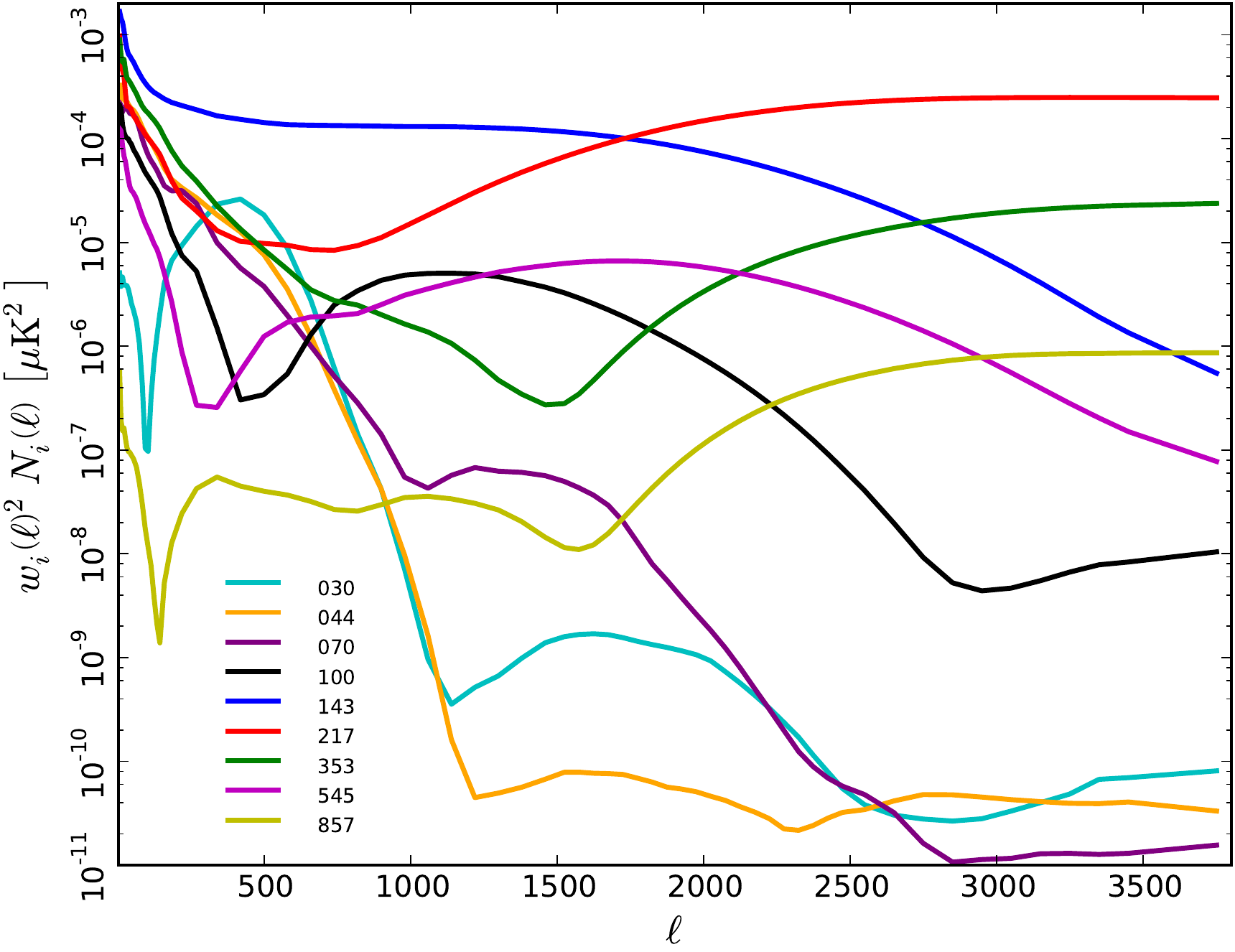} \\
  \end{center}
  \caption{Contribution of each input channel to the noise in the \smica\ map.}
  \label{fig:smica:noisebychan}
\end{figure}

\smica\ (Spectral Matching Independent Component Analysis) reconstructs
a CMB map as a linear combination in harmonic space of
$N_\mathrm{chan}$ input frequency maps with weights that depend on
multipole $\ell$.  Given the $N_\mathrm{chan}\times 1$ vector
$\mathbf{x}_\lm$ of spherical harmonic coefficients for the input
maps, it computes coefficients $\hat s_\lm$ for the CMB map as
\begin{equation}\label{sec:smica:Hfiltering}
  \hat s_\lm = \mathbf{w}_\ell\adj \mathbf{x}_\lm,
\end{equation}
where the \mbox{$N_\mathrm{chan}\times 1$} vector $\mathbf{w}_\ell$
containing the multipole-dependent weights is chosen to give unit gain
to the CMB with minimum variance.  This is achieved with
\begin{equation}\label{eq:smica:app:wl}
  \mathbf{w}_\ell = \frac
  {\mathbf{C}_\ell\inv \mathbf{a}}
  {\mathbf{a}\adj \mathbf{C}_\ell\inv \mathbf{a}},
\end{equation}
where vector $\mathbf{a}$ is the spectrum of the CMB evaluated at each
channel (allowing for possible inter-channel re-calibration factors)
and $\mathbf{C}_\ell$ is the $N_\mathrm{chan}\times N_\mathrm{chan}$
spectral covariance matrix of $\mathbf{x}_\lm$.  Taking
$\mathbf{C}_\ell$ in Eq.~(\ref{eq:smica:app:wl}) to be the sample
spectral covariance matrix $\widehat{\mathbf{C}}_\ell$ of the
observations,
\begin{equation}\label{eq:smica:hR}
  \widehat{\mathbf{C}}_\ell = \frac{1}{2\ell+1}\sum_m \mathbf{x}_\lm \mathbf{x}_\lm\adj,
\end{equation}
would implement a simple harmonic-domain ILC similar to
\citep{Tegmark2003}.  At the largest scales, we instead use a model,
$\mathbf{C}_\ell(\theta)$, and determine the covariance matrix to be
used in Eq.~(\ref{eq:smica:app:wl}) by fitting
$\mathbf{C}_\ell(\theta)$ to $\widehat{\mathbf{C}}_\ell$.  This is
done in the maximum likelihood sense for stationary Gaussian fields,
that is, the best fit matrices, $\mathbf{C}_\ell(\hat\theta)$, are
obtained for
\begin{equation}\label{eq:smica:app:crit}
  \hat\theta = \arg\min_\theta \sum_\ell (2\ell+1) \left(
  \widehat{\mathbf{C}}_\ell\mathbf{C}_\ell(\theta)\inv + \log\det
  \mathbf{C}_\ell(\theta) \right).
\end{equation}
Equations \ref{sec:smica:Hfiltering}-\ref{eq:smica:app:crit} summarize
the basic principles of \smica; its actual operation depends on a
choice for the spectral model $\mathbf{C}_\ell(\theta)$, and on
several implementation-specific details, which we briefly describe
below.

We model the data as a superposition of CMB, noise and foregrounds.
The latter are not parametrically modelled; instead, we represent the
total foreground emission by $d$ templates with arbitrary frequency
spectra, angular spectra and correlations.  In the spectral domain,
this is equivalent to modelling the covariance matrices as
\begin{equation}\label{eq:smica:app:Rlmodel}
  \mathbf{C}_\ell(\theta) 
  = \mathbf{a}\mathbf{a}\adj \,C_\ell 
  + \mathbf{A}\mathbf{P}_\ell\mathbf{A}\adj + \mathbf{N}_\ell,
\end{equation}
where $C_\ell$ is the angular power spectrum of the CMB, $\mathbf{A}$
is a $N_\mathrm{chan}\times d$ matrix, $\mathbf{P}_\ell$ is a positive
definite $d\times d$ matrix, and $\mathbf{N}_\ell$ is a diagonal
matrix representing the noise power spectra of the data.  The
parameter vector $\theta$ contains all or part of the quantities
in~(\ref{eq:smica:app:Rlmodel}).

The decomposition~\ref{eq:smica:app:Rlmodel} reflects the fact that
CMB, foregrounds and noise are independent components of the signal.
Thus, \smica\ is an ICA (independent component analysis) method.  It
operates by matching the observations $\widehat{\mathbf{C}}$ to the
spectral model (\ref{eq:smica:app:Rlmodel}) using the
criterion~(\ref{eq:smica:app:crit}).

The maximal flexibility in a \smica\ fit of model
(\ref{eq:smica:app:Rlmodel}) is obtained with all the parameters free,
that is without any constraint on the spectrum $C_\ell$, on the
diagonal entries of $\mathbf{N}_\ell$, on $\mathbf{a}$, or on
$\mathbf{A}$ and $\mathbf{P}_\ell$.  One would ideally fit all those
parameters (except for obvious degeneracies, like that between a scale
factor in $\mathbf{a}$ and the overall normalization of the CMB
spectrum $C_\ell$) over the whole multipole range.  In practice, this
turns out to be too difficult given the large dynamic range both over
the sky and over multipoles.  We resort to a pragmatic three-step
approach in which the criterion (\ref{eq:smica:app:crit}) is minimized
by first fitting $\mathbf{a}$, then $\mathbf{A}$, and finally the
linear parameters $C_\ell$ and $\mathbf{N}_\ell$.  Each fit is
conducted over the multipole ranges and the sky fraction most
appropriate for the parameter of interest, as follows.
 
We first estimate the CMB spectral law $\mathbf{a}$ by fitting all
model parameters (that is, without constraint) over a clean fraction
of sky ($f_{sky}$=40\,\%) in the range $100\leq\ell\leq 680$ where the
signal is CMB-dominated in most of the channels and the beam window
functions are accurately known.  In this fit, which is done over a
clean part of the sky, we use a foreground emission matrix,
$\mathbf{A}$, with only four columns.  From this step, we only retain
the best fit value for vector $\mathbf{a}$.  In the second step, we
estimate the foreground emissivity by fixing $\mathbf{a}$ to its value
from the previous step and fitting all the other parameters over a
large fraction of sky ($f_{sky}$=97\,\%) in the range $4\leq\ell\leq
150$ where the signal is dominated by the Galactic emission in all
channels. From this second step, we retain the best fit value for the
matrix $\mathbf{A}$ which, again, is adjusted without constraint other
than having $d=6$ columns.  In the last step, we fit all power
spectrum parameters: we fix $\mathbf{a}$ and $\mathbf{A}$ to their
previously found values and fit for $C_\ell$ and $\mathbf{P}_\ell$ at
each $\ell$.
 
Note that the first step (fitting $\mathbf{a}$) amounts to
re-calibrating the input maps on the basis of CMB anisotropies.  For
the maps in thermodynamics units, we find
\begin{equation}
\label{eq:smica:recalibnumbers}
\begin{array}{rcl}
  \hat{\mathbf{a}} &=& [0.9900, 1.0000, 1.0020, 0.9990, 1.0000, \\
    && \quad 1.0004, 0.9920, 1.0457, 1.0000]
\end{array}
\end{equation}
The value at 857\GHz\ is not accurately recovered by \smica, so we
have set $a_\mathrm{857}=1$. Since the norm of $\mathbf{a}$ is
degenerate with a global scale factor for the CMB angular spectrum, it
can only be recovered by \smica\ up to a scale factor.  This
degeneracy is fixed here by taking $a_\mathrm{143}=1$. The
re-calibration step could have been omitted since $\hat{\mathbf{a}}$
is very close the unit vector.  However, we found that using
$\hat{\mathbf{a}}$ improved the behavior of \smica\ over using
$\mathbf{a}=[ 1,\ldots, 1]$.

Before describing implementation details, we explain how \smica\ deals
with the varying resolution of the input channels, since the
discussion thus far assumed that all input maps had the same
resolution.  Since \smica\ works in the harmonic domain, it is a
simple matter to account for the beam transfer function, $b_i(\ell)$,
of the $i$-th input map.  The CMB sky multipoles $s_\lm$ contribute
$s_\lm a_i b_i(\ell) p_i(\ell)$ to the harmonic coefficient $x_\lm^i$
of the $i$-th map (where $p_i(\ell)$ is the pixel window function for
the \healpix\ map at $\nside^i$).  Therefore, in order to produce a
final CMB map at 5\arcm resolution, close to the highest
resolution of \Planck, we only need to work with input spherical
harmonics re-beamed to 5\arcm; that is, to apply \smica\ on vectors
$\tilde{\mathbf{x}}_\lm$ with entries $\tilde x_\lm^i = x_\lm^i
b_5(\ell) / b_i(\ell) / p_i(\ell)$, where $b_5(\ell)$ is a
five-arcminute Gaussian beam function.  By construction, \smica\ then
produces an CMB map with an effective Gaussian beam of
5\arcm\ (without the pixel window function).

We now give further details on the actual implementation of \smica:

\begin{itemize}

\item{Inputs:} \smica\ uses all nine \Planck\ frequency channels from
  30 to 857\GHz, harmonically transformed up to $\ell=4000$.

\item{Pre-processing of point sources:} \smica\ is applied on input
  maps in which point sources are subtracted or masked.  We start by
  fitting the PCCS point sources with $\mathrm{SNR}>5$ to a Gaussian
  shape where the source amplitude is estimated together with its
  position and a constant factor representing the background variance.
  If the fit is successful ($\chi^2 \leq 2$), the fitted point source
  is removed from the map; otherwise it is masked in all channels and
  the hole is inpainted by a simple diffusive filling process.  This
  is done at all frequencies except 545 and 857\GHz, where all point
  sources with $\mathrm{SNR}>7.5$ are masked and inpainted.

\item{Beams:} When the harmonic coefficients of the input maps are
  re-beamed at 5\arcm, we do not apply exactly the expression $\tilde
  x_\lm^i = x_\lm^i b_5(\ell) / b_i(\ell) / p_i(\ell)$ mentioned above
  because the factor $1/b_i(\ell)$ would diverge at high $\ell$ for
  the lowest resolution input channels.  That may not be a problem in
  infinite precision arithmetic, but would lead to matrices
  $\widehat{\mathbf{C}}(\ell)$ with extremely large condition numbers.
  Instead, we re-beam with the factor $1/b_i(\ell)$ replaced by
  $\mathrm{min}(1/b_i(\ell), 1000)$.  The re-beaming of the CMB modes
  then is no longer perfect, but this is of course irrelevant because
  the thresholding occurs in a regime where the signal is completely
  dominated by the noise, so that the contribution of the
  corresponding channel is already highly attenuated by the
  \smica\ weights (as shown in Fig.~\ref{fig:smica:filtersDX9}).

\item{Masking:} In practice, \smica\ operates on a masked sky, the
  mask being applied after the point source processing.  The mask is
  obtained by thresholding a heavily smoothed version of the point
  source mask.  The threshold is chosen to leave about 97\,\% of the
  sky.  Because of the heavy smoothing, the mask has smooth contours
  and is only sensitive to large aggregates of point sources: the
  masked areas mostly lie in the Galactic plane, but include also a
  few bright regions like the Large Magellanic Cloud.

\item{Inpainting:} The \smica\ map used in this paper has no real
  power in the masked region described above.  However, for
  convenience, an inpainted \smica\ map has also been produced by
  replacing the masked pixels with a constrained Gaussian realization
  obtained by the method of \citet{BenoitLevy13}.  That map appears in
  \citet{planck2013-p01}.

\item{Binning:} In our implementation, we use binned spectra.

\item{Processing at fine scales:} Since there is little point trying to
  model the spectral covariance at high multipoles, because the sample
  estimate is sufficient, \smica\ implements a simple harmonic ILC at
  $\ell > 1500$; that is, it applies the
  filter~(\ref{eq:smica:app:wl}) with $\mathbf{C}_\ell =
  \widehat{\mathbf{C}}_\ell$.
  
\item{Confidence mask:} A confidence mask
  (Fig.~\ref{fig:cmb_mask_union}) is provided with \smica, constructed
  in the following way.  The \smica\ CMB map is bandpass filtered
  through a spectral window $v(\ell) =\exp[-((\ell-1700)/200)^2/2]$.
  The result is squared and smoothed at two-degree resolution,
  yielding a map of the (bandpassed) variance of the CMB map.  That
  variance is corrected for the noise contribution by subtracting the
  variance map for the noise obtained by the same procedure applied to
  the \smica\ HRHD map.  If the \smica\ map contained only CMB and
  noise, the variance map would have a uniform value $\sum_\ell
  v(\ell)^2 b_5(\ell)^2 C(\ell) (2\ell+1)/4\pi =
  31.3\,\mu\textrm{K}^2$ over the sky.  The confidence map is obtained
  by thresholding the noise-corrected variance map at 70\microK$^2$.

\end{itemize}

Viewed as a filter, \smica\ can be summarized by the weights
$\mathbf{w}_\ell$ applied to each input map as a function of
multipole.  In this sense, \smica\ is strictly equivalent to co-adding
the input maps after convolution by specific axisymmetric kernels
directly related to the corresponding entry of $\mathbf{w}_\ell$.

The \smica\ weights used here are shown in
Figure~\ref{fig:smica:filtersDX9} (for input maps in units of
$K_\mathrm{RJ}$).  We see, in particular, the (expected) progressive
attenuation of the lowest resolution channels with increasing
multipole.
Figure~\ref{fig:smica:noisebychan} shows the contribution of each
input channel to the noise in the \smica\ CMB map as a function of
multipole.  The spectral noise contribution from channel $i$ is simply
obtained as $w(\ell)^2 N_i(\ell)$, where $w_i(\ell)$ is the $i$-th
entry of the weight vector $\mathbf{w}(\ell)$ and $N_i(\ell)$ is the
angular spectrum of the $i$-th noise map.

More details about \smica\ are given in \citet{smicaIEEE}, as well as
in applications to the analysis of \textit{WMAP}
\citep{2005MNRAS.364.1185P} and Archeops data
\citep{2005A&A...436..785T}.  An application to the measurement of the
tensor to scalar ratio using CMB B-modes is discussed in
\citet{2009A&A...503..691B}.  Within the Planck collaboration,
\smica\ is used to define the \Plik\ high-$\ell$ likelihood
\citep{planck2013-p08}, but physical models of foreground emission are
used there instead of the non-parametric foreground model used here.
\smica\ is also used to cross-check the HFI calibration
(see~\citet{planck2013-p03f}).

\section{\ffp\ simulation results}
\label{sec:ffp6_appendix}

\begin{figure}
  \begin{center}
    \includegraphics[width=\columnwidth]{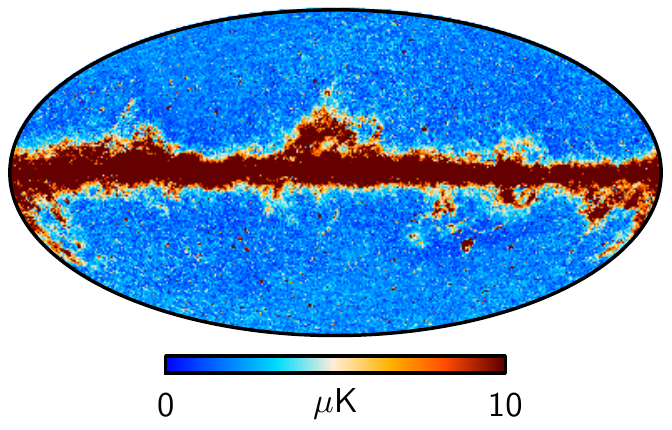}
  \end{center}
  \caption{Standard deviation of the four foreground-cleaned CMB maps
    from the \ffp\ simulation. All maps have been downgraded to a
    \healpix\ resolution of $\nside=128$.}
  \label{fig:ffp6_cmb_std}
\end{figure}

\begin{figure*}
  \begin{center}
    \includegraphics[width=\textwidth]{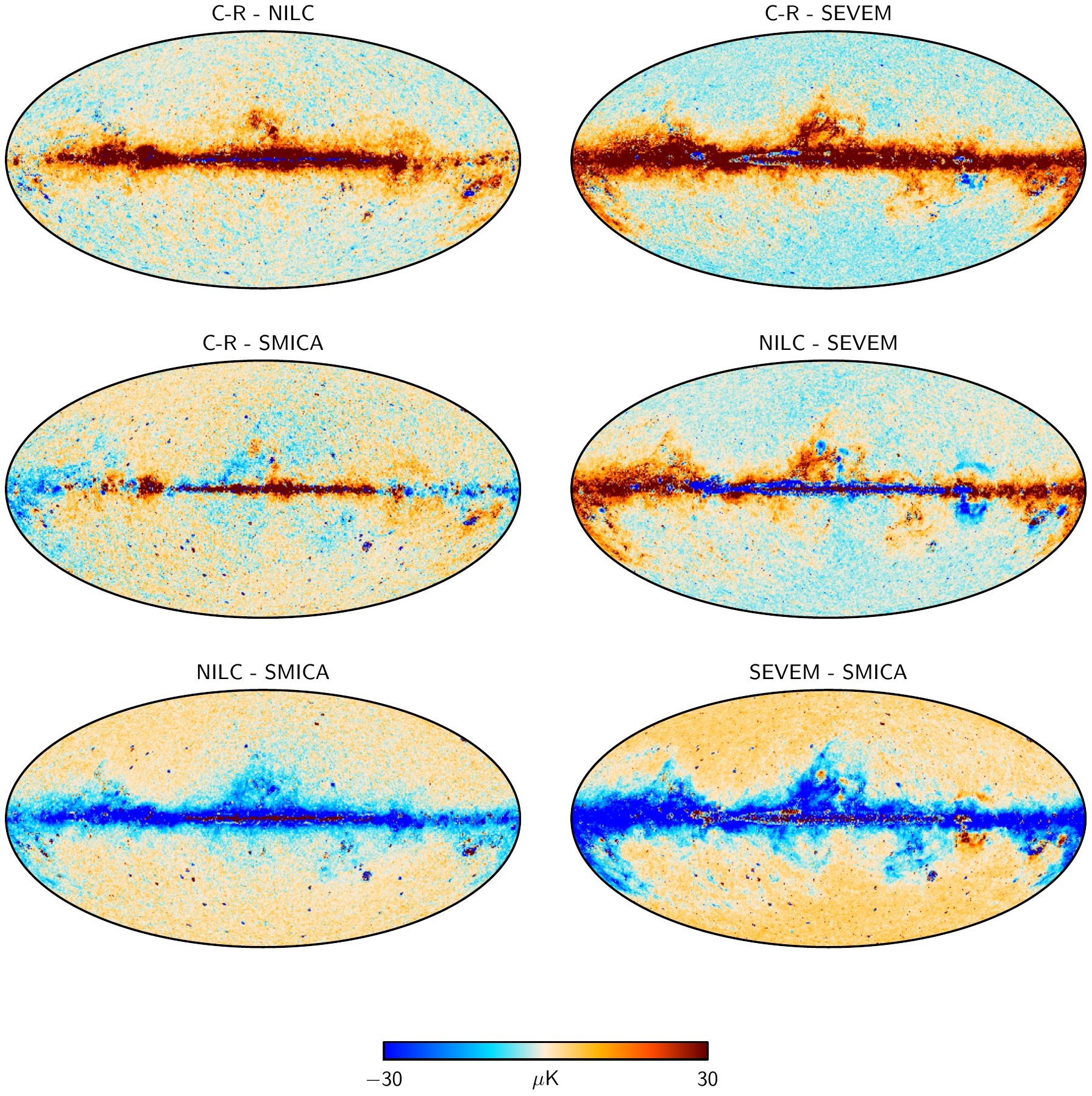}
  \end{center}
  \caption{Pairwise differences between foreground-cleaned CMB maps
    from the \ffp\ simulation. All maps have been downgraded to a
    \healpix\ resolution of $\nside=128$ to show the large-scale
    differences.}
  \label{fig:ffp6_cmb_diff}
\end{figure*}

\begin{figure}
  \begin{center}
    \includegraphics[width=\columnwidth]{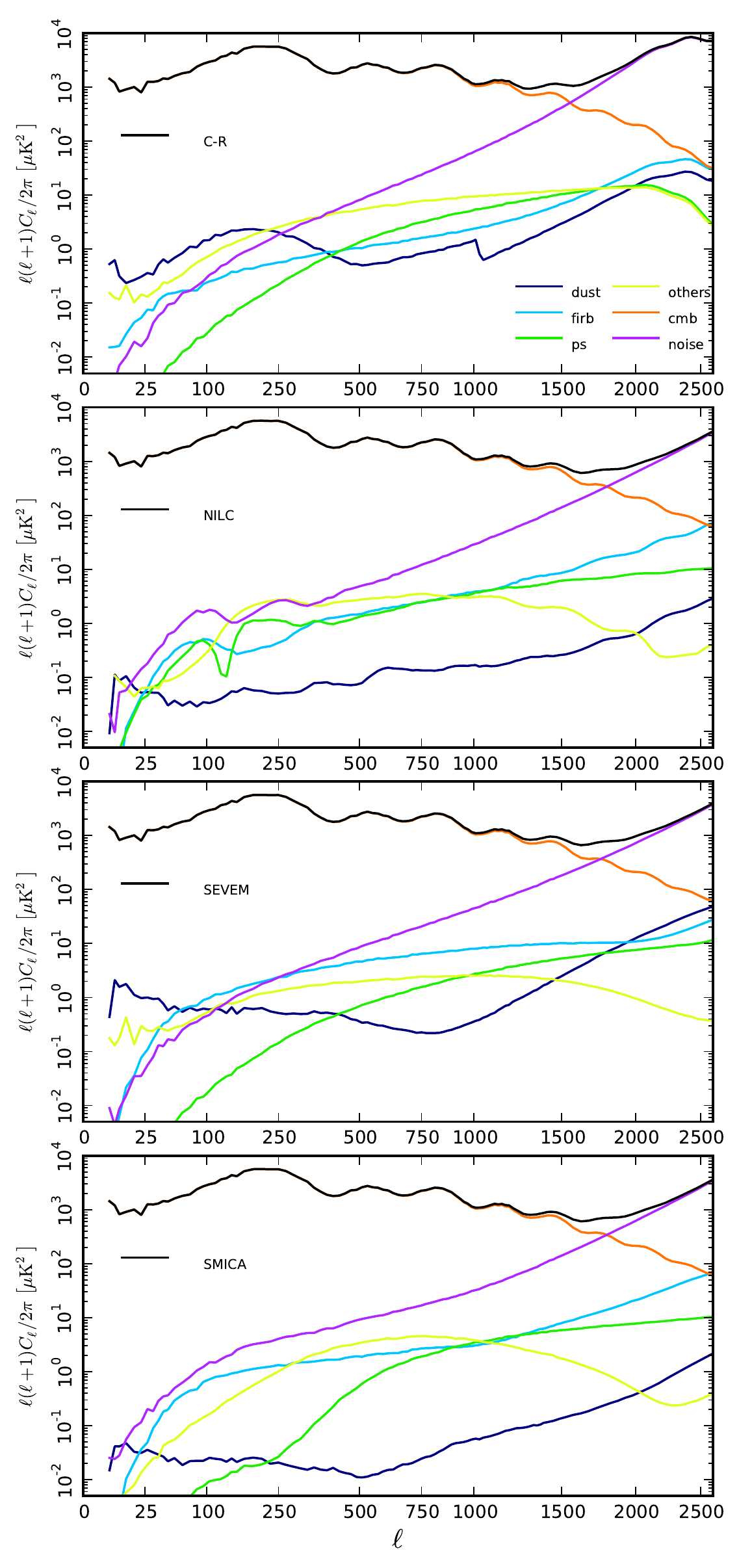}
  \end{center}
  \caption{Angular power spectra of residual foreground emission in
    the CMB maps from the \ffp\ simulation. The components shown are:
    thermal dust, cosmic infrared background fluctuations, point
    sources, CMB, noise, and the sum of all others.  From top to
    bottom, the panels show the results for \comrul, \nilc, \sevem,
    and \smica.}
  \label{fig:ffp6comp:decompcommon}
\end{figure}

\begin{figure}
  \begin{center}
    \includegraphics[width=\columnwidth]{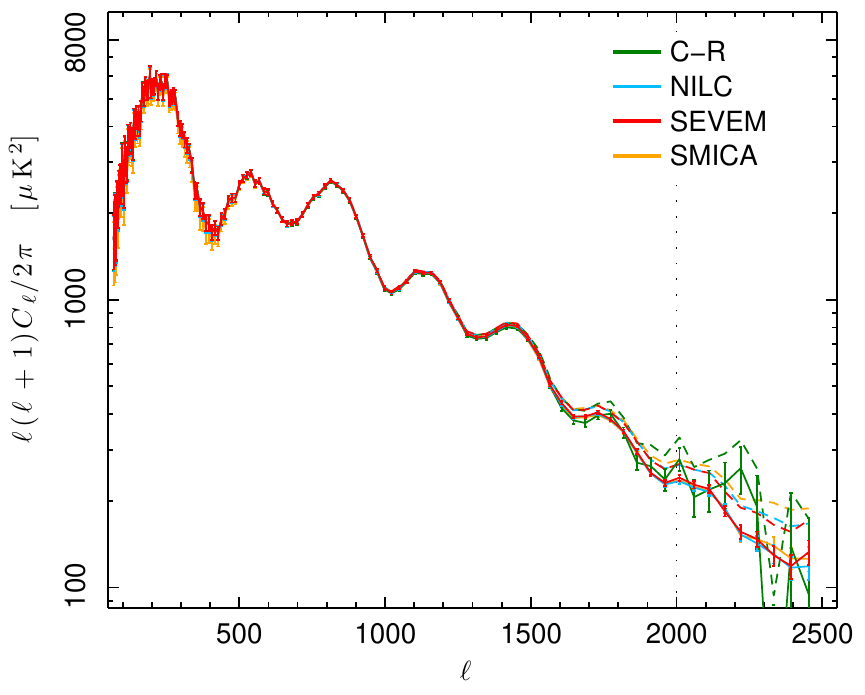}
  \end{center}
  \caption{Estimates of the CMB power spectra from the
    foreground-cleaned \ffp\ maps, computed by \XFaster.  The solid
    lines show the spectra after subtracting the best-fit model of
    residual foregrounds. The vertical dotted line shows the maximum
    multipole ($\ell = 2000$) used in the likelihood for fitting the
    foreground model and cosmological parameters (see
    Sect.~\ref{sec:xfaster_ffp6} for further details).  The dashed
    lines show the spectra before residual foreground subtraction.}
  \label{cb}
\end{figure}

\begin{figure*}
  \begin{center}
    \includegraphics[width=\textwidth]{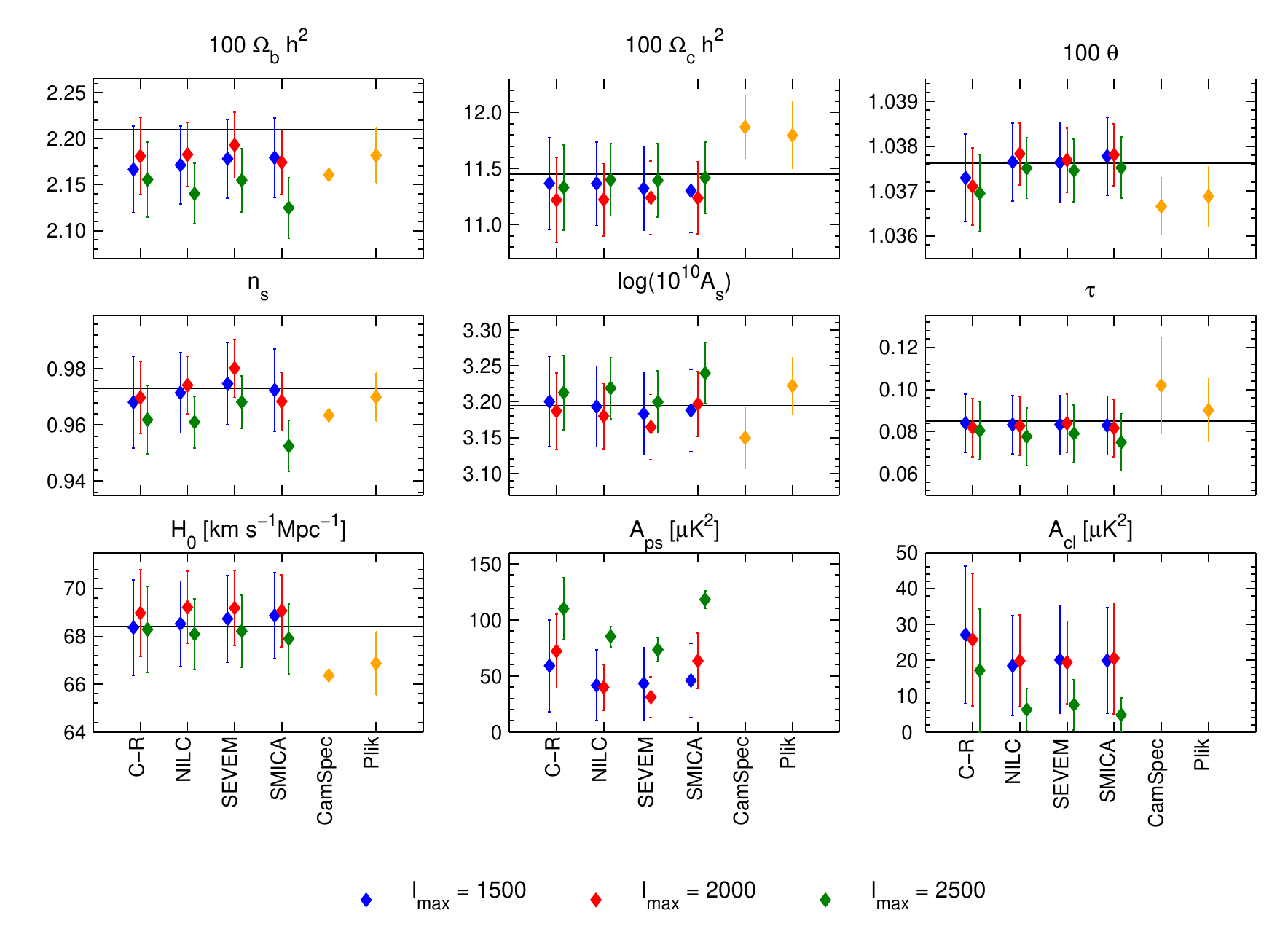}
  \end{center}
  \caption{Comparison of cosmological parameters derived from the
    \ffp\ simulation using different methods. The parameters shown as
    blue, red and green points indicate results obtained with
    $\ell_{\textrm{max}} =$ 1500, 2000 and 2500, respectively, and the
    yellow points show the results derived by \CamSpec\ and
    \Plik\ using cross-spectra. The black horizontal lines mark the
    input parameter values. The values of the foreground parameters
    are not shown for \CamSpec\ or \Plik\ since they use a different
    model. The matter power spectrum pivot scale was
    $k_{\mathrm{pivot}} = 0.002$ for all likelihoods, except
    \CamSpec\ for which $k_{\mathrm{pivot}} = 0.05$ was used.}
  \label{par-summ}
\end{figure*}

\begin{figure}
  \begin{center}
    \includegraphics[width=\columnwidth]{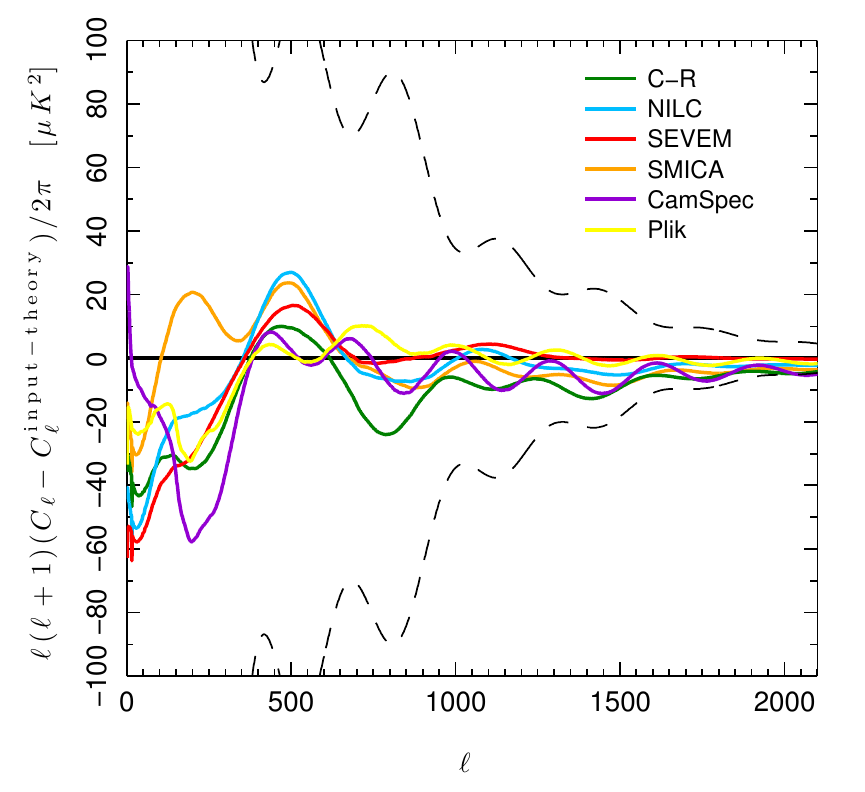}
  \end{center}
  \caption{Residuals of map-based and spectrum-based best-fit models
    relative to the \ffp\ simulation input \LCDM\ spectrum, shown for
    each algorithm up to $\ell_{\textrm{max}}=2000$.  Cosmic variance
    is shown as the black dashed line.}
  \label{bfm}
\end{figure}

We study the performance of the component separation algorithms on the
\ffp\ simulation described in Sect.~\ref{sec:data}, providing
additional information beyond that in the body of the paper.  Much of
the analysis presented here mirrors that shown for the data in
Sects.~\ref{sec:cmb_maps} and \ref{sec:powspec_params}, allowing a
direct comparison between the two analyses.

\subsection{CMB maps}

First, we show in Fig.~\ref{fig:ffp6_cmb_std} the standard deviation
between the four foreground-cleaned \ffp\ maps, similar to that shown
in Fig.~\ref{fig:cmb_rms} for the data.
Figure~\ref{fig:ffp6_cmb_diff} shows all pairwise differences between
the same maps, mirroring Fig.~\ref{fig:cmb_diff} for the data.  These
two plots highlight an important point concerning the \ffp\ analysis
already mentioned in Sect.~\ref{sec:cmb_maps}, namely that in
near-Galactic regions, where the foregrounds are important, the
internal differences between the four algorithms are larger in the
\ffp\ simulation than in the real data. This is due to the fact that
each component separation algorithm has been optimized with the real
data in terms of model definition, localization, etc. Then, the same
models have been used for the \ffp\ simulation without change. Only
the parameters within those models are refitted to the new data
set. This implies in fact that we expect each method to perform better
on the data than the simulations in terms of absolute residuals, to
the extent that the simulation matches the real sky. In other words,
the \ffp\ simulation provides a conservative estimate of the residual
errors in the real data.

\subsection{Power spectrum residuals from individual components}

In Fig.~\ref{fig:ffp6comp:decompcommon} we show the residual effect of
some of the individual components on the foreground-cleaned CMB map.
The thermal dust emission, CIB fluctuations, point sources, and noise
have been processed individually with each algorithm.  All other
components (free-free, synchrotron, spinning dust, CO, thermal SZ, and
kinetic SZ) are shown as a single, composite residual component.

\subsection{CMB power spectra and cosmological parameters}
\label{sec:xfaster_ffp6}

We assess the performance of our component separation techniques by
evaluating cosmological constraints from the foreground-cleaned CMB
maps derived from the \ffp\ simulation.

Figure~\ref{cb} shows the estimates of the angular power spectra of
the CMB maps.  Figure~\ref{par-summ} compares the cosmological
parameters derived from the four foreground-cleaned CMB maps,
\texttt{CamSpec}\footnote{For \texttt{CamSpec},
  $k_{\textrm{pivot}}=0.05$ was adopted for this test, while all
  others, input parameters and input CMB realization included, use
  $k_{\textrm{pivot}}=0.002$.}, and \texttt{Plik} to the input
(theoretical) parameters for different $\ell$-ranges. The parameter
space is defined by the same model applied to the real data in
Sect.~\ref{sec:powspec_params}, including six $\Lambda$CDM and two
foreground parameters.  All deviations from input parameters are small
and within $1\,\sigma$ up to $\ell=2000$, verifying that all methods
work well in this multipole range. However, for
$\ell_{\textrm{max}}=2500$ we start to see significant shifts, e.g.,
for $\Omega_{\textrm{b}}h^2$ and $n_{\textrm{s}}$. Further, the point
source foreground parameter, $A_{\textrm{ps}}$, reaches large values,
implying that assumptions concerning the high-$\ell$ foreground model
become important. For these reasons, we consider
$\ell_{\textrm{max}}=2000$ as the maximum recommended $\ell$-range for
these maps in the current data release.

Still, the overall agreement is excellent between all codes and all
$\ell$-ranges. In particular, we see that differences in the band power
spectra at high $\ell$ between the different codes are mostly absorbed
by the two-parameter foreground model.  For instance, the
\comrul\ band power spectrum has more power at high $\ell$ due to noise
or residual point sources, but this excess is well fitted by the
two-parameter foreground model, and mostly interpreted in terms of a
residual point source component; this is expected, given the lower
angular resolution of this map.

As mentioned above, $n_{s}$ and and $A$ are to some extent sensitive
to $\ell_{\textrm{max}}$. These parameters are degenerate with the
foreground parameters.  This may suggest that our $C_{\ell}$
foreground templates deviate more from the shape of the Poissonian and
clustered component in the CMB map.  This is a limitation of the
simple foreground templates used here.  To properly describe the
foreground residuals in the reconstructed maps, we should use a
foreground power spectrum template tailored to each method. For
instance, such templates may be constructed by processing simulated
foreground maps though each of the four pipelines. The templates are
then given by the pseudo-$C_{\ell}$ of each of the processed
foreground map.  However, our analysis shows that the current simple
model provides accurate results when restricting the analysis to
$\ell_{\textrm{max}}=2000$.

Figure~\ref{bfm} shows the best-fit power spectrum residuals for the
CMB map, \texttt{CamSpec} and \texttt{Plik} relative to the input CMB
$\Lambda$CDM model estimated up to $\ell=2000$. These plots show that
the residuals of the CMB map-based best-fit models are comparable to
the \texttt{CamSpec} and \texttt{Plik} residuals, and smaller than
$40\,\mu\textrm{K}^{2}$ for most of the $\ell$ range with larger
deviations observed for \texttt{CamSpec} at $\ell\sim 200$.  At higher
$\ell$s the residuals are smaller than $10\,\mu\textrm{K}^{2}$ for
both approaches, all showing similar trends. Thus, both the map- and
spectrum-based likelihoods recover input parameters reasonably well,
with the latter yielding slightly larger deviations from the best-fit
model of the input CMB realization.

\end{appendix}

\raggedright
\end{document}